\documentclass{article}
\usepackage[table,xcdraw]{xcolor}
\usepackage{iclr2025_conference,times}
\usepackage{amsmath,amsfonts,bm}

\def\eqref#1{equation~\ref{#1}}
\def\1{\bm{1}}

\DeclareMathAlphabet{\mathsfit}{\encodingdefault}{\sfdefault}{m}{sl}
\SetMathAlphabet{\mathsfit}{bold}{\encodingdefault}{\sfdefault}{bx}{n}

\iclrfinalcopy

\usepackage[utf8]{inputenc} % allow utf-8 input
\usepackage{fontawesome}
\usepackage[T1]{fontenc}    % use 8-bit T1 fonts
\usepackage{hyperref}       % hyperlinks
\usepackage{url}            % simple URL typesetting
\usepackage{booktabs}       % professional-quality tables
\usepackage{amsfonts}       % blackboard math symbols
\usepackage{nicefrac}       % compact symbols for 1/2, etc.
\usepackage{microtype}      % microtypography
\usepackage[most]{tcolorbox}
\usepackage{mdframed}
\usepackage{multirow}
\usepackage{graphicx}
\usepackage{enumitem}
\usepackage{tabularx}
\usepackage{rotating}
\usepackage{subfigure}
\usepackage{longtable}
\usepackage{caption}
\usepackage{makecell}
\usepackage{lscape}
\usepackage{tablefootnote}
\usepackage[toc,page,header]{appendix}

\newcommand{\titlebname}{\textsc{CodeMMLU}}

\definecolor{elegantcolor}{RGB}{0, 112, 192}
\newcommand*{\new}{\textcolor{black}}

\title{\texttt{\textcolor{elegantcolor}{\titlebname}}: A Multi-Task Benchmark for \\Assessing Code Understanding \& Reasoning \\Capabilities of CodeLLMs}

\author{%
	\bf Dung Nguyen Manh$^{\dagger}$\thanks{Corresponding authors: Dung Nguyen Manh (\texttt{dungnm31@fpt.com}) and Nghi D. Q. Bui (\texttt{bdqnghi@gmail.com}).}, 
            Thang Phan Chau$^{\dagger}$, 
            Nam Le Hai$^{\ddagger}$, 
            Thong T. Doan$^{\dagger}$,
            Nam V. Nguyen$^{\dagger}$, \\
        \bf
            Quang Pham$^{\diamondsuit}$\thanks{The author contributed to this work while working with FPT Software AI Center.},
            Nghi D. Q. Bui$^{\dagger *}$
        \\
	{$^{\dagger}$FPT Software AI Center, Viet Nam}, \\
	{$^{\ddagger}$Hanoi University of Science and Technology}, \\
        {$^{\diamondsuit}$Independent Researcher}\\
}

\begin{document}

% \input{abstract}

%%%%%%%%%%%%%%%%%%%%%%%%%%%%%%

\maketitle
\begin{abstract}
% Recent advancements in Code Large Language Models (CodeLLMs) have predominantly focused on open-ended code generation tasks, often neglecting the critical aspect of code understanding and comprehension. To bridge this gap, we present CodeMMLU, a comprehensive multiple-choice question-answer benchmark designed to evaluate the depth of software and code understanding in LLMs. CodeMMLU includes nearly 20,000 questions sourced from diverse domains, encompassing tasks such as code analysis, defect detection, and software engineering principles across multiple programming languages. Unlike traditional benchmarks, CodeMMLU assesses models’ ability to reason about program for concrete tasks such as code repair, reasoning about execution results, fill in the blank, etc. rather than merely generate code, providing deeper insights into their grasp of complex software concepts and systems. Our extensive evaluation reveals that even state-of-the-art models face significant challenges with CodeMMLU, highlighting deficiencies in comprehension beyond code generation. By underscoring the crucial relationship between code understanding and effective generation, CodeMMLU serves as a vital resource for advancing AI-assisted software development, ultimately aiming to create more reliable and capable coding assistants.

Recent advances in Code Large Language Models (CodeLLMs) have primarily focused on open-ended code generation, often overlooking the crucial aspect of \textbf{code understanding \& reasoning}. To bridge this gap, we introduce CodeMMLU, a comprehensive multiple-choice benchmark designed to evaluate the depth of software and code comprehension in LLMs. CodeMMLU includes nearly 20,000 questions spanning diverse domains, including code analysis, defect detection, and software engineering principles across multiple programming languages. Unlike traditional benchmarks that emphasize code generation, CodeMMLU assesses a model’s ability to reason about programs across a wide-range of tasks such as code repair, execution reasoning, and fill-in-the-blank challenges. Our extensive evaluation reveals that even state-of-the-art models struggle with CodeMMLU, highlighting significant gaps in comprehension beyond generation. By emphasizing the essential connection between code understanding and effective AI-assisted development, CodeMMLU provides a critical resource for advancing more reliable and capable coding assistants. CodeMMLU is publicly available at:  \href{https://github.com/FSoft-AI4Code/CodeMMLU}{\faGithub \hspace{0.5mm}CodeMMLU}

%Our experiments reveal significant insights into model performance, highlighting the influence of factors such as model size, family, and prompting strategies. 

% Notably, we observe that while the GPT-4 model consistently achieves the highest average performance, there are noticeable inconsistencies related to model architecture and training data. These findings underscore the importance of robust, diverse benchmarks like CodeMMLU in advancing the evaluation and development of CodeLLMs.
\end{abstract}
\section{Introduction}

% \todo{What capabilities does CodeLLMs shows? and why it is urgent to have such a benchmark for the mentioned task? - CodeLLMs shows capabilities through coding application.}

% \todo{What and why does the existing benchmarks are not completely showcase CodeLLMs ability? Benchmark often pose a potential to data leakage, how can CodeMMLU overcome this}

% \improve{The uncertainty of execution-based benchmark are raising a question of is whether CodeLLM comprehen their solution completely, or it remmber a seen solution from elsewhere. In that case, we grounded the answer probability to MCQs format to force the LLM to comprehen the given solution. On the other hand, MCQs format also offers to scale the evaluation to massive.}

Recent advancements in Code Large Language Models (CodeLLMs)~\citep{wang2021codet5,wang2023codet5+, feng2020codebert, allal2023santacoder, li2023starcoder, lozhkov2024starcoder, guo2024deepseek, pinnaparaju2024stable, zheng2024opencodeinterpreter, roziere2023code, nijkamp2022codegen, luo2023wizardcoder, xu2022systematic, bui2023codetf, hui2024qwen2, bui2022detect, dau2024xmainframe, dau2024docchecker} have demonstrated impressive capabilities across various software engineering (SE) tasks~\citep{bui2022detect,dau2024docchecker,to2023better,white2024chatgpt,sobania2023analysis,phan2024hyperagent, sun2023automatic, nguyen2022hierarchynet, bui2019towards, zhang2022overwatch, wang2023deepvd}. However, existing benchmarks often fall short to provide rigorous and reliable evaluations, largely due to outdated methodologies and the risk of data leakage \citep{matton2024leakage}.
Moreover, practical applications of CodeLLMs reveal limitations such as bias and hallucination \citep{rahman2024code, liu2024exploring} that current benchmarks fail to adequately address. 

The predominant focus of coding-related benchmarks has been on open-ended, free-form generation tasks, such as code generation/code completion \citep{iyer2018mapping, lu2021codexglue, chen2021evaluating, austin2021program, lai2023ds, hendrycksapps2021, DBLP:conf/nips/DingWADTJRNBRX23, zhuo2024bigcodebench} and other SE tasks like program repair \cite{ouyang2024benchmarking, xia2023automated} (Table \ref{tab:benchmark_comparison}). While appealing, these benchmarks struggle to discern whether CodeLLMs truly understand code or merely reproduce memorized training data \citep{carlini2022quantifying, nasr2023scalable}. Additionally, the reliance on test cases and executability for evaluation limits the quantity and diversity of these benchmarks across domains, potentially leading to biased and limited generalizations.
Recent efforts to improve evaluation through free-form question answering \citep{liu2021codeqa, li2024inficoder} have introduced new challenges, often requiring less rigorous metrics or LLM-as-a-judge approaches \citep{zheng2023judging}. However, LLMs-as-a-judge methods are susceptible to adversarial attacks \citep{raina2024llm}, raising concerns about the reliability of such evaluation pipelines for coding tasks.

To address the aforementioned shortcomings, we introduce CodeMMLU, a novel benchmark designed to evaluate CodeLLMs' ability to comprehend and reason about code through multiple-choice question answering (MCQ). This approach enables a deeper assessment of how CodeLLMs grasp coding concepts, moving beyond the mere generation capabilities. Inspired by the MMLU dataset \citep{hendrycks2020measuring} from natural language understanding, CodeMMLU offers a robust and easy evaluation with the following key features. CodeMMLU comprises nearly 20,000 questions, facilitating a robust and comprehensive evaluation. Its large-scale data curation process mitigates potential biases and improves statistical reliability in measuring CodeLLMs’ performance across many capabilities. CodeMMLU covers over 50 software engineering disciplines and more than 10 programming languages, providing a holistic evaluation of CodeLLMs. The MCQ format is highly scalable, allowing for an accurate and straightforward assessment using precision-based metrics. Moreover, by incorporating permutations of answer choices, CodeMMLU emphasizes the model's code understanding capabilities rather than memorizing the training datasets. Consequently, CodeMMLU provides a robust and accurate assessment of the models capabilities to understand software tasks.
\if0
\begin{itemize}[leftmargin=*]
\item \textbf{Comprehensiveness:} \new{CodeMMLU comprises nearly 20,000 questions, facilitating a robust and comprehensive evaluation. Its large scale and data curation process mitigate potential biases and enhances statistical reliability in measuring CodeLLMs’ performance across many capabilities.}
\item \textbf{Diversity:} \new{CodeMMLU covers over 50 software engineering topics, and more than 10 programming languages, allowing for a holistic evaluation of CodeLLMs.}
\item \textbf{Scalability and robustness:} \new{The MCQ format offers a highly scalable evaluation pipeline, allowing for accurate and straightforward assessment using precision-based metrics. Moreover, by incorporating permutations of answer choices, CodeMMLU emphasizes the model's code understanding capabilities rather than memorizing the training datasets. Consequently, CodeMMLU necessitates more sophisticated and resilient models to understand the software domains more comprehensively, pushing the boundaries of LLM robustness.}
\end{itemize}
\fi
\begin{figure}
    \centering
    \includegraphics[width=0.7\linewidth]{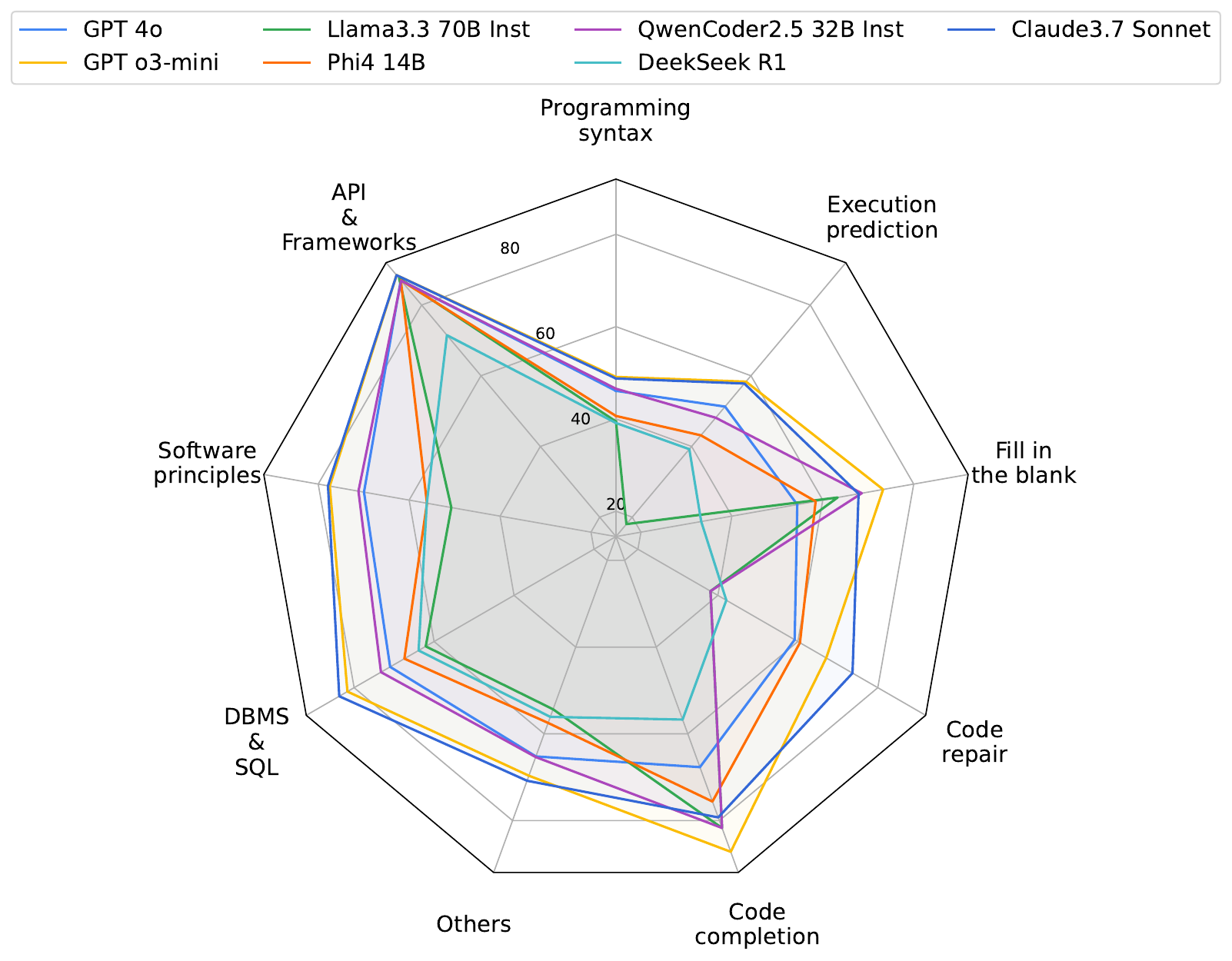}
    \caption{\textbf{Summary performance of LLMs on the CodeMMLU benchmark.} This radar chart presents the evaluation results (accuracy \%) of different models across various CodeMMLU tasks.}
    \label{fig:spiderfigure-model}
\end{figure}

CodeMMLU evaluates LLMs’ abilities in coding and software problem-solving from a fresh perspective, extending beyond conventional code generation and completion tasks. Our analysis uncovers several key insights: (1) previously unidentified bias issues in CodeLLMs, consistent with those observed in natural language MCQA tasks; (2) proprietary models (GPT-4o and Claude 3.5 Sonnet) consistently deliver the highest average performance; (3) among open-source models, the DeepSeek and Meta-Llama families achieve the greatest accuracy; (4) scaling laws tied to model size hold partially within the same model family but not across families, highlighting the critical roles of pre-training, post-training, and model architecture; (5) advanced prompting strategies, such as Chain-of-Thought (CoT), consistently impair performance, casting doubt on CodeLLMs’ reasoning capabilities for complex, multi-step tasks; and (6) when code completion benchmarks (e.g., HumanEval) are reframed from open-ended generation to MCQA format, LLMs exhibit reduced performance, questioning their true understanding of code. Notably, even strong reasoning models like DeepSeek-R1 underperform on CodeMMLU, suggesting significant room for improvement in future iterations. These findings highlight the shortcomings of CodeLLMs in truly understanding code and reveal several promising future research avenues.

% Moreover, the performance and advancements in prompting techniques (such as chain-of-thought) show limited improvement, raising concerns about potential hallucinations in the reasoning abilities of these models. These results not only highlight the current state of CodeLLMs but also underscore the complex interplay between model design, training data, and evaluation methodologies in determining model performance on software-related tasks.
In summary, this work makes the following contributions:
\begin{enumerate}[leftmargin=*]
    \item We present the first MCQ benchmark for software and coding-related tasks, addressing the need for a comprehensive and large-scale evaluation in the code domain. CodeMMLU enables the evaluation of LLMs' alignment with human inference in the software knowledge domain, similar to advances in the NLP field.
    % \item We present CodeMMLU, the first multiple-choice question (MCQ) benchmark tailored for software and coding-related knowledge. CodeMMLU addresses the need for diverse and rigorous evaluation scenarios in the code domain, enabling the assessment of LLMs’ alignment with human-like reasoning in software-related tasks, similar to advancements achieved in the NLP domain.
    
    \item CodeMMLU provides a thorough assessment of LLM capabilities, ensuring a substantial number of samples and the diversity across tasks, domains, and languages. This enables a more nuanced understanding of an LLM's strengths and weaknesses, facilitating the development of models better aligned with the complexities and demands of the software domain.
    % \item CodeMMLU offers a robust assessment of LLM capabilities by providing a large-scale dataset with nearly 20,000 questions. It ensures diversity across tasks, domains, and programming languages, allowing for a nuanced evaluation of an LLM’s strengths and weaknesses. This facilitates the development of models that better address the complexities and demands of real-world software engineering scenarios.
    
    % \item Our experiments offer critical insights into LLM performance, highlighting the impact of factors such as model size, model family, and prompting techniques. \new{Through our extensive experiment, LLM also indicates a performance gap between their code generation and comprehension ability when dealing with software-related task. We also find out the sensitivity of LLM when dealing with selection bias introduce by MCQ format} This provides essential information to the community on effectively utilizing and improving LLMs for robustness and specific tasks and domains in software engineering.

    \item Our experiments offer critical insights into LLM performance, highlighting the impact of factors such as model size, model family, and prompting techniques. Notably, CodeMMLU unveils a performance gap between LLMs’ code generation and comprehension abilities. Additionally, we identify LLMs' sensitivity to the selection biases introduced by the MCQ format. These findings provide valuable guidance for the community to enhance the robustness, adaptability, reliability, and domain-specific capabilities of LLMs in real-world software engineering.
\end{enumerate}

\section{Related Work}
\paragraph{Code Intelligence Benchmarks.} The rapid development of Large Language Models (LLMs) for code-related tasks necessitates the development of diverse benchmarks to evaluate their performance. Algorithm-focused benchmarks, such as HumanEval \citep{chen2021evaluating}, MBPP \citep{austin2021program}, and their extended versions (e.g., HumanEval+, MultiPL, MBPP+) \citep{liu2024your}, focus on small-scale code generation tasks but lack the depth needed to assess broader comprehension. More challenging tasks, such as those in CodeContests \citep{li2022competition} and LiveCodeBench \citep{jain2024livecodebench}, provide competitive programming problems but remain primarily generative (Table \ref{tab:benchmark_comparison}). Comprehensive evaluation frameworks, such as CodeXGLUE \citep{lu2021codexglue}, XLCoST \citep{zhu2022xlcost}, and XCodeEval \citep{khan2023xcodeeval}, provide versatility through multi-task assessments. However, these benchmarks are either dependent on metrics like BLEU and ROUGE, or testcase execution, which limits their reliability and scalability for a large-scale, comprehensive evaluation of LLM.

In contrast, multiple-choice question (MCQ) benchmarks offer a more standardized, scalable, and reliable evaluation method, as demonstrated in popular general-purpose benchmarks such as MMLU \citep{hendrycks2020measuring} and TruthfulQA \citep{lin2022truthfulqameasuringmodelsmimic}. Although MCQs facilitate large-scale assessments, recent studies highlight their susceptibility to biases, such as sensitivity to the answer choice orders~\citep{wang2023largelanguagemodelsfair, robinson2023leveraginglargelanguagemodels}. Existing MCQ benchmarks also lack focus on software engineering, limiting their applicability to code-related evaluations. In contrast, we curated data from a wide range of tasks and applied various filtering and debiasing techniques to improve CodeMMLU's comprehensiveness, while minimizing data leakage and biases.
% \paragraph{Benchmarks for Code Generation \& Understanding}
% The development of Large Language Models (LLMs) for code-related tasks has been accompanied by the creation of diverse benchmark datasets. These benchmarks span a wide range of programming challenges, from basic algorithms to complex software development scenarios. Algorithm-focused benchmarks include HumanEval \citep{chen2021evaluating} and MBPP \citep{austin2021program}, along with their extended versions HumanEval+, MultiPL, and MBPP+ \citep{liu2024your}. 

% More advanced algorithmic tasks are represented by CodeContests \citep{li2022competition} and LiveCodeBench \citep{jain2024livecodebench}, which draw from competitive programming problems. Specialized benchmarks like DS-1000 \citep{lai2023ds} target data manipulation and analysis tasks, while MathQA-Python \citep{austin2021program} focuses on mathematical problem-solving in Python. Repository-level benchmarks such as RepoBench \citep{liu2023repobench}, RepoEval \citep{zhang2023repocoder}, and SWE-Bench \citep{jimenez2023swe} simulate real-world software development scenarios. Comprehensive evaluation frameworks like XCodeEval \citep{khan2023xcodeeval}, CRUXEval \citep{gu2024cruxeval}, and CodeXGLUE \citep{lu2021codexglue} assess LLMs across multiple dimensions of software development, providing a holistic view of model capabilities. \todo{Shrink the citation, summary their strength, detail list their drawback and what motivated CodeMMLU. This part can merge into MCQs benchmarks. Edit the comparison table}

\paragraph{Understanding \& Reasoning on Code}
There is a large body of research leveraging AI models for reasoning about code \cite{gu2024cruxeval,chen2024reasoning,liu2025tool,dehghan2024assessing,le2024visualcoder,le2024learning,bieber2020learning,shi2019learning,li2021learning}. Reasoning about code requires AI models to comprehend both its syntactic and semantic aspects. In the early days, static analysis was the primary approach to reasoning about code. With the rise of deep learning, methods that model code structures~\cite{mou2016convolutional,bui2021treecaps,bui2021infercode} emerged as promising alternatives. Subsequently, graph-based representations of code have gained traction, enabling reasoning about program properties and execution using graph neural networks (GNNs)~\cite{le2024learning,bieber2020learning,shi2019learning}. 
With the advent of large language models (LLMs), these models have been directly applied to reasoning about program behavior in downstream tasks~\cite{gu2024cruxeval,chen2024reasoning,liu2025tool,dehghan2024assessing,le2024visualcoder}. CodeMMLU provides a comprehensive benchmark that reflects real-world reasoning requirements by offering ground-truth answers to programming-related questions, allowing LLMs to derive final solutions through various reasoning paths.

% Research into modeling programmer behavior and cognitive processes has been ongoing since the early days of software development \citep{1421034, article, 8453126}. Cognitive models aim to describe the mental structures and processes involved in programming, encompassing knowledge, concepts, and techniques used during comprehension and problem-solving. \cite{article} introduced a multi-level model of cognitive structures, distinguishing between semantic and syntactic knowledge. Semantic knowledge includes programming concepts and techniques (e.g., dynamic programming, recursion, sorting methods), while syntactic knowledge relates to programming language grammar (e.g., iteration formats, conditional statements, library functions). To measure programmer comprehension in terms of cognitive processes, \cite{article} proposed five core programming tasks: composition, comprehension, debugging, modification, and learning. This model architecture addresses two crucial questions in programmer comprehension: (1) what knowledge is available to programmers, and (2) what processes do programmers undergo during solution design.

\vspace{0.5cm}
\vspace{-5mm}
\begin{table}[htbp]
  \centering
  \caption{\textbf{Comparison between common code understanding benchmarks for LLMs in terms of coverage of foundation tasks of programming comprehension model.}}
  \scalebox{0.7}{%
    \begin{tabular}{l|cc|cccc|c}
    \toprule
    \multicolumn{1}{c}{\multirow{2}[2]{*}{\textbf{Benchmark}}} & \multicolumn{2}{c}{\textbf{Question}} & \multicolumn{4}{c}{\textbf{Programming Task}} & \multirow{2}[2]{*}{\textbf{Test size}} \\
    \multicolumn{1}{c}{} & \textit{Open-end} & \multicolumn{1}{c}{\textit{MCQ}} & \multicolumn{1}{p{3.5em}}{\textit{SWE knowledge}} & \multicolumn{1}{p{3.585em}}{\textit{Code composition}} & \multicolumn{1}{p{3.415em}}{\textit{Code comprehension}} & \multicolumn{1}{p{3.335em}}{\textit{Code debugging}} &  \\
    \midrule
    \midrule
    APPS \cite{hendrycksapps2021} & \checkmark &       &       & \checkmark &       &       & 5000 \\
    MBPP \cite{austin2021program} & \checkmark &       &       & \checkmark &       &       & 974 \\
    HumanEval \cite{chen2021evaluating} & \checkmark &       &       & \checkmark &       &       & 164 \\
    CRUXEval \cite{gu2024cruxevalbenchmarkcodereasoning} & \checkmark &       &       &       & \checkmark &       & 800 \\
    LiveCodeBench \cite{jain2024livecodebench} & \checkmark &       &       & \checkmark & \checkmark & \checkmark & 880\tablefootnote{https://github.com/LiveCodeBench/LiveCodeBench} \\
    CodeApex \cite{Fu2023CodeApexAB} & \checkmark & \checkmark & \checkmark & \checkmark &       & \checkmark & 2.056 \\
    \midrule
    \textbf{CodeMMLU} &       & \textbf{\checkmark} & \textbf{\checkmark} & \textbf{\checkmark} & \textbf{\checkmark} & \textbf{\checkmark} & \textbf{19.912} \\
    \bottomrule
    \end{tabular}%
  \label{tab:benchmark_comparison}%
  }
\end{table}%

% \paragraph{Multi-task question answering benchmarks}

% Multiple Choice Questions (MCQs) have emerged as a powerful tool for evaluating the capabilities of Large Language Models (LLMs) across various domains. Recent trends in MCQ-based benchmarks focus on testing advanced reasoning, domain-specific knowledge, and robustness in LLMs, particularly as their capabilities continue to expand \citep{hendrycks2020measuring, lin2022truthfulqameasuringmodelsmimic, zellers2019hellaswagmachinereallyfinish, talmor2019commonsenseqaquestionansweringchallenge}. The benefits of MCQs extend beyond enabling large-scale evaluation; they also provide highly reliable results. Compared to open-ended assessment methods \citep{chiang2024chatbotarenaopenplatform} that heavily rely on LLM judges or human annotation, MCQs enhance reliability by grounding the knowledge, problem context, and defining possible answers.
% Despite their convenience, recent studies have revealed that LLMs display significant sensitivity to the order of answer options in MCQs \citep{wang2023largelanguagemodelsfair, robinson2023leveraginglargelanguagemodels}. This finding underscores the need for appropriate debiasing methods in MCQ benchmarks to address LLM selection bias \citep{zheng2024largelanguagemodelsrobust, inproceedingsMCQ}. By implementing such methods, researchers can ensure more accurate and fair assessments of LLM performance.

% \todo{Code Foundation Models}
\section{CodeMMLU: Data Curation}

% The CodeMMLU benchmark is constructed to assess large language models' (LLMs) comprehension of programming tasks. We are inspired by programmer comprehension behavior models and integrate multi-leveled cognitive structures and processes to measure the understandability of LLMs on software problems \citep{article}. CodeMMLU is divided into two primary categories: (i) knowledge-based test sets containing syntactic and semantic tasks, and (ii) real-world programming problems. The overall CodeMMLU structure, as presented in Figure \ref{fig:overview}, includes distinct approaches for data collection, filtering, and validation in both test sets.

\new{The CodeMMLU benchmark is structured into two primary categories: (i) knowledge-based tests, designed to evaluate programming knowledge through questions addressing both syntactic and semantic aspects, and (ii) fundamental coding-skill tests, created by transforming high-quality codebase seeds into task-specific challenges. CodeMMLU includes nearly 20,000 questions spanning 52 diverse topics (Table \ref{tab:summaryCodeMMLU}). We design the knowledge-based tests to probe multi-level cognitive structures, assessing an LLM’s understanding of software knowledge at both semantic and syntactic levels. In contrast, the fundamental coding-skill test sets align with the cognitive process model of \citep{article}, focusing on core programming tasks that mimic real-world problem-solving scenarios.}

\subsection{Knowledge-based task creation}
\label{sub:knowledge-task}

The knowledge-based test sets are designed to cover a wide range of topics and follow the multi-level cognitive structures model \citep{article} which combines syntactic and semantic knowledge. The subset target is to measure the LLM's coding capability and comprehensibleness of programming concepts. We collected raw programming-related \new{questions and their corresponding multiple-choices answer from W3School \citep{w3schools-no-date} and Common Crawl project\footnote{https://commoncrawl.org/} (See more license detail in Appendix \ref{appendix:license}). The knowledge-based test set include:}
% \todo{The foreshadow of diverse task is unnecessary. Directly into the task}
% \quang{Some sentences to connect the syntactic and semantic sets to the previous paragraph.}
\begin{itemize}[leftmargin=*]
    \item \textbf{Syntactic subset.} Focused on programming language grammar and structural correctness, such as condition statement, format of iteration, common library usage.
    \item \textbf{Semantic subset.} Targeted more abstract programming concepts, such as algorithms, data structures, object-oriented principles.
\end{itemize}

We maintain a high-quality evaluation set by filtering the raw data that undergoes a rigorous formatting and deep-learning-based filter in which we remove any instances that do not meet our quality criteria (see in section \ref{sub:cleaning} and Appendix \ref{appendix:filtering}). \new{Resulting in an evaluation set (Table \ref{tab:summaryCodeMMLU}) that contains more than 11,000 instances, lying in 52 topics classified to 5 main subjects (categorized by source tag).}

% \paragraph{}
% Knowledge subset creation process involves collecting a diverse set of programming-related multiple-choices question answering (MCQ) data from reputable online sources (Figure \ref{fig:crawling}). We specifically target platforms known for their high-quality content, namely GeeksforGeeks, W3Schools, and Sanfoundry \cite{geeksforgeeks-no-date, w3schools-no-date, sanfoundry-no-date}. We maintain a high-quality evaluation set by filtering the raw data undergoes a rigorous formatting and deep-learning based filter in which we remove any instances that do not meet our quality criteria (detail criteria description in Appendix \ref{appendix:filter1}).

\vspace{0.2cm}
% \usepackage{booktabs}
% \usepackage{multirow}
% \usepackage{graphicx}
\definecolor{Mymint}{HTML}{E4E7EB}
\vspace{-5mm}
\begin{table}[ht]
\centering
\caption{\textbf{Summary of CodeMMLU Subject Categories and Task Distribution.}}
\resizebox{\columnwidth}{!}{%
\begin{tabular}{@{}cllcc@{}}
\toprule
 & \multicolumn{1}{c}{\textbf{Subject}} & \multicolumn{1}{c}{\textbf{Topic}} & \textbf{Source} & \textbf{Testsize} \\ \midrule
 \midrule
\multicolumn{1}{c|}{\multirow{5}{*}{\textit{\begin{sideways}\begin{tabular}[c]{@{}c@{}}Syntactic\\ knowledge\end{tabular}\end{sideways}}}} & \multicolumn{1}{l|}{\multirow{3}{*}{API \& Frameworks usage}} & \multicolumn{1}{l|}{Jquery, Django,   Pandas, Numpy, Scipy,} & \multicolumn{1}{c|}{\multirow{16}{*}{\rotatebox{90}{\makecell{W3Schools, Geeksforgeeks, \\ \new{CommonCrawl}}}}} & \multirow{3}{*}{740} \\
\multicolumn{1}{c|}{} & \multicolumn{1}{l|}{} & \multicolumn{1}{l|}{Azure, Git, AWS, svg, xml,} & \multicolumn{1}{c|}{} &  \\
\multicolumn{1}{c|}{} & \multicolumn{1}{l|}{} & \multicolumn{1}{l|}{Bootstrap, NodeJS, AngularJS, React, Vue.} & \multicolumn{1}{c|}{} &  \\
\multicolumn{1}{c|}{} & \multicolumn{1}{l|}{\cellcolor{Mymint!50}Programming language syntax} & \multicolumn{1}{l|}{\cellcolor{Mymint!50}\begin{tabular}[c]{@{}l@{}}C, C\#, C++, Java, Javascript, PHP, Python, \\ R, Ruby, MatLab, HTML, CSS, TypeScript.\end{tabular}} & \multicolumn{1}{c|}{} & 6,220 \\ \cmidrule(r){1-3} \cmidrule(l){5-5} 
\multicolumn{1}{c|}{\multirow{10}{*}{\rotatebox{90}{\textit{Semantic knowledge}}}} & \multicolumn{1}{l|}{DBMS \& SQL} & \multicolumn{1}{l|}{DBMS, MySQL,   PostgreSQL, SQL.} & \multicolumn{1}{c|}{} & 393 \\
\multicolumn{1}{c|}{} & \multicolumn{1}{l|}{} & \multicolumn{1}{l|}{\cellcolor{Mymint!50}Data structure \& Algorithm,} & \multicolumn{1}{c|}{} & \multirow{6}{*}{3,246} \\
\multicolumn{1}{c|}{} & \multicolumn{1}{l|}{} & \multicolumn{1}{l|}{\cellcolor{Mymint!50}Object-oriented programming,} & \multicolumn{1}{c|}{} &  \\
\multicolumn{1}{c|}{} & \multicolumn{1}{l|}{\cellcolor{Mymint!50}} & \multicolumn{1}{l|}{\cellcolor{Mymint!50}Compiler design,} & \multicolumn{1}{c|}{} &  \\
\multicolumn{1}{c|}{} & \multicolumn{1}{l|}{\cellcolor{Mymint!50}} & \multicolumn{1}{l|}{\cellcolor{Mymint!50}Computer organization and Architecture,} & \multicolumn{1}{c|}{} &  \\
\multicolumn{1}{c|}{} & \multicolumn{1}{l|}{} & \multicolumn{1}{l|}{\cellcolor{Mymint!50}Software Development \& Engineering,} & \multicolumn{1}{c|}{} &  \\
\multicolumn{1}{c|}{} & \multicolumn{1}{l|}{\multirow{-6}{*}{Software principles}} & \multicolumn{1}{l|}{\cellcolor{Mymint!50}System Design.} & \multicolumn{1}{c|}{} &  \\
\multicolumn{1}{c|}{} & \multicolumn{1}{l|}{Others} & \multicolumn{1}{l|}{\begin{tabular}[c]{@{}l@{}}Program accessibility, Computer networks, \\ Computer science, Cybersecurity, Linux, \\ Web technologies, AWS.\end{tabular}} & \multicolumn{1}{c|}{} & 1,308 \\ \midrule
\multicolumn{1}{c|}{\multirow{4}{*}{\textit{\begin{sideways}\begin{tabular}[c]{@{}c@{}}\new{Fundamen}\\\new{-tal task}\end{tabular}\end{sideways}}}} & \multicolumn{2}{c|}{Code completion} & \multicolumn{1}{l|}{HumanEval} & 163 \\
\multicolumn{1}{c|}{} & \multicolumn{2}{c|}{Fill in the blank} & \multicolumn{1}{l|}{LeetCode} & 2,129 \\
\multicolumn{1}{c|}{} & \multicolumn{2}{c|}{Code repair} & \multicolumn{1}{l|}{QuixBugs} & 76 \\
\multicolumn{1}{c|}{} & \multicolumn{2}{c|}{\new{Execution Prediction}} & \multicolumn{1}{l|}{IBM CodeNet} & 6,006 \\
% \multicolumn{1}{c|}{} & \multicolumn{2}{c|}{Input/output prediction} & \multicolumn{1}{l|}{LeetCode} & 2,313 \\ 
\bottomrule
\end{tabular}%
}
\label{tab:summaryCodeMMLU}%
\end{table}

% The final knowledge evaluation set involves a manual review for ensuring the question not only meets our quality standards (Appendix \ref{appendix:filter2}) but is also challenging enough to provide a meaningful assessment of LLM capabilities. Resulting a $D_{knowledge}$ set contains approximately 11,000 instances, lie in 52 topics classified to 5 main subjects (Table \ref{tab:summaryCodeMMLU}). 

\subsection{Fundamental Test Construction}
\label{sub:funda-task}
Our benchmark encompasses \new{four} distinct MCQ programming tasks designed to assess the foundational capabilities outlined in the cognitive process model of programmer comprehension, namely: composition, comprehension, debugging, and modification.

% \todo{This is four tasks, not five. CodeMMLU fundamental benchmark cover core diverse task to provide a complete vision of cognitive capabilities} 

% These tasks cover the core capabilities that any cognitive model of programmer behavior must address: composition, comprehension, debugging, and modification.

\textbf{Code Completion} evaluates a model's composition ability by requiring it to complete partially written code based on provided requirements. We adapted HumanEval \citep{chen2021evaluating}, originally designed for code generation, into an MCQ format. From its 164 unique programming problems, we employed Large Language Models (LLMs) to generate plausible but incorrect solutions as distractors. All options, including correct solutions migrated from HumanEval and generated incorrect ones, were tested for executability. Some incorrect solutions were designed to pass certain test cases but fail others, adding complexity and challenging models to distinguish between correct and nearly-correct solutions based on semantic and syntactic understanding. 
% \todo{This might need to dig into detail in Appendix. Report number of fail sample. The coverage of distractor.}

\textbf{Code Repair} assesses a model's debugging capability by requiring it to identify and fix errors in provided code snippets. We built this task upon QuixBugs \citep{Lin2017QuixBugsAM}, which was originally designed for debugging algorithmic programs. We used a "diff" operation on buggy and corrected versions in QuixBugs (Python and Java) to identify specific fixes, which served as correct solutions. To create plausible distractors, we targeted components frequently involved in bugs (e.g., return statements, loop conditions, if/else/switch expressions) and guided LLMs to generate alternative fixes. These alternatives were designed to seem plausible but not fully resolve the bug. Each distractor was verified for incorrectness, and all options were made executable to ensure that models needed a deep understanding of the code to identify and apply the correct fix.

\textbf{\new{Execution Prediction}} evaluates a model's ability to identify and understand defects within code snippets, focusing on both logical and syntactical errors. This task measures the comprehension and debugging capabilities of LLMs by requiring them to predict the execution outcome of given code. It includes two sub-tasks: detecting any defects/flaws in the provided code and comprehending the output of a certain test sample. We derived this task set from IBM CodeNet \citep{NEURIPS_DATASETS_AND_BENCHMARKS2021_a5bfc9e0}, a large-scale benchmark for algorithmic coding tasks. We focused on Python and Java subsets, collecting both accepted and buggy versions of code. After filtering out duplicates, we created a diverse set of code samples. For each snippet, we provide the correct execution result (golden answer) and three distracting options, \new{which could be one of several possible outcomes: (i) Compile Error, (ii) Time Limit Exceeded, (iii) Memory Limit Exceeded, (iv) Runtime Error, or (v) No abnormally found.}

\begin{figure}[htbp]
    \centering
    \includegraphics[width=\linewidth]{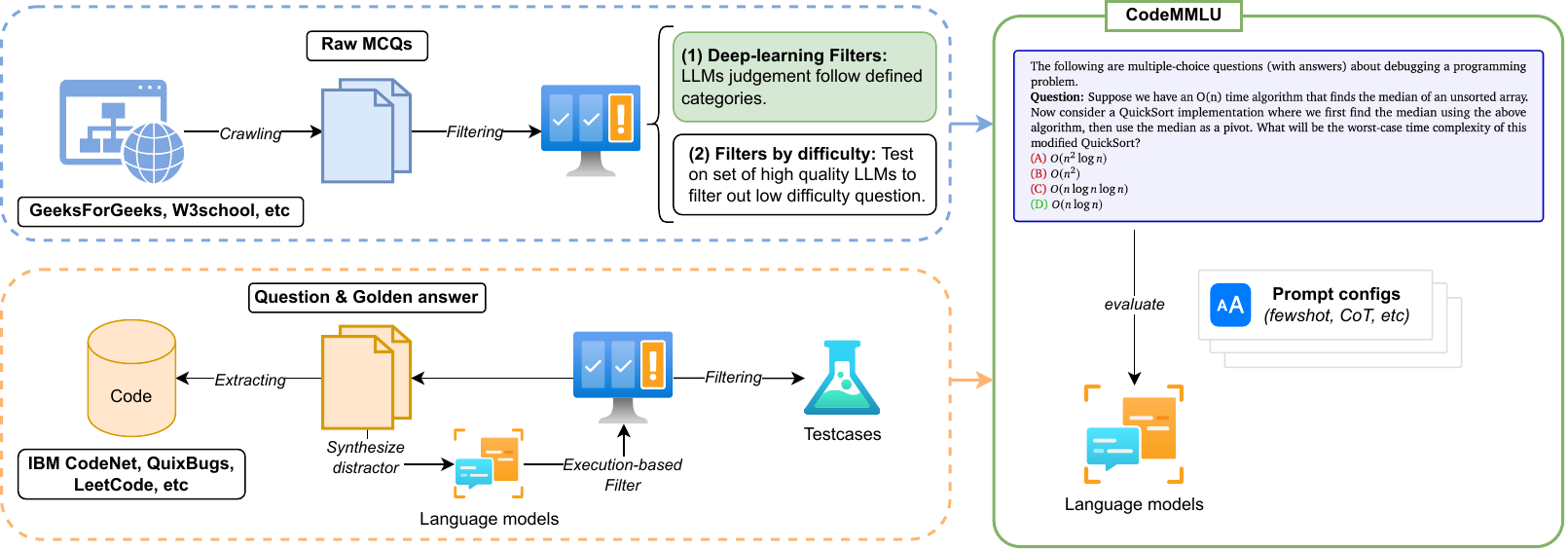}
    \caption{\textbf{Overview of CodeMMLU data creation pipeline.} The blue diagram describe the process of collecting raw multiple-choice questions (MCQs) from open source internet for a knowledge testset. Otherwise, the pipeline of real-world problem indicated in orange area.}
    \label{fig:data-creation}
\end{figure}
% \todo[inline]{R1: the authors fail to clearly explain the process of constructing multiple-choice questions from various data forms after filtering. Moreover, the inclusion of the LLMs evaluation in Figure 3 is not well-explained}

\textbf{Fill in the Blank} evaluates a model's code comprehension ability by requiring it to complete missing parts of a code snippet, given documentation and an incomplete code sample. This task assesses not only the model's ability to fill gaps but also its understanding of both high-level programming concepts and low-level grammatical structures. We collected approximately 2,000 coding problems from LeetCode \footnote{https://leetcode.com/}, covering solutions in three widely-used programming languages (Python, Java, C++). From each problem's solution, \new{we parsed and randomly selected key components (i.e.  crucial elements of the program's logic and flow like loop conditions, expression statements, conditional statements) to be blanked out.} To create plausible but incorrect options for the multiple-choice question (MCQ) format, we employed LLM to generate alternative solutions for the blanked-out components. These distractors were designed to be contextually relevant but incorrect, adding complexity to the task. We executed all generated options to verify their incorrectness, ensuring they do not solve the problem as intended. 

\subsection{Data cleaning}
\label{sub:cleaning}

% \quang{Quang: Add 1-3 sentences to introduce the data cleaning process and its purposes. Also need to explain a bit more on how you combined the scores from several LLMs.}

The preprocessing process (described in Figure \ref{fig:data-creation}) includes a deep learning-based filtering and execution-based filtering to ensure that each question met the desired quality standards, including clarity, lack of ambiguity, and difficulty.

\paragraph{LLM-based Filtering}
\new{To begin, we employed an LLM-based filter to assess the instances in the crawled knowledge test set. Each instance was evaluated based on three criteria: Completeness, Coherence and clarity, and Coding relevance. The models utilized for this evaluation included GPT-3.5, Llama3.1-8B Instruct, and Mixtral-8×7B Instruct. We averaged the scores and used them to select a filtering threshold for each criteria (see discussion in Appendix \ref{appendix:filtering}). To detect and handle duplications, we applied the MinHash LSH algorithm (\citep{eric_zhu_2023_8402527}), configured with 256 permutations, to cluster near-duplicate questions. We remove all false positive instances in each cluster with 0.8 as the similarity threshold. To verify the efficacy of the LLM-based filter, we randomly selected 100 instances from each subject area for manual verification against the three criteria.}

% \new{First we employ an LLM-based filtering process on the crawled knowledge test set by requesting the deep learning neural to judge the given instance in 3 criteria: (i) completeness, (ii) coherence and clarity, and (iii) coding relevance. The models we used are GPT-3.5, Llama3.1-8B Instruct, and Mixtral8$\times$7B Instruct (Detail discussion in Appdendix \ref{appendix:filtering} and prompt used in \ref{appendix:filtering_prompts}). To identify duplication, we employ the MinHash LSH \cite{eric_zhu_2023_8402527} with 256 permutations to cluster near-duplicate questions. Then, we remove all false positive instances in each cluster with 0.8 as the similarity threshold. To verify the effectiveness of LLM-based, we randomly subset 100 instances from every subject to manually check the triple criteria.}

\paragraph{Execution-based Filtering}
% \new{To further ensure the correctness of questions and enhance the difficulty of the CodeMMLU dataset, we applied execution-based filtering to the fundamental test sets. We combined the distractor of (i) code completion, (ii) fill-in-the-blank, and (iii) code repair and execute with their corresponding test cases.}

\new{To ensure the question correctness, we apply an execution-based filtering in the fundamental test sets. We merge the distractor of (i) code completion, (ii) fill-in-the-blank, and (iii) code repair with their codebase and execute with their corresponding test cases. The distractor is designed to bring challenge since it requires LLM to comprehend their correctness without executing it, we select distractors that are executable with 0 to few (less than 50\%) test cases passed in their execution result. In the other hand, the task Execution Prediction's groundtruth are collected from executing process, the distractor are randomly pick from common executing scenarios.}

% The final knowledge-based subset was refined using both manual and deep learning-based filtering to ensure that each question met the desired quality standards, including clarity, lack of ambiguity, and difficulty in evaluating both the semantic and syntactic understanding of LLMs.  This filtering process resulted in a knowledge-based subset containing approximately 6,000 syntactic questions and over 3,000 semantic questions, covering 52 topics classified into 5 main subjects (Table \ref{tab:summaryCodeMMLU}).

% We manually classified the questions into subjects based on the topics of the collected data. A deep learning-based filtering model was then applied to automatically eliminate low-quality or irrelevant questions. 

% \todo{In the 1st sentence mentioned manually classify the question, in the 2nd state a deep learning-based filter} 

% For instance, duplicate or trivial questions that did not sufficiently challenge the code comprehension abilities of LLMs were removed (Appendix  \ref{appendix:mcq_filtering}). 

% \todo{Lack of detail: what is question not meet the requirement, what is the requirement? Which model was being use? What is the data analysis after collecting from open-source} 

\section{Experimental Results}
% We demonstrate the unique advantages of the CodeMMLU benchmark, particularly its scalability, comprehensive coverage, and ability to yield valuable insights into model performance across a range of software-related tasks. We present key findings that highlight the superiority of our benchmark in terms of cost-efficiency, task diversity, and accuracy in ranking large language models (LLMs), providing a more nuanced understanding of model capabilities.

\subsection{Setup} 

\paragraph{Model selection.} We evaluate CodeMMLU on 40 popular open-source LLMs, covering a wide range of parameter sizes and architectures. The models were selected from 13 different families, with parameters ranging from 1 billion to over 70 billion. Each family included base and instructed/chat versions. In addition to open-source models, we also included several proprietary models from OpenAI and Anthropic to ensure a comprehensive coverage of the state-of-the-art in language modeling. \new{All model information can be found at \ref{appendix:models}.}

% : GPT-3.5/GPT-4 \citep{openai2024gpt4technicalreport}, Claude-3-opus/Claude-3.5-sonnet \citep{TheC3}. 

% MetaLlama3.1/8B/70B \citep{dubey2024llama3herdmodels}, MetaLlama3/8B/70B \citep{llama3modelcard}, CodeLlaMA/7B/13B/34B \citep{rozière2024codellamaopenfoundation}, DeepSeek-ai/6.7B/7B/33B \citep{guo2024deepseek}, MistralAI/8x7B \citep{jiang2024mixtralexperts}, Qwen2/7B \citep{qwen2}, CodeQwen1.5/7B \cite{qwen}, Yi/6B/9B/34B \citep{ai2024yiopenfoundationmodels}, StarCoder2/7B/15B \citep{lozhkov2024starcoder2stackv2}.

\paragraph{Answer extraction.} CodeMMLU leverages the MCQ format for scalability and ease of evaluation. In order to maintain this advantage, we only apply simple regex methods to extract the selection answer (i.e., extract by directly answering (A|B|C|D) or containing the pattern ``answer is {A|B|C|D}''). The model response is required to be parsable; otherwise, it will be marked as unanswered. 

\new{In the following, we present key findings of CodeMMLU on (i) knowledge and fundamental test correlation; (ii) MCQ bias evidance; (iii) Disagreement between code-generation alike benchmark and MCQ format. Due to space constraints, we provide detail experimental results, additional discussions and analyses in the appendix, including assessing data leakage (Appendix \ref{appendix:data_leakage}); MCQs analysis (Appendix \ref{appendix:mcq_bias}); Chain-of-thought technique analysis  \ref{appendix:cot_study} and full 43 LLMs results \ref{appendix:full_result}.}

% \new{We employ various prompt strategies to test model performance across different technique, namely: Zeroshot, Fewshot, CoT, and CoT Fewshot. (Detail discussed in Appendix \ref{appendix:prompt-collection})}

\subsection{Key Insights}
\paragraph{Overall performance}
CodeMMLU revealed significant performance differences across models, as shown in Table \ref{tab:codemmlu_res}. OpenAI's GPT-4o outperformed all models on CodeMMLU, demonstrating its quality across diverse tasks (Figure \ref{fig:spiderfigure-model}). Notably, despite not being the latest model, the instructed version of Llama3.1 70B from Meta achieved the highest score among open-source models from 13 families. While LLMs perform well on knowledge-based tasks, they struggle with real-world problems, particularly in execution prediction tasks \new{(see Appendix \ref{appendix:full_result} for all the experiment details.)}

\begin{figure}[h]
\centering
\includegraphics[width=\linewidth]{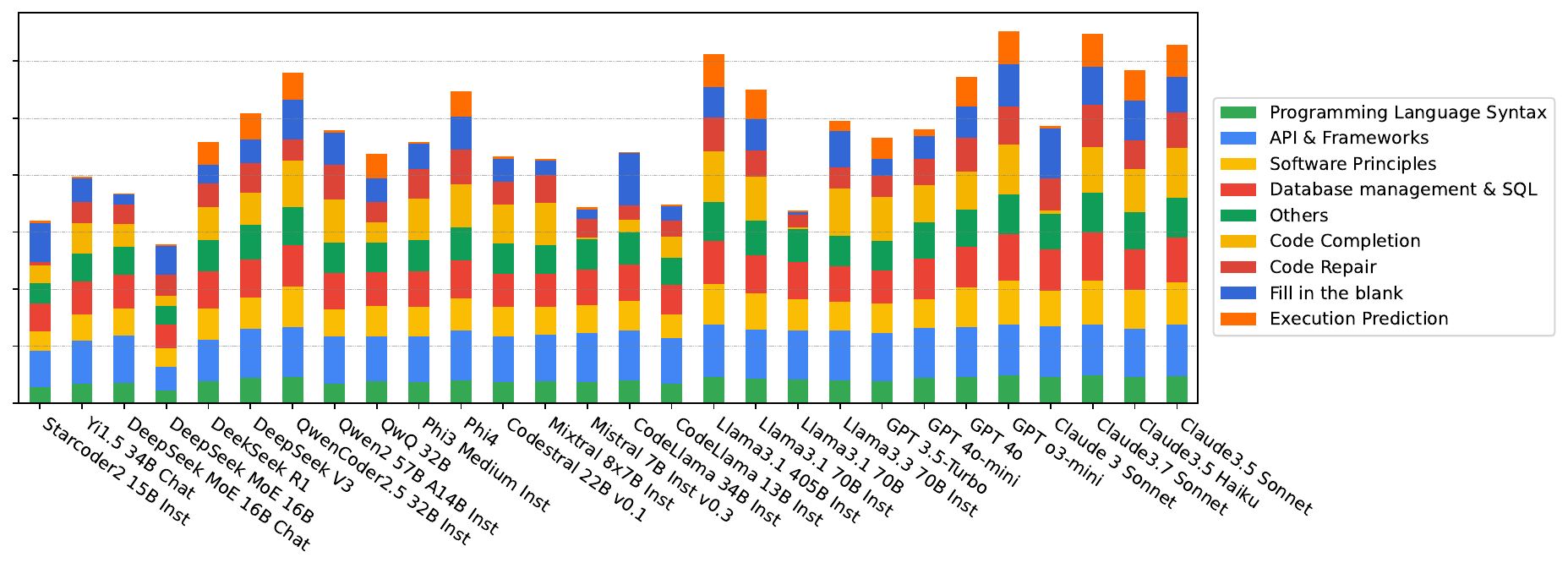}
\caption{\textbf{CodeMMLU accuracy by task on LLMs.} While knowledge tasks are following the scaling law, real-world tasks offer more challenges to LLMs which indicate the performance of instruction tuning and data quality when evaluating on CodeMMLU.}
\label{fig:benchmark-vis}
\vspace{-5mm}
\end{figure}

Figure \ref{fig:benchmark-vis} illustrates CodeMMLU's capability to measure LLMs' coding knowledge and skills across a wide range of subjects. Our benchmark provides clear, distinct rankings that establish a higher hierarchy of models compared to other benchmarks \new{(see the result in Table \ref{tab:appendix:ben_compare})}. Interestingly, the results do not strictly adhere to scaling laws \citep{kaplan2020scalinglawsneurallanguage}, where larger parameter sizes typically outperform smaller ones. This highlights the impact of data quality in the LLM pretraining process, as recently released models often achieve comparable performance to larger models from previous versions.
CodeMMLU also indicates the importance of instruction tuning in improving model performance on complex tasks. Models with instruction tuning substantially outperform their non-instructed counterparts, for example, DeepSeek-Coder-33b surpasses its base model by approximately 29\%. 

\begin{figure}
    \centering
    \includegraphics[width=0.85\linewidth]{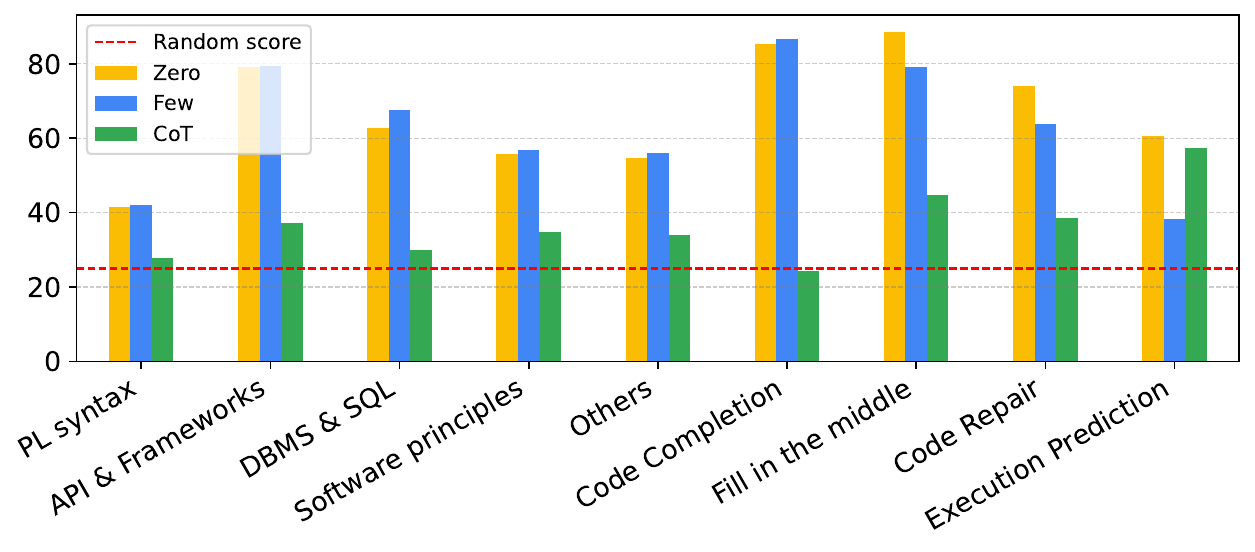}
    \caption{\textbf{Comparison of prompt configuration on GPT-4o.} The experiment exposes the drawback of Chain-of-Thought prompting technique in term of boosting performance on task that not require logic or reasoning.}
    \label{fig:prompt-compare}
\end{figure}

\begin{figure}[h]
    \centering
    \begin{minipage}{0.4\linewidth}
        \centering
        % \vspace{-5mm}
        \includegraphics[width=0.9\linewidth]{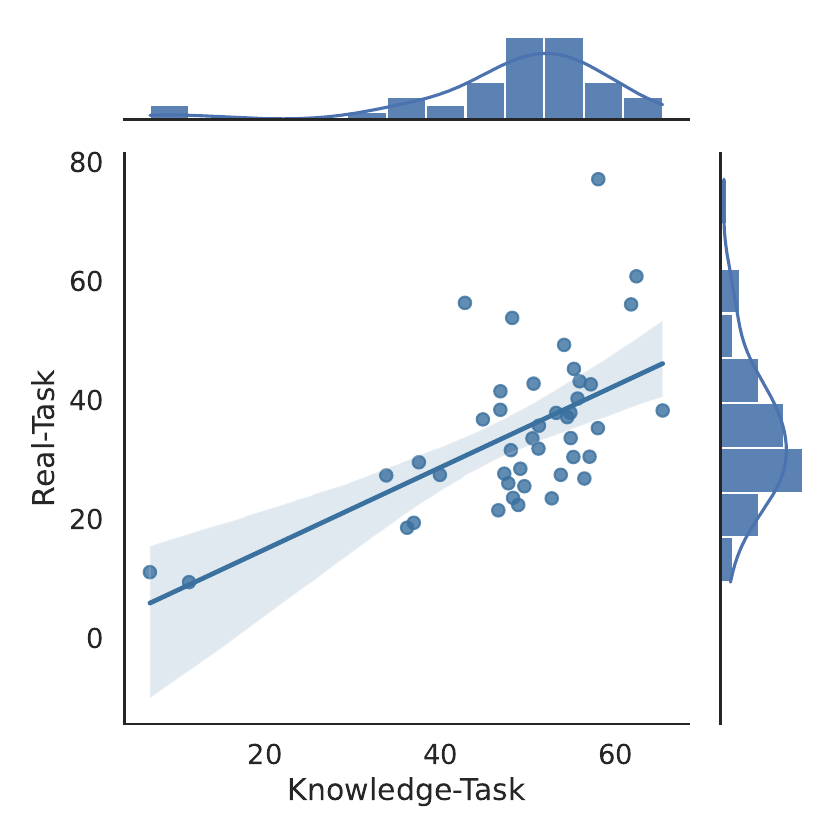}
        % \vspace{-7mm}
        \caption{\textbf{Correlation between knowledge tests and fundamental skill tests.} Experiments on 10 LLM families show a clear alignment between models with a strong understanding of software knowledge and their performance on diverse problem-solving tasks in the CodeMMLU fundamental skill tests.}
        \label{fig:correlation}
    \end{minipage}
    \hfill
    \begin{minipage}{0.55\linewidth}
        % \vspace{5mm}
        \includegraphics[width=\linewidth]{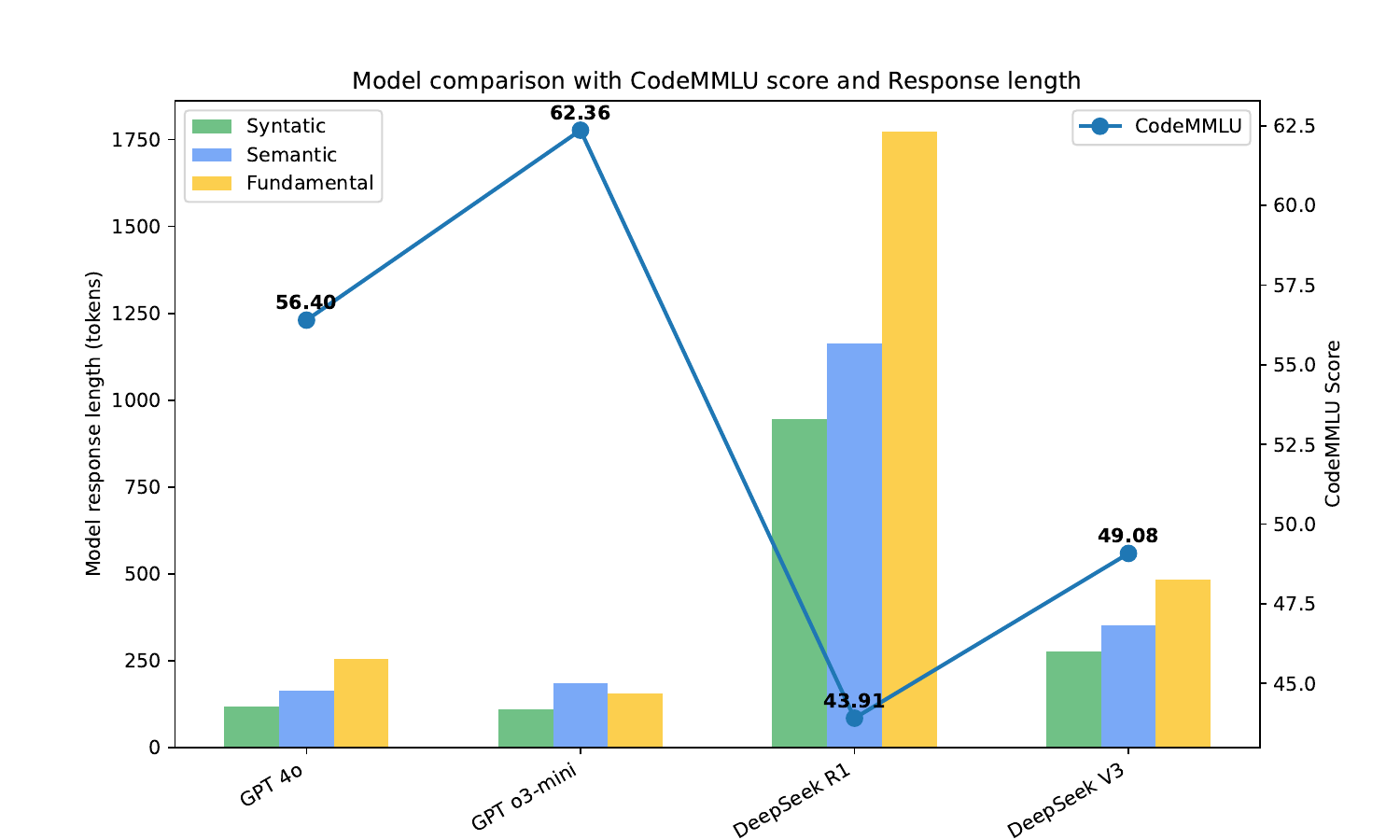}
        % \includegraphics[width=\linewidth]{img/humaneval_mcq.png}
        % \vspace{-2mm}
        \caption{\textbf{Comparison Between LLM Response Length and CodeMMLU Score.} While reasoning models (e.g., DeepSeek-R1) generate longer, reasoning-heavy responses, this does not necessarily correlate with higher accuracy on CodeMMLU.}
        \label{fig:response_length}
    \end{minipage}
\end{figure}

\paragraph{The impact of model reasoning and Chain-of-Thought (CoT) prompting} Although CoT prompting \citep{wei2023chainofthoughtpromptingelicitsreasoning} is often expected to enhance performance by eliciting deeper reasoning, our experiments reveal that CoT and reasoning models may not always offer improvements in CodeMMLU. 
Table \ref{tab:codemmlu_res} shows that DeepSeek R1 performs significantly worse than its base model, DeepSeek V3, despite being designed for reasoning tasks.
Meanwhile, GPT o3-mini, a native reasoning model, achieves the best results. Our analysis suggests that, apart from GPT o3-mini, other reasoning models (e.g, GPT o1, DeepSeek R1) tend to overreason across all CodeMMLU tasks. In contrast, o3-mini demonstrates the ability to decide when to apply reasoning versus directly answering knowledge-seeking tasks, such as in syntactic and semantic evaluations (Figure \ref{fig:appendix:reasoning_1} \ref{fig:reasoning_2} ). These findings align with \citep{chen2025think23overthinkingo1like}, which observes that current reasoning models often overreason even on simple questions. This suggests the need for a more effective decision-making mechanism to determine when reasoning is beneficial. Table \ref{fig:response_length} further highlights that while R1 generates the longest responses (in tokens), it underperforms compared to the latest top-tier LLMs.

We further investigate the impact of different prompting techniques and report the results in Figure \ref{fig:prompt-compare}. The results show a significant decline in GPT-4o’s performance with CoT, suggesting that the additional complexity introduced by step-by-step reasoning does not align well with knowledge-seeking tasks (see Appendix \ref{appendix:cot_study} for more discussions). In contrast, few-shot prompting consistently emerges as the most reliable and effective strategy across various tasks, offering a balanced approach without overwhelming the models. Overall, with the exception of GPT o3-mini, we found no improvements from reasoning models or CoT prompting, suggesting that CodeMMLU presents a challenging benchmark to test the models reasoning capabilities in code domains.

% Table generated by Excel2LaTeX from sheet 'Sheet2'
\begin{table}[htbp]
  \centering
  \caption{\textbf{Summary of LLM Family Performance on CodeMMLU}. The evaluation results (accuracy \%) of different language models across the CodeMMLU task \new{(CodeMMLU column represents the accuracy average among all subject).}}
  \resizebox{0.95\linewidth}{!}{
    \begin{tabular}{cl|c|cc|c|c}
\toprule    \multirow{2}[2]{*}{\textbf{Family}} & \multicolumn{1}{c}{\multirow{2}[2]{*}{\textbf{Model name}}} & \multicolumn{1}{c}{\multirow{2}[2]{*}{\textbf{Size (B)}}} & \multicolumn{2}{c}{\textbf{Knowledge test}} & \multicolumn{1}{c}{\multirow{2}[2]{*}{\textit{\textbf{Fundamental test}}}} & \multicolumn{1}{c}{\multirow{2}[2]{*}{\textbf{CodeMMLU}}} \\
          & \multicolumn{1}{c}{} & \multicolumn{1}{c}{} & \multicolumn{1}{c}{\textit{Syntactic}} & \multicolumn{1}{c}{\textit{Semantic}} & \multicolumn{1}{c}{} &  \\
\midrule
\midrule
          & \multicolumn{6}{c}{Closed-source models} \\
\cmidrule{2-7}    
    % \multicolumn{1}{r}{Claude} 
    \multirow{4}[1]{*}{Claude}
    & Claude3.7 Sonnet  & -      & 52.78 &	\textbf{76.26} &	\textit{\underline{60.92}} &	\textit{\underline{61.65}} \\
    & Claude3.5 Sonnet  & -      & 52.23 & 73.45 & 58.56 & 59.81 \\
    & Claude3.5 Haiku  & -      & 49.24 &	68.20 &	57.83 &	57.25 \\
    & Claude3 Sonnet  & -      & \textbf{67.22} & 66.08 & 38.26 & 53.97 \\
    \cmidrule{2-7}
    \multirow{4}[1]{*}{GPT} & GPT o3-mini &   -   & 53.08 & \textit{\underline{75.50}} & \textbf{62.77} & \textbf{62.36} \\
          & GPT 4o  & -      & 50.63 & 69.61 & 53.89 & 56.40 \\
          & GPT 4o-mini & -     & 48.66 & 55.90 & 20.33 & 38.43 \\
          & GPT-3.5-turbo  & -      & \underline{\textit{61.68}} & 53.65 & 45.26 & 51.70 \\
\cmidrule{2-7}          & \multicolumn{6}{c}{Open-source models} \\
\cmidrule{2-7}    \multirow{4}[2]{*}{Llama} & Llama3.3 70B Inst & 70    & 44.31 & 52.76 & 30.96 & 40.66 \\
          & Llama3.1 405B Inst & 405   & 50.82 & \textbf{71.41} & \textbf{57.10} & \textbf{58.23} \\
          & Llama3 70B Inst & 70    & 46.94 & 62.64 & 53.15 & 53.19 \\
          & CodeLlama34B Inst & 34    & 56.81 & 46.93 & 23.55 & 38.73 \\
\cmidrule{2-7}    \multirow{3}[2]{*}{DeepSeek} & DeepSeek R1 & 671   & 42.39 & 56.77 & 38.08 & 43.91 \\
          & DeepSeek V3 & 685   & 48.30 & 57.53 & 45.06 & 49.08 \\
          & DeepSeekCoder 33B Inst & 33    & 53.65 & 45.43 & 21.46 & 36.60 \\
          & DeepSeekMoE 16B Chat  & 16.4  & 31.74 & 35.42 & 27.32 & 31.01 \\
\cmidrule{2-7}    \multirow{2}[2]{*}{Mistral} & Mistral7B Inst (v0.3)  & 7     & 54.42 & 51.25 & 31.85 & 43.33 \\
          & Mixtral 8$\times$7B Inst & 46.7  & \underline{\textit{61.17}} & 54.89 & 24.09 & 42.96 \\
          & Codestral 22B  & 22    & 60.34 & 52.10 & 37.85 & 47.60 \\
\cmidrule{2-7}    \multirow{1}[2]{*}{Phi} & Phi4  & 14    & 45.34 & 57.46 & 47.82 & 49.19 \\
          & Phi4 Mini Inst & 12    & 41.94 & 51.59 & 19.75 & 34.85 \\
\cmidrule{2-7}          
    \multirow{5}[0]{*}{Qwen} 
          & Qwen2.5 14B Inst & 14    & 46.38 & 58.70 & 51.49 & 51.38 \\
          & QwQ 32B Preview & 57    & \textbf{61.34} & 57.48 & 30.48 & 46.34 \\
          & QwenCoder2.5 32B Inst & 32    & 50.63 & \textit{\underline{69.61}} & \textit{\underline{53.89}} & \textit{\underline{56.40}} \\

    \bottomrule
    \end{tabular}%
    \label{tab:codemmlu_res}%
}
\end{table}%

% \vspace{-5mm}

\paragraph{Correlation Between Software Knowledge and Real-World Performance}
Our experiments revealed a strong correlation between performance on knowledge-based tasks and real-world coding challenges. Specifically, the Pearson correlation score r = 0.61 between model rankings on the knowledge test set and their performance on real-world problems, derived from the accuracy of 43 LLMs across 15 model families, indicates a moderate alignment (Figure \ref{fig:correlation}). This suggests that models demonstrating a deeper understanding of software principles consistently excel in real-world coding tasks, highlighting the importance of foundational knowledge for practical coding performance.

% \begin{figure}
%     \centering
%     \begin{minipage}{0.35\linewidth}
%     \includegraphics[width=\linewidth]{img/joint_plot_corr.pdf}
%     \caption{\textbf{Task-Specific Accuracy and Performance Fluctuations Across Answer Options}
% Models exhibit marked fluctuations in accuracy depending on the position of the correct answer in Code Completion in CodeMMLU. Revealing the bias and inconsistencies in related coding multiple-choice question (MCQ) task and how sensitive LLMs are to answer ordering.}
%     \end{minipage}
%     \label{fig:correlation}
    
%     \hfill
    
%     \begin{minipage}{0.5\linewidth}
%     \includegraphics[width=\linewidth]{img/humaneval_mcq.png}
%     % \vspace{-8mm}
%     \caption{\textbf{Task-Specific Accuracy and Performance Fluctuations Across Answer Options}
% Models exhibit marked fluctuations in accuracy depending on the position of the correct answer in Code Completion in CodeMMLU. Revealing the bias and inconsistencies in related coding multiple-choice question (MCQ) task and how sensitive LLMs are to answer ordering.}
%     \end{minipage}
%     \label{fig:intro_bias}
% \end{figure}

\vspace{-4mm}

% This observation suggests that LLMs with superior software knowledge are better equipped to tackle practical coding challenges, supporting the claim that these models possess an understanding beyond mere probability-based guessing. This highlights the potential of CodeMMLU to reveal deeper comprehension in LLMs, compared to benchmarks that assess isolated tasks without connecting them to underlying knowledge.

\paragraph{Selection bias in MCQs format}

\begin{figure}[ht]
\centering
\includegraphics[width=0.8\linewidth]{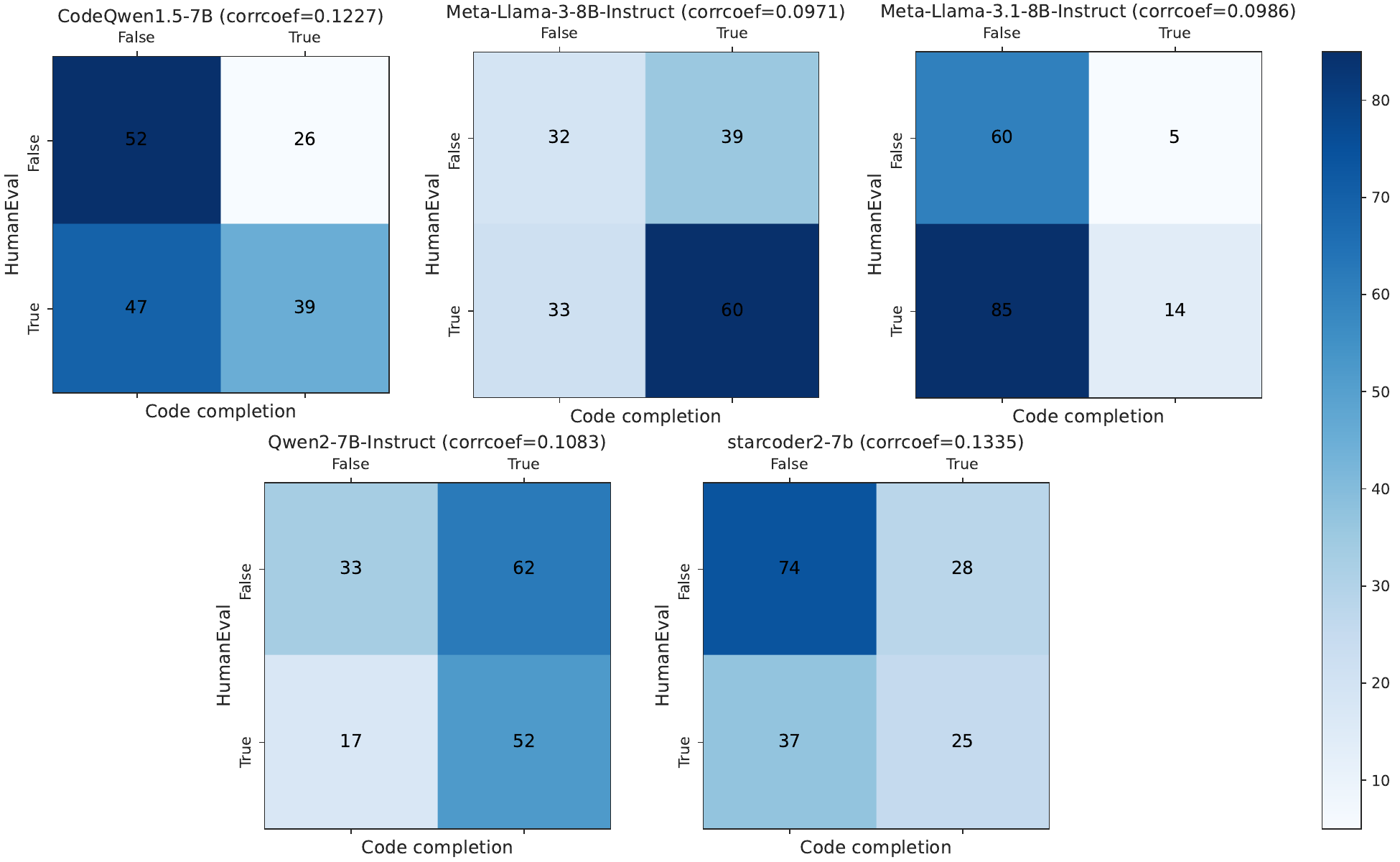}
\caption{\textbf{Comparison of CodeMMLU’s code completion task and HumanEval.} Many LLMs show a performance discrepancy between the two tasks, where models that successfully passed the HumanEval code generation test often failed to select the correct answer in the multiple-choice (MCQ) format, or vice versa, for the same question.}
\label{fig:humaneval-alignment}
\vspace{-5mm}
\end{figure}

We experimented with multiple answer order permutations (follow \citep{llm-mcq-bias}), the result displayed significant inconsistent behavior exhibited by LLMs when swapping golden answer positions. As presented in Table \ref{fig:humaneval-alignment}, the model's performance changes dramatically in each answer order configuration, which is based on the correct answer's position. The LLM's accuracy fluctuates between different permutations (i.e.  DeepSeek-Coder-34B $\Delta\sigma=36.66$), demonstrating how sensitive it can be to the structure and order of answers (Figure \ref{fig:intro_bias}).  \new{However, Table \ref{tab:appendix_mcq_bias} indicate the different of MCQ bias between strong models (e.g GPT-4o, Claude3-orpus) and others, which highlight the consistency and robustness among them (see discussion in \ref{appendix:mcq_bias}).}

\begin{table}[htbp]
    \centering
    \caption{\textbf{Performance Comparison between HumanEval and MCQ Code Completion Tasks.} The performance fluctuation highlights the selection biases observed when the correct (golden) answer is moved to positions A, B, C, or D.}
    % \vspace{4mm}
    \scalebox{0.7}{
    \begin{tabular}{c|c|cccc}
    \toprule
    \multirow{2}[2]{*}{\textbf{Models}} & \multicolumn{1}{c|}{\multirow{2}[2]{*}{\textbf{HumanEval}}} & \multicolumn{4}{c}{\textbf{Code Completion MCQ}} \\
          &       & A     & B     & C     & D \\
    \midrule
    \midrule
    \multirow{2}[1]{*}{CodeLlama-7B-Python} & \multirow{2}[1]{*}{40.48} & 0.00  & 90.24 & 14.02 & 0.61 \\
          &       & \textcolor[rgb]{ 1,  0,  0}{(-40.48)} & \textcolor[rgb]{ .259,  .522,  .957}{(+49.76)} & \textcolor[rgb]{ 1,  0,  0}{(-26.46)} & \textcolor[rgb]{ 1,  0,  0}{(-39.87)} \\
    \multirow{2}[0]{*}{CodeLlama-7B-Instruct} & \multirow{2}[0]{*}{45.65} & 3.66  & 1.22  & 93.90 & 15.85 \\
          &       & \textcolor[rgb]{ 1,  0,  0}{(-41.99)} & \textcolor[rgb]{ 1,  0,  0}{(-44.43)} & \textcolor[rgb]{ .259,  .522,  .957}{(+48.25)} & \textcolor[rgb]{ 1,  0,  0}{(-29.80)} \\
    \multirow{2}[0]{*}{CodeLlama-13B-Python} & \multirow{2}[0]{*}{42.89} & 0.61  & 54.88 & 70.12 & 12.20 \\
          &       & \textcolor[rgb]{ 1,  0,  0}{(-42.28)} & \textcolor[rgb]{ .259,  .522,  .957}{(+11.99)} & \textcolor[rgb]{ .259,  .522,  .957}{(+27.23)} & \textcolor[rgb]{ 1,  0,  0}{(-30.69)} \\
    \multirow{2}[0]{*}{CodeLlama-13B-Instruct} & \multirow{2}[0]{*}{50.6} & 2.44  & 68.29 & 72.56 & 29.88 \\
          &       & \textcolor[rgb]{ 1,  0,  0}{(-48.16)} & \textcolor[rgb]{ .259,  .522,  .957}{(+17.69)} & \textcolor[rgb]{ .259,  .522,  .957}{(+21.96)} & \textcolor[rgb]{ 1,  0,  0}{(-20.72)} \\
    \multirow{2}[0]{*}{CodeLlama-34B-Python} & \multirow{2}[0]{*}{45.11} & 0.61  & 77.44 & 70.73 & 49.39 \\
          &       & \textcolor[rgb]{ 1,  0,  0}{(-44.50)} & \textcolor[rgb]{ .259,  .522,  .957}{(+32.33)} & \textcolor[rgb]{ .259,  .522,  .957}{(+25.62)} & \textcolor[rgb]{ .259,  .522,  .957}{4.28} \\
    \multirow{2}[1]{*}{CodeLlama-34B-Instruct} & \multirow{2}[1]{*}{50.79} & 9.15  & 84.76 & 65.24 & 46.34 \\
          &       & \textcolor[rgb]{ 1,  0,  0}{(-41.64)} & \textcolor[rgb]{ .259,  .522,  .957}{(+33.97)} & \textcolor[rgb]{ .259,  .522,  .957}{(+14.45)} & \textcolor[rgb]{ 1,  0,  0}{(-4.45)} \\
    \midrule
    \multirow{2}[1]{*}{Deepseek-Coder-7B-base-v1.5} & \multirow{2}[1]{*}{43.2} & 40.85 & 74.39 & 64.02 & 39.02 \\
          &       & \textcolor[rgb]{ 1,  0,  0}{(-2.35)} & \textcolor[rgb]{ .259,  .522,  .957}{(+31.19)} & \textcolor[rgb]{ .259,  .522,  .957}{(+20.82)} & \textcolor[rgb]{ 1,  0,  0}{(-4.18)} \\
    \multirow{2}[0]{*}{DeepSeek-Coder-33B-base} & \multirow{2}[0]{*}{56.1} & 1.22  & 82.32 & 75.00 & 56.10 \\
          &       & \textcolor[rgb]{ 1,  0,  0}{(-54.88)} & \textcolor[rgb]{ .259,  .522,  .957}{(+26.22)} & \textcolor[rgb]{ .259,  .522,  .957}{(+18.90)} & \textcolor[rgb]{ 1,  0,  0}{(0.00)} \\
    \midrule
    \multirow{2}[0]{*}{Phind-CodeLLama-34B-v2} & \multirow{2}[0]{*}{71.95} & 6.10  & 90.85 & 75.00 & 46.34 \\
          &       & \textcolor[rgb]{ 1,  0,  0}{(-65.85)} & \textcolor[rgb]{ .259,  .522,  .957}{(+18.90)} & \textcolor[rgb]{ .259,  .522,  .957}{(+3.05)} & \textcolor[rgb]{ 1,  0,  0}{(-25.61)} \\
    \multirow{2}[1]{*}{Mixtral-8x7B-Instruct-v0.1} & \multirow{2}[1]{*}{40.2} & 22.56 & 74.39 & 71.95 & 63.41 \\
          &       & \textcolor[rgb]{ 1,  0,  0}{(-17.64)} & \textcolor[rgb]{ .259,  .522,  .957}{(+34.19)} & \textcolor[rgb]{ .259,  .522,  .957}{(+31.75)} & \textcolor[rgb]{ .259,  .522,  .957}{(+23.21)} \\
    \bottomrule
    \end{tabular}%
    }
    \label{tab:move_attack}%
\end{table}

\paragraph{Disagreement between Open-ended generation benchmark and MCQ Code completion}
A notable finding from our experiments is the discrepancy in model performance between open-ended benchmarks and multiple-choice formats. Specifically, when comparing the original HumanEval questions with their multiple-choice equivalents in our CodeMMLU code completion set, we found that models performing well on HumanEval do not consistently replicate their success in CodeMMLU. For instance, when evaluating identical questions across the formats, the number of cases where models answered both correctly or incorrectly was unexpectedly low. The correlation scores in Figure \ref{fig:humaneval-alignment} further illustrate the weak alignment of success between these two benchmarks, revealing that performance in open-ended tasks does not reliably predict performance in multiple-choice coding tasks. 
This lack of alignment suggests that traditional benchmarks might overestimate a model's understanding by focusing too narrowly on code generation, \new{which is highly susceptible to data leakages. In contrast, CodeMMLU requires the models to engage in complex reasoning to understand code and solve software engineering problems.}
%The MCQ format in CodeMMLU forces models to engage with more complex reasoning and contextual understanding, exposing weaknesses that remain hidden in generative tasks. 

% A notable finding from our experiments is the discrepancy between performance on open-ended benchmarks when comparing side-by-side the question from HumanEval, and our code completion set (i.e. multiple-choice question of HumanEval). Many LLMs that perform well on HumanEval do not consistently translate their success to CodeMMLU. For instance, when comparing sample questions from HumanEval with their MCQ equivalents, the number of cases where models answered both correctly or both incorrectly was surprisingly low. The correlation scores in Figure \ref{fig:humaneval-alignment} indicate a surprisingly low alignment of success case between these two benchmarks.

% While capable of generating code, LLMs may still struggle with maintaining a coherent understanding of software principles when posed with different types of questions. These MCQs result calls for more rigorous evaluation methods, as traditional benchmarks tend to overestimate model capabilities by focusing narrowly on code generation.
%This lack of alignment suggests that traditional benchmarks might overestimate a model's understanding by focusing too narrowly on code generation. The MCQ format in CodeMMLU forces models to engage with more complex reasoning and contextual understanding, exposing weaknesses that remain hidden in generative tasks. 

% \begin{figure}
%     \centering
%     \includegraphics[width=0.5\linewidth]{img/prompt-setting.png}
%     \caption{Caption}
%     \label{fig:enter-label}
% \end{figure}

% \subsection{Error Analysis}
% \input{content/table/5_humaneval_mcq}
% \paragraph{Selection in MCQ} In MCQ format, LLMs in somecase tend to predict only one charecter answer without reasoning and trying to explain why this answers is True. In Table \ref{fig:intro_bias} demonstrates that when the positions of the correct answers are altered, there is a significant fluctuation in the model's performance, indicating a bias towards options B and C.

\par
\section{Conclusions}

In this work, we introduced CodeMMLU, a comprehensive and scalable benchmark designed to evaluate large language models’ (LLMs) capabilities across a wide range of software knowledge and real-world programming tasks. Our experiments highlighted the benchmark’s key advantages, including its cost-effectiveness, scalability, and extensive task coverage. The insights gained revealed a strong correlation between software knowledge and real-world task performance, demonstrating that models with deeper comprehension outperform those relying purely on probabilistic generation.

Additionally, CodeMMLU provides more accurate and detailed rankings of LLMs, particularly in open-source models, where significant reordering of performance was observed. The benchmark also revealed inconsistencies in model comprehension when compared to traditional evaluations like HumanEval, emphasizing the need for more robust benchmarks that go beyond simple code generation.

% \newpage
% \section*{Limitations, and Future Work}
% \paragraph{Limitations.} While CodeMMLU offers a broad and diverse evaluation, there are some limitations. First, the MCQ format, though effective at testing comprehension, might not fully capture creative aspects of code generation or models' ability to optimize code. Second, the current scope of languages and tasks could be expanded to include more specialized domains or additional programming languages to better assess models' versatility.

% \paragraph{Future Work.} Looking forward, we plan to release CodeMMLU as an open-source benchmark for the research community. This release will include the full dataset, along with tools for automated evaluation, allowing for widespread adoption and further improvements. Future updates will focus on adding more complex tasks, refining the balance between real-world scenarios and theoretical knowledge, and incorporating user feedback to make the benchmark even more robust for next-generation LLMs.

%%%%%%%%%%%%%%%%%%%%%%%%%%%%%%
\newpage
% \section*{Acknowledgements}
% All of the authors gratefully acknowledge FPTSoftware AI Center, ...

% REMOVE THIS
% \listoftodos[Notes]
\newpage
% REMOVE THIS

\bibliographystyle{iclr2025_conference}
\bibliography{references}

%%%%%%%%%%%%%%%%%%%%%%%%%%%%%%
\newpage
% \addcontentsline{toc}{section}{Appendix} % Add the appendix text to the document TOC
% \part{Appendix} % Start the appendix part
% \parttoc % Insert the appendix TOC
% \appendixpage
\appendix
\section{Dataset}

% \subsection{Data analysis}
% - Dataset subtopic distribution
% - Data length
% - Answer position distribution

\subsection{Data cleaning}
\label{appendix:filtering}

\paragraph{Rule-based filtering} We prefer questions that contain code when collecting data; therefore, MCQs often contain noisy patterns and low-quality questions. In the cleaning process, we defined a heuristic rule-based filter to eliminate incomplete data and non-textual content.  First, we detect and eliminate non-textual questions by filtering questions that contain hrefs, image URLs, links to other questions or media. We also applied BeautifulSoup to remove unwanted HTML tags.

\begin{figure}[htbp]
    \centering
    \includegraphics[width=0.5\linewidth]{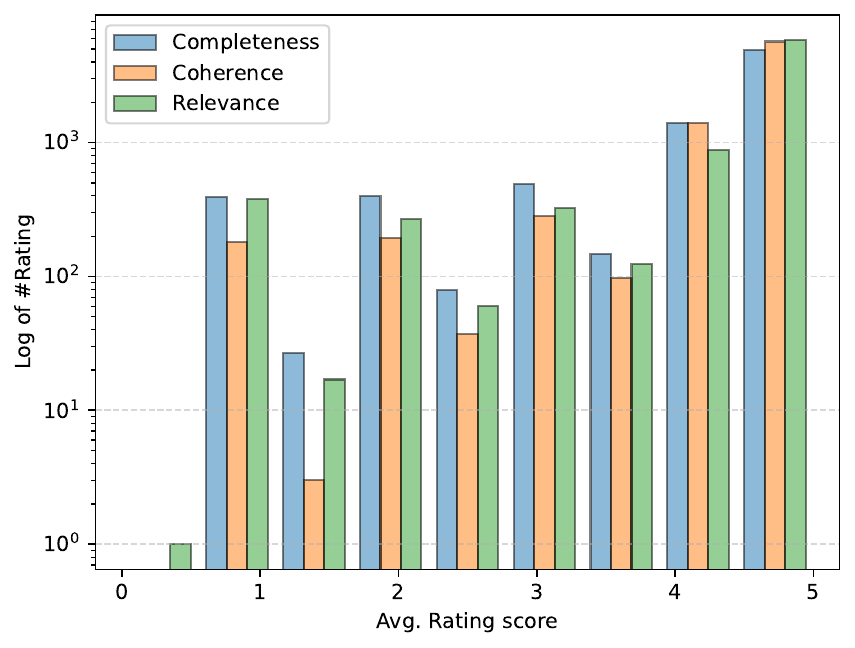}
    \caption{\textbf{LLM-based filter score distribution.}}
    \label{fig:appendix:rating_distribution}
\end{figure}

\paragraph{Deep learning-based filtering} To ensure the CodeMMLU is fully targeted on coding and software-related task, we employed models from OpenAI (GPT-3.5-turbo), Mistral (Mixtral 8$\times$7B Instruct), and Llama (Llama3.1 8B) as our annotators to judge the triple criteria: \textbf{Completeness}; \textbf{coherence and clarity}; and \textbf{coding relevance} (check appendix \ref{appendix:prompts} to see the prompt). We averaged LLM ratings by category and selected a threshold of 4 in 3 aspects. Result of removing $\approx$ 25.6\% of raw data. On the other hand, we simultaneously sampled a subset of 100 instances in each subject to update our filter rule. (Figure \ref{fig:appendix:rating_distribution}

\paragraph{Execution-based filtering} After synthesizing the fundamental task's distractor (i.e., false answer), we concatenated and executed them as a complete function in an isolated environment. The code completion and fill-in-the-blank tasks have their original test cases, while code repair needs to synthesize new test cases. Therefore, we extracted the method signature (using the code-text parser toolkit from \citep{manh2023vaultcomprehensivemultilingualdataset}) and synthesized the function input, which later was executed to create test cases for the corresponding function. We ran in parallel the distractor executing on the testcase and synthesized a new distractor, ensuring the distractor collection is executable and able to pass 0-50\% test cases.

\subsection{Data Contamination}
\label{appendix:data_leakage}

The development of large language models (LLMs) often involves crawling data from diverse sources across the internet, with limited transparency regarding their preprocessing. Given the vast and often proprietary nature of these training datasets, it is widely acknowledged that creating a fully leakage-free benchmark is virtually impossible. While recent benchmarks have recognized this issue and generally accept that avoiding data leakage entirely is extremely difficult, one common mitigation strategy involves filtering data based on its timeline \citep{jain2024livecodebench, wang2024testevalbenchmarkinglargelanguage, zhang2023repocoderrepositorylevelcodecompletion}.

In our efforts to address this challenge, we acknowledge the complexity of completely eliminating data leakage. To enhance the reliability of CodeMMLU, we adopt proactive measures during the data creation process. Specifically, we transform seed data into multiple-choice question formats and introduce synthetic distractors. For tasks like code repair and execution prediction, the test sets were extracted from codebase seeds and modified to align with specific task requirements.

To further assess and quantify potential data leakage, we employ the methodology outlined in \citep{xu2024benchmarkingbenchmarkleakagelarge}. This includes calculating \textit{perplexity} and conducting \textit{n-gram} analysis on several well-known models from diverse families (e.g., Mistral, DeepSeek, Llama). The results, presented in Tables \ref{tab:appendix_ppl} and \ref{tab:appendix_ngram}, highlight a significant margin between CodeMMLU and other coding benchmarks, reinforcing the reliability of CodeMMLU as a robust evaluation tool.

% Developing LLM processes often involves crawling data all over the internet, and their preprocessing is little to unknown in detail. It is known to be impossible to create a leakage-free benchmark, especially when LLMs are trained with massive corpora and most of them remain closed-source. On one hand, recent benchmarks acknowledge the leakage and generally believe it is extremely difficult to avoid the problem. On the other hand, one solution for the data leakage problem is filtering data based on timeline \cite{jain2024livecodebench, wang2024testevalbenchmarkinglargelanguage, zhang2023repocoderrepositorylevelcodecompletion}.

% We also find it extremely difficult to overcome the challenge. However, to ensure the reliability of CodeMMLU, we try to early prevent data leakage from the data creation process by involving the data seed transformation into multiple-choice question format and by creating synthesis distractors. Code repair and Execution prediction test sets are extracted from the codebase seed and modified to fit the task purpose. Furthermore, we apply the method from \cite{xu2024benchmarkingbenchmarkleakagelarge} of calculating \textit{perplexity} and \textit{n-gram} on well-known models from different families (Mistral, DeepSeek, Llama) to demonstrate the data leakage level on CodeMMLU. Results in Table \ref{tab:appendix_ppl} and Table \ref{tab:appendix_ngram} indicate a significant margin between CodeMMLU when comparing it with other coding benchmarks, showcasing the reliability of the benchmark itself.

\paragraph{Perplexity} measures the uncertainty of a language model when predicting the next token in a sequence \citep{jelinek1977perplexity}. Therefore, as low as the perplexity score indicates, the model is confident in predicting the evaluating sequence and the more likely that the model was encountered during the training process. Perplexity is expressed as the exponentiated average negative log-likelihood of a sequence:
\begin{equation}
\newcommand{\lt}{\ensuremath <}
    \text{PPL}(\textbf{X})=\text{exp}\left( -\frac{1}{t}\sum_{t=0}^{t}\text{log}p_\theta(x_i|x_{\lt i}) \right)
\end{equation}
where $\textbf{X}=\left[ x_0, x_1, ...,x_t \right]$ denotes a tokenized sequence.

\paragraph{N-gram Accuracy}\citep{xu2024benchmarkingbenchmarkleakagelarge} 
is a metric designed to detect fine-grained data leakage at the instance level by combining the question and answer into a single text (X), uniformly sampling starting points, and predicting the next n-grams based on the given prompts. If most n-grams are accurately predicted, it suggests the model may have encountered the data during training. The N-gram accuracy can be expressed as:

\begin{equation}
\text{N-gram Accuracy}(X) = \frac{1}{S \cdot K} \sum_{i=0}^{S} \sum_{j=0}^{K} I(X_{start_j:start_j+n}, \hat{X}_{start_j:start_j+n}),
\end{equation}

where  $S$  is the dataset size,  $K$  is the number of sampled starting points,  $X_{start_j:start_j+n}$  is the actual n-gram,  $\hat{X}_{start_j:start_j+n}$  is the predicted n-gram, and  $I$  checks for exact matches. Author add ROUGE-L and edit distance similarity to provide robustness for augmented datasets. A high accuracy for each n-gram in a prediction indicates a strong likelihood that the sample was seen during the training process. \citep{xu2024benchmarkingbenchmarkleakagelarge}

\begin{table}[htbp]
  \centering
  \caption{\textbf{Perplexity score comparison between coding benchmark.} \textit{(higher is better)}}
    \begin{tabular}{c|cc|c}
    \toprule
    \textbf{Models} & \textbf{CodeScope} & \textbf{CodeApex} & \textbf{CodeMMLU} \\
    \midrule
    Mistral7B-v0.3 & 9.32  & 16.08 & 16.32 \\
    DeepSeekCoder7B-v1.5 & 5.26  & 9.39  & 57.36 \\
    DeepSeekV2-Lite & 6.89  & 11.99 & 1419.48 \\
    Llama-3.1-8B & 10.05 & 123.20 & 197.31 \\
    \bottomrule
    \end{tabular}%
  \label{tab:appendix_ppl}%
\end{table}%
% Table generated by Excel2LaTeX from sheet 'Sheet2'
\begin{table}[htbp]
  \centering
  \caption{\textbf{5-gram accuracy comparison between coding benchmark.} \textit{(lower is better)}}
  % Higher accuracy for each n-gram of an example's prediction suggests a high probability the sample was encountered during the training process. 
    \begin{tabular}{c|cc|c}
    \toprule
    \textbf{Models} & \textbf{CodeScope} & \textbf{CodeApex} & \textbf{CodeMMLU} \\
    \midrule
    Mistral7B-v0.3 & 0.2510 & 0.1702 & 0.1365 \\
    DeepSeekCoder7B-v1.5 & 0.2818 & 0.1680 & 0.1416 \\
    DeepSeekV2-Lite & 0.2492 & 0.1587 & 0.0687 \\
    Llama-3.1-8B & 0.2219 & 0.1309 & 0.0652 \\
    \bottomrule
    \end{tabular}%
  \label{tab:appendix_ngram}%
\end{table}%

% \subsection{Extended Description of Tasks}
% \label{appendix:task_detail_extend}

% \subsubsection{Code Completion}

% \subsubsection{Fill-in-the-blank}

% \subsubsection{Execution Prediction}

% Refer to Figure \ref{fig:appendix_execute_pred}.

% \begin{figure}[http]
%     \centering
%     \includegraphics[width=0.7\linewidth]{img/execution_prediction.pdf}
%     \caption{\textbf{Execution Prediction test set portion.}}
%     \label{fig:appendix_execute_pred}
% \end{figure}

% \subsubsection{Code Repair}

\subsection{License}
\label{appendix:license}
In the construction of CodeMMLU, we collect only the multiple-choice questions, problem descriptions, code solutions, and test cases from the publicly visible parts of W3School and Geeksforgeeks quizzes/puzzles and LeetCode. We avoid any data collection that requires login or interaction with these websites. On one hand, most of our knowledge test set ($\approx61\%$) are collected from Common Crawl (from portion tagged CC-MAIN-2021-41 to CC-MAIN-2024-46). On the other hand, the fundamental tasks were created on a permissively licensed codebase, namely IBM Project CodeNet (Apache 2.0), HumanEval, QuixBugs (MIT). For data crawled from websites such as W3Schools (fair use for research purposes) and GeeksforGeeks (under the Copyright Act 1957), we fully complied with their copyrights or sought their permission to use such data for this project. CodeMMLU will be published and distributed under the MIT license.
\section{All Experimental Results}

\begin{table}[htbp]
  \centering
  \caption{\textbf{CodeMMLU and other coding benchmarks comparison.} The ranking reorder comparison between CodeMMLU (CM) and other benchmarks (namely HumanEval (HE)).}
  \scalebox{0.68}{%
    \begin{tabular}{c|l|c|cccc|c|c}
    \toprule
    \multicolumn{1}{c}{\textbf{Family}} & \multicolumn{1}{c}{\textbf{Model}} & \multicolumn{1}{c}{\textbf{Size (B)}} & \textbf{MMLU} & \textbf{GSM8k} & \textbf{HumanEval} & \multicolumn{1}{c}{\textbf{MBPP}} & \multicolumn{1}{c}{\textbf{CodeMMLU}} & \multicolumn{1}{c}{\textbf{HE$\rightarrow$CM}} \\
    \midrule
    \midrule
    \multicolumn{9}{c}{\textit{Closed-source models}} \\
    \midrule
    Anthropic & Claude-3 Sonnet & -     & \textbf{88.70} & \textbf{96.40} & \textbf{92.00} & 76.6  & 55.48 & \textbf{1}$\rightarrow$4 \\
    \multirow{2}[1]{*}{OpenAI} & GPT-4o & -     & \textbf{88.70} & 95.80 & 90.20 & \textbf{\underline{81.4}} & \textbf{64.96} & \textbf{\underline{2}}$\rightarrow$\textbf{1} \\
          & GPT-3.5-turbo & -     & 61.90 & 73.80 & 61.40 & 78.5  & 51.59 & 10$\rightarrow$6 \\
    \midrule
    \multicolumn{9}{c}{\textit{Open-source models}} \\
    \midrule
    \multirow{5}[2]{*}{MetaLlama} & Llama3.1 70B Instruct & 70    & \textbf{\underline{83.60}} & \textbf{\underline{95.10}} & 80.50 & 75.4  & 59.68 & 6$\rightarrow$3 \\
          & Llama3.1 70B & 70    & 79.30 & 83.70 & 58.50 & 66.2  & 40.45 & 11$\rightarrow$20 \\
          & Llama3 70B & 70    & 79.50 & 83.00 & 48.20 & 70.4  & 49.7  & 14$\rightarrow$8 \\
          & Llama3 70B Instruct & 70    & 82.00 & 93.00 & 81.70 & \textbf{82.3} & \textbf{\underline{61.79}} & 4$\rightarrow$\textbf{\underline{2}} \\
          & CodeLlama 34B Instruct & 34    & -     & -     & 41.50 & 57    & 39.27 & 17$\rightarrow$21 \\
    \midrule
    \multirow{3}[2]{*}{Mistral} & Mistral 7B Instruct (v0.3) & 7     & 62.50 & 50.00 & 26.20 & 50.2  & 44.14 & 21$\rightarrow$17 \\
          & Mixtral 8x7B Instruct & 46.7  & 70.60 & 74.40 & 40.20 & 60.7  & 42.74 & \multicolumn{1}{c}{18} \\
          & Codestral 22B & 22    & -     & -     & 81.10 & 78.2  & 47.61 & 5$\rightarrow$13 \\
    \midrule
    \multirow{2}[2]{*}{Phi} & Phi3 Medium 128k Instruct & 14    & 78.00 & 91.00 & 62.20 & 75.2  & 48.65 & \multicolumn{1}{c}{9} \\
          & Phi3 Mini 128k Instruct & 3.8   & 68.80 & 82.50 & 58.50 & 70    & 39.22 & 11$\rightarrow$22 \\
    \midrule
    \multirow{3}[2]{*}{Qwen} & Qwen2 7B Instruct & 7     & 70.50 & 82.30 & 79.90 & -     & 51.86 & 7$\rightarrow$5 \\
          & Qwen2 57B-A14B Instruct & 57    & 76.50 & 80.70 & 53.00 & 71.9  & 47.34 & 12$\rightarrow$14 \\
          & CodeQwen1.5 7B Chat & 7     & -     & -     & \textbf{\underline{83.50}} & 77.7  & 47.71 & 3$\rightarrow$12 \\
    \midrule
    \multirow{2}[2]{*}{Yi} & Yi1.5 34B Chat & 34    & 67.62 & 71.70 & 23.20 & 41     & 50.03 & 22$\rightarrow$7 \\
          & Yi1.5 9B Chat & 9     & 68.40 & 52.30 & 39.00 & 54.4  & 48.15 & 19$\rightarrow$10 \\
    \midrule
    \multirow{4}[2]{*}{DeepSeek} & DeepSeek Coder 7B Instruct (v1.5) & 7     & 49.20 & 41.00 & 42.10 & 60.7  & 41.59 & 16$\rightarrow$19 \\
          & DeepSeek Coder 33B Instruct & 33    & -     & 60.70 & 79.30 & 70    & 37.45 & 8$\rightarrow$23 \\
          & DeepSeek Moe 16B Chat & 16.4  & 45.00 & 18.80 & 26.80 & 39.2  & 31.45 & 20$\rightarrow$24 \\
          & DeepSeek CoderV2 Lite Instruct & 16    & 60.10 & 86.40 & 81.10 & -     & 47.12 & 5$\rightarrow$15 \\
    \midrule
    InternLM & InternLM2.5 20B Chat & 20    & 66.50 & 79.60 & 48.80 & 63    & 46.15 & 13$\rightarrow$16 \\
    \midrule
    StarCoder & StarCoder2 15B Instruct & 15    & -     & -     & 46.3  & 66.2  & 47.76 & 15$\rightarrow$11 \\
    \bottomrule
    \end{tabular}%
    }
    \label{tab:appendix:ben_compare}%
\end{table}%

\subsection{Selection bias in MCQs format}
\label{appendix:mcq_bias}

Building on the findings from \citep{llm-mcq-bias}, which investigated the effects of reordering answer options in multiple-choice questions (MCQs), we observe inconsistent behavior among large language models (LLMs) when performing the same code completion task. Table \ref{tab:appendix_mcq_bias} highlights the sensitivity of LLMs to the order of answers, even for models renowned for their high performance (e.g., GPT, Claude, MetaLlama). Specifically, the results reveal that most models experience significant performance degradation when the correct answer is positioned as ``A'', with an average performance drop of \textbf{25\%}. In contrast, placing the correct answer in position ``B'' leads to a marked performance improvement, with an average increase of \textbf{15.49\%}.

\begin{figure}[htbp]
    \centering
    \includegraphics[width=0.7\linewidth]{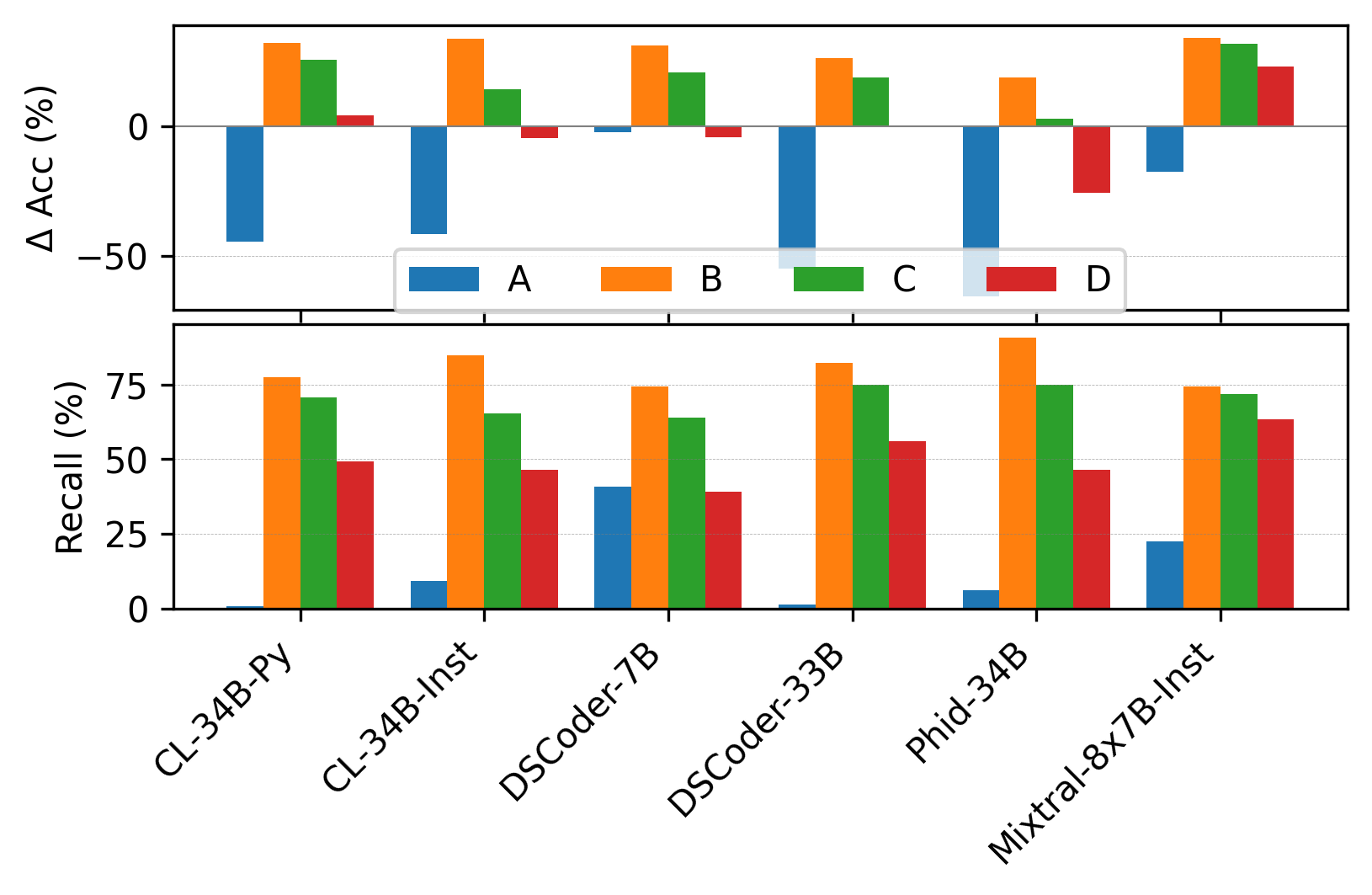}
    \caption{\textbf{Task-Specific Accuracy and Performance Fluctuations Across Answer Options} Models exhibit marked fluctuations in accuracy depending on the position of the correct answer in Code Completion in CodeMMLU. Revealing the bias and inconsistencies in related coding multiple-choice question (MCQ) task and how sensitive LLMs are to answer ordering.}
    \label{fig:intro_bias}
\end{figure}

The standard deviation (STD) further illustrates how differently models respond to answer reordering. For instance, models such as CodeLlama-7B/13B/34B and DeepSeekCoder-33B exhibit substantial dependency on the arrangement of options, whereas models like GPT-4o/3.5, Claude-3, and Claude-3.5 show greater resilience to such selection bias. Interestingly, instruction-tuned models, which are generally expected to demonstrate increased robustness, show minimal to no improvement over their base versions in this regard.

These findings suggest that higher-quality models are more resistant to MCQ biases, reflecting a human-like ability to maintain performance irrespective of answer order. We believe that introducing this MCQ bias into the CodeMMLU benchmark adds an extra layer of difficulty for LLMs, encouraging the research community to prioritize enhancing the consistency and robustness of LLMs.

% Draft
% Suggesting by \cite{llm-mcq-bias}'s experiment of reordering the answer position permutations, we observe an inconsistent behavior of LLM in performing the same code completion task. Table \ref{tab:appendix_mcq_bias} demonstrates how sensitive LLM can be to the order of answers, even models that are well-known for their high quality (e.g GPT, Claude, MetaLlama). The full result shows that most models suffer from degradation when the correct answer was in position 'A’, with an average of \textbf{25\%} performance drop, while placing it in position ‘B’ resulted in a sharp increase to \textbf{15.49\%} (on average).

% Standard deviation (STD) indicates that while models like CodeLlama-7B/13B/34B and DeepSeekCoder-33B heavily depend on how the options are presented, models like GPT-4o/3.5, Claude-3, and MetaLlama-3.1 are less affected by the selection bias. Interestingly, models with instruction tuning are expected to be more robust than the base version, which actually shows little to no difference. We believe this experiment suggests models with high quality are more robust and have stronger resistance to MCQ bias since this mimics the fact that answer order switching in MCQ format does not gain any difficulty from a human perspective. This MCQ bias introduced into CodeMMLU is hoped to gain more layers of difficulty for LLM in general and encourage the community to pay more attention to keeping LLM consistent and robust.

\begin{table}[htbp]
  \centering
  \caption{\textbf{Selection bias effect comparison on LLMs.} The performance fluctuation trends show a significant margin of model with high quality and the other. STD stands for standard deviation.}
    \begin{tabular}{c|c|cccc|c}
    \toprule
    \textbf{Models} & \textbf{Instructed} & \textbf{A} & \textbf{B} & \textbf{C} & \textbf{D} & \textbf{STD} \\
    \midrule
    \midrule
    GPT-4o & \checkmark & 80.49 & 78.05 & 71.34 & 70.12 & 4.38 \\
    GPT-3.5-turbo & \checkmark & 51.22 & 43.29 & 47.56 & 54.88 & 4.30 \\
    \midrule
    Claude3.5 Sonnet & \checkmark & 90.24 & 81.1  & 85.37 & 79.27 & \textbf{\underline{4.23}} \\
    Claude3.5 Haiku & \checkmark & 86.59 & 69.51 & 72.56 & 68.29 & 7.30 \\
    \midrule
    Claude3 Opus & \checkmark & 79.27 & 77.44 & 82.32 & 84.76 & \textbf{2.81} \\
    Claude3 Sonnet & \checkmark & 62.8  & 64.02 & 73.17 & 73.78 & 5.06 \\
    Claude3 Haiku & \checkmark & 56.1  & 75    & 73.78 & 76.83 & 8.34 \\
    \midrule
    Mixtral 8x7B & \checkmark & 22.56 & 74.39 & 71.95 & 63.41 & 20.91 \\
    DSCoder 33B & -     & 1.22  & 82.32 & 75.00 & 56.10 & 31.75 \\
    DSCoder 7B & -     & 40.85 & 74.39 & 64.02 & 39.02 & 15.10 \\
    Phind-CL 34B & \checkmark & 6.10  & 90.85 & 75.00 & 46.34 & 32.21 \\
    \midrule
    CL 34B Python & -     & 0.61  & 77.44 & 70.73 & 49.39 & 30.09 \\
    CL 34B Instruct & \checkmark & 9.15  & 84.76 & 65.24 & 46.34 & 27.91 \\
    CL 13B Python & -     & 0.61  & 54.88 & 70.12 & 12.20 & 28.85 \\
    CL 13B Instruct & \checkmark & 2.44  & 68.29 & 72.56 & 29.88 & 28.85 \\
    CL 7B Python & -     & 0.00  & 90.24 & 14.02 & 0.61  & 37.39 \\
    CL 7B Instruct & \checkmark & 3.66  & 1.22  & 93.90 & 15.85 & 38.07 \\
    \bottomrule
    \end{tabular}%
  \label{tab:appendix_mcq_bias}%
\end{table}%

% \subsection{Impact of model scaling and data on CodeMMLU}
% \label{appendix:scaling_law}

% Follow up on the insight that scaling law is aligned with the CodeMMLU experiment, and we dig deeper inside the advancement of the larger model compared to the smaller one by analyzing use cases.

\subsection{CoT might not be all you need}
\label{appendix:cot_study}

% \begin{figure}
%     \centering
%     \includegraphics[width=0.5\linewidth]{img/prompt-setting.png}
%     \caption{Caption}
%     \label{fig:appendix-prompt-setting-summary}
% \end{figure}

\begin{figure}[ht]
    \centering 
    \includegraphics[width=\linewidth]{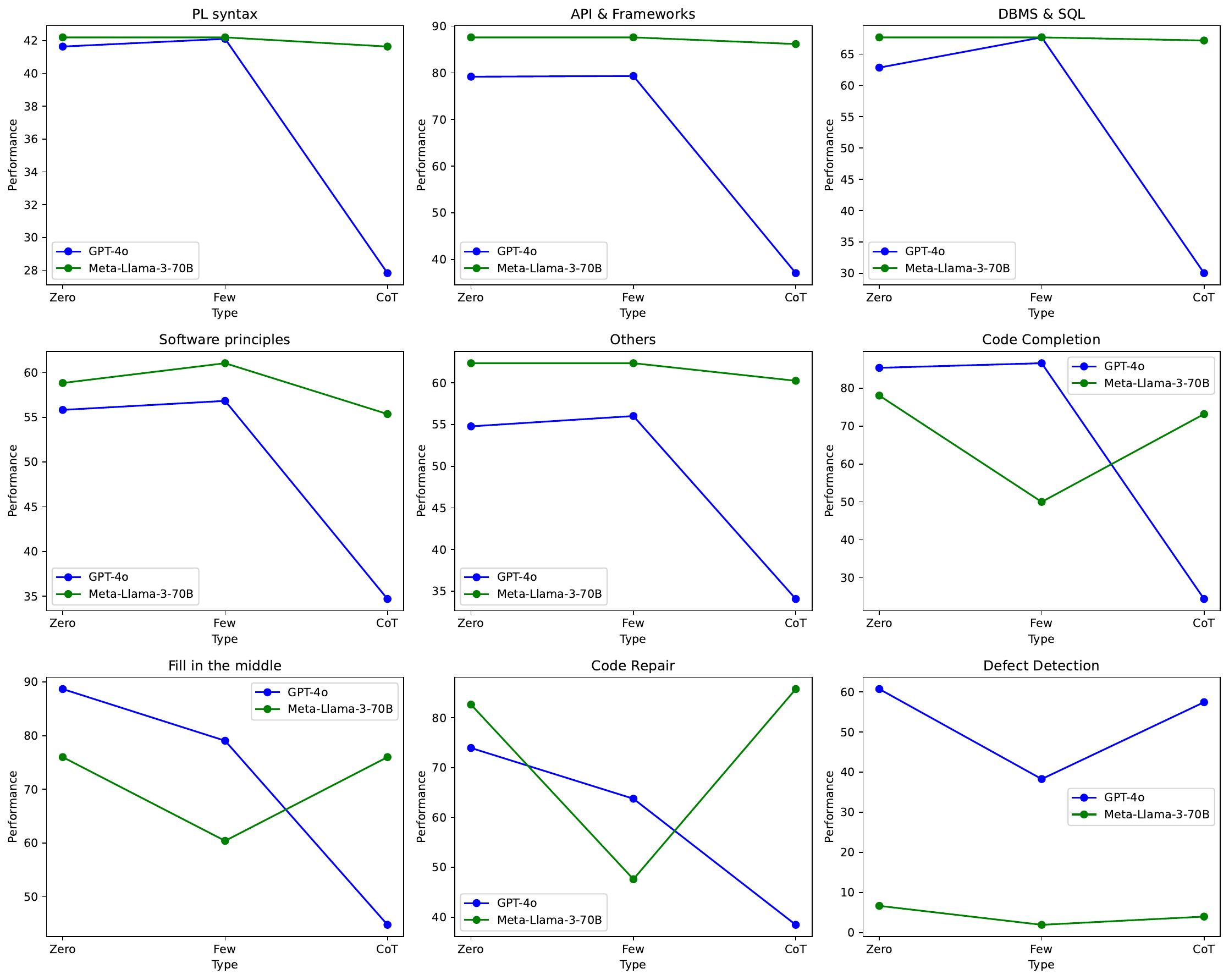}
    \caption{\textbf{Comparison between GPT4o and Meta Llama-3 70B on various prompt settings}. We experiment with zero-shot, 1-shot, and CoT prompt configuration, where the result indicates the ineffectiveness of CoT in boosting the models' performance. Comparing to zeroshot config, 1-shot prompt slightly increase the performance in knowledge tasks but falls shorter in real tasks.}
    \label{fig:appendix-prompt-settings}
\end{figure}

In our experiments with models from over 15 families, we evaluated CodeMMLU under two different prompt settings: standard zero-shot and few-shot, as well as Chain-of-Thought (CoT) with short and long prompts. The detailed results, provided in \ref{appendix:full_result}, reveal a consistent trend of decreased performance in the CoT setting compared to zero-shot and few-shot configurations. Even strong models like GPT-4o and Llama3 70B, known for their robust reasoning and comprehensive capabilities, exhibited this pattern, as illustrated in Figure \ref{fig:appendix-prompt-settings}.

\begin{figure}[htbp]
    \centering
    \includegraphics[width=0.8\linewidth]{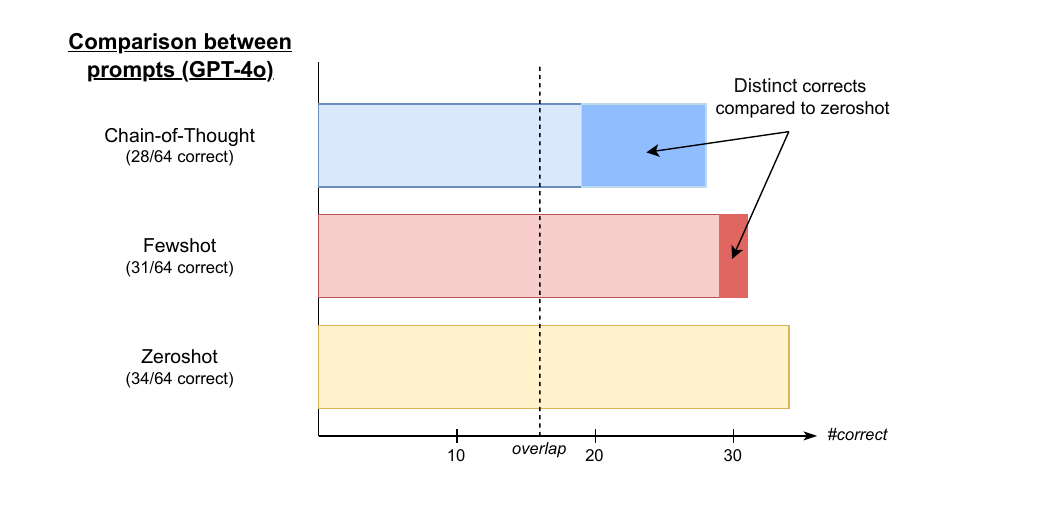}
    \caption{\textbf{Comparison experiment of different prompts by GPT-4o in OOP subject.}}
    \label{fig:appendix:cot_analysis}
\end{figure}

\begin{figure}[htbp]
    \centering
    \includegraphics[width=\linewidth]{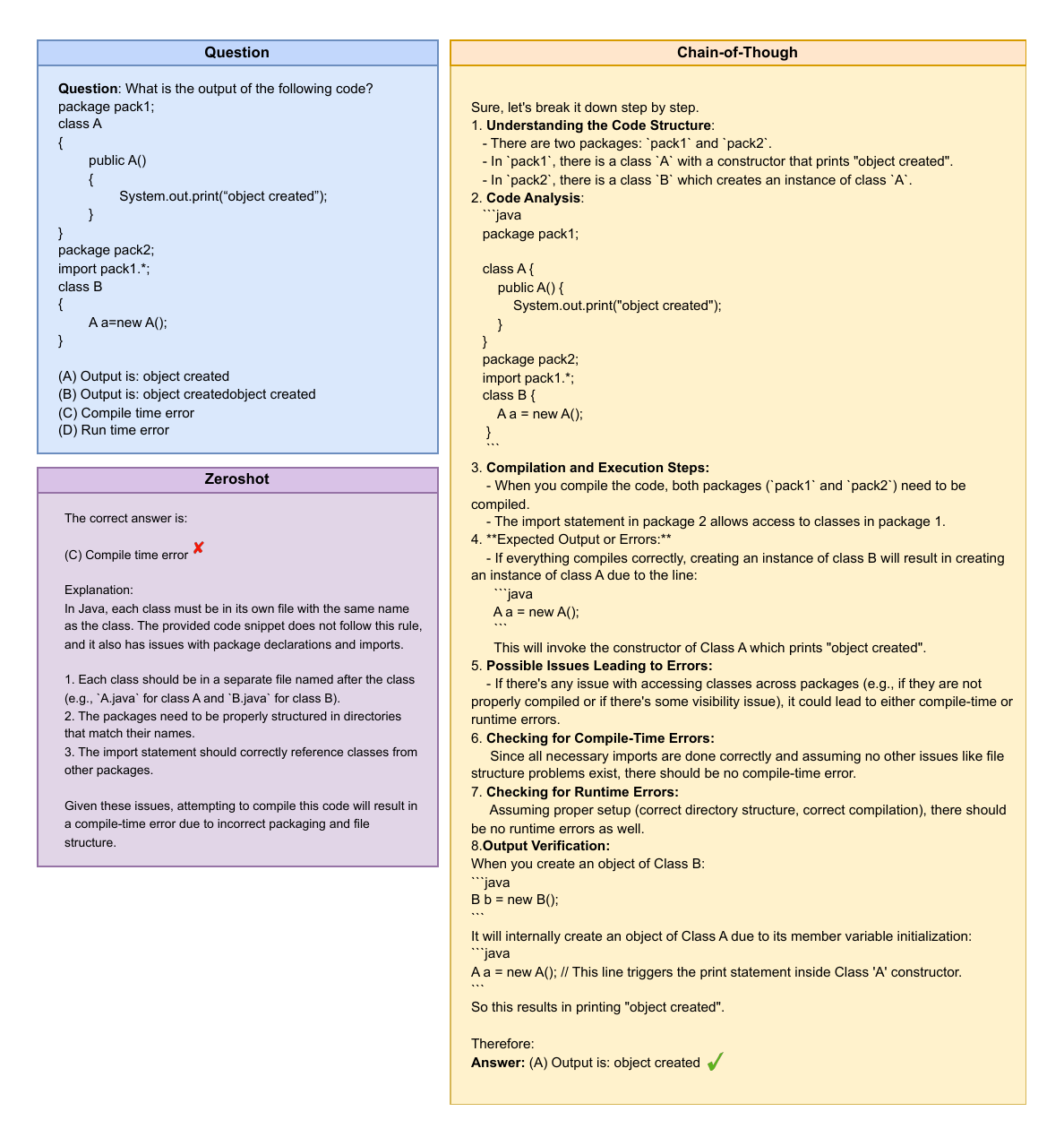}
    \caption{\textbf{Examples of Chain-of-Thought and zeroshot in Object-oriented programming (OOP) subset.}}
    \label{fig:appendix:corr_cot_example}
\end{figure}

A closer analysis of results for zero-shot, few-shot, and CoT prompts, particularly in the GPT-4o experiment on the object-oriented programming (OOP) subject (Figure \ref{fig:appendix:cot_analysis}), highlights the effects of an additional “thinking” step introduced by CoT. While this step intended to enhance reasoning, the thought often diverged from the original focus or brought up another challenge for the task. For instance, example 1 in Figure \ref{fig:appendix:corr_cot_example} showcases a case where the CoT approach incorporated an unnecessary definitional analysis but failed to arrive at the correct answer. Similarly, examples 2 and 3 (Figure \ref{fig:appendix:corr_cot_example}) demonstrate misinterpretations during the reasoning process. This additional “thinking” step often introduced new challenges that might lead to incorrect conclusions, resulting in a significant performance gap between zero-shot and CoT prompts.

Interestingly, when visualizing the newly addressed questions by CoT settings (compared to 0-shot), we do not observe a clear pattern of these new corrects. Questions that CoT addressed were neither inherently tricky nor required deeper interpretive reasoning. For example, example 4 (Figure \ref{fig:appendix:corr_cot_example}) asks for a prediction of program outcomes within the OOP domain, comparable in scope to example 3, yet the CoT approach showed inconsistent performance.

As concluded by the findings in \citep{sprague2024cotcotchainofthoughthelps}, the CoT strategy introduces an additional reasoning step that is effective primarily for tasks involving mathematical or logical problem-solving. However, for benchmarks like MMLU, the results with and without CoT prompting are nearly identical. The observed performance decline in the CodeLLM knowledge test set aligns with \citep{sprague2024cotcotchainofthoughthelps}, further underscoring the inefficiency of CoT for non-reasoning tasks.

\begin{figure}[htbp]
    \centering
    \includegraphics[width=\linewidth]{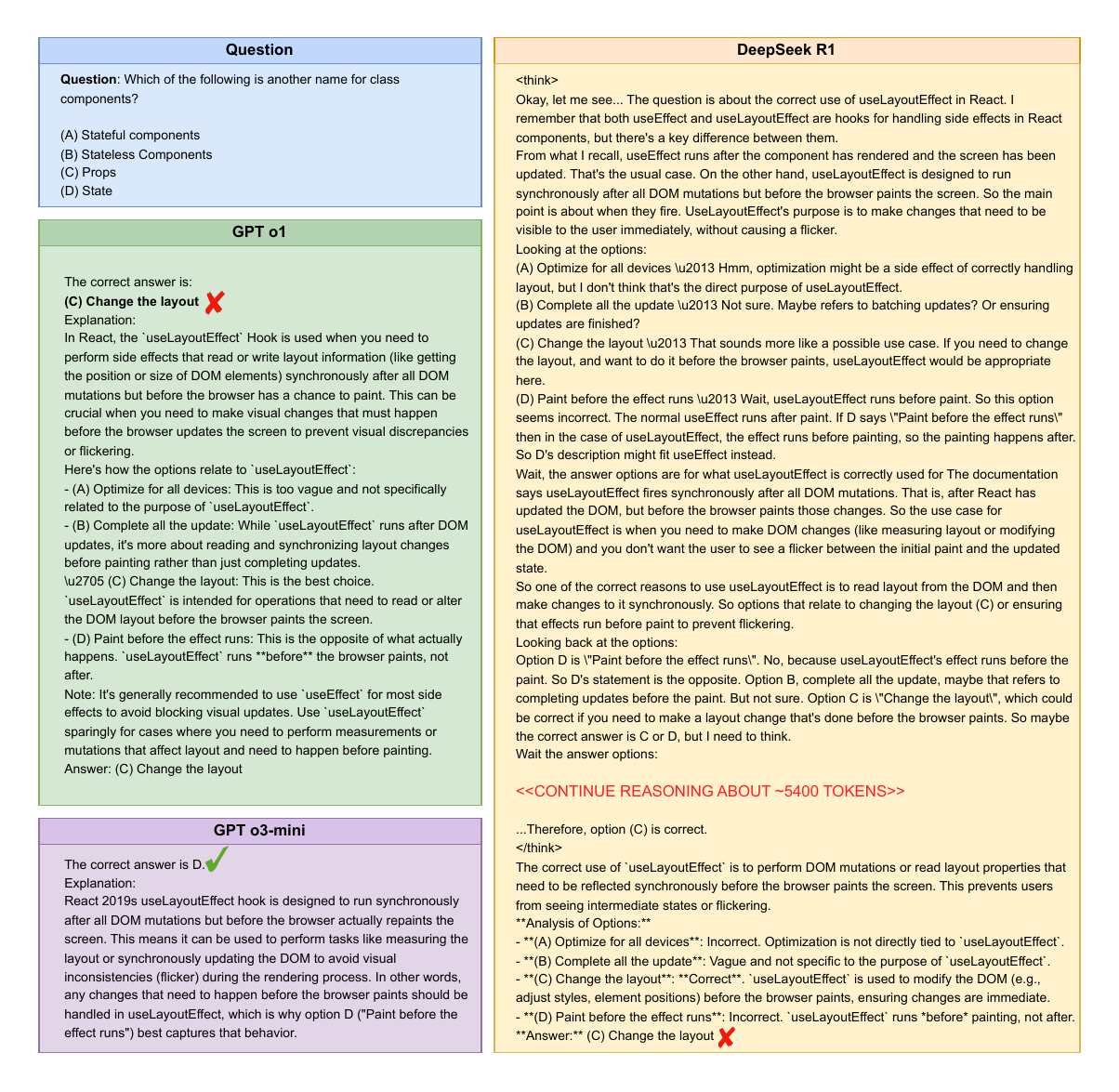}
    \caption{Comparison reasoning process of GPT o1, GPT o3-mini and DeepSeek R1 on knowledge-seeking task.}
    \label{fig:appendix:reasoning_1}
\end{figure}

\begin{figure}
    \centering
    \includegraphics[width=\linewidth]{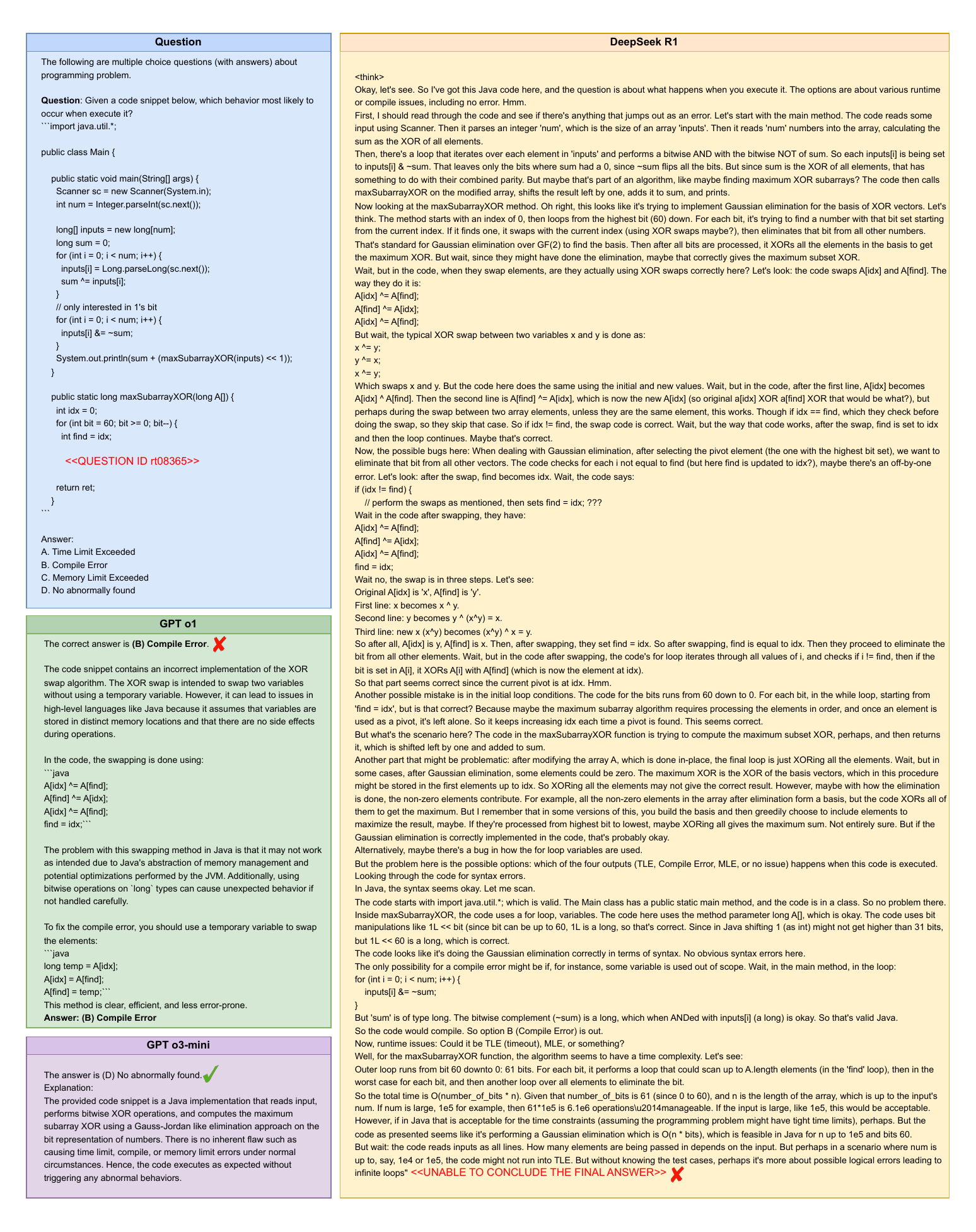}
    \caption{Comparison reasoning process of GPT o1, GPT o3-mini and DeepSeek R1 on problem solving task}
    \label{fig:reasoning_2}
\end{figure}

\begin{figure}[htbp]
    \centering
    \includegraphics[width=\linewidth]{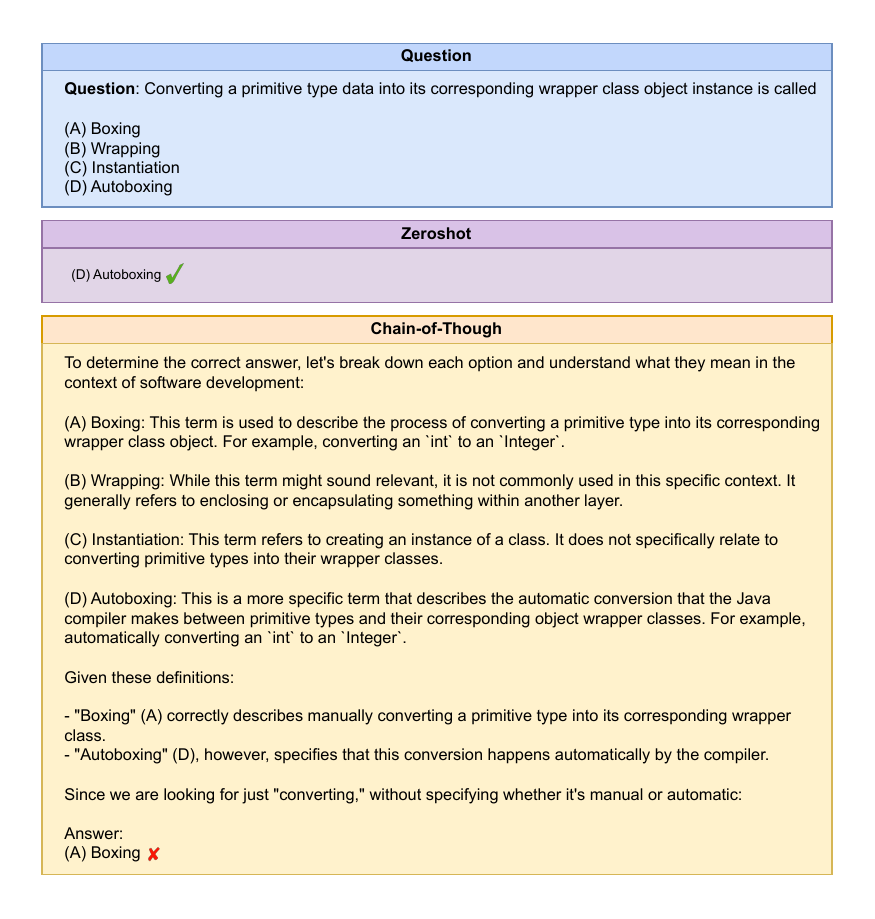}
    \label{fig:appendix:cot_example_2}
\end{figure}

\begin{figure}[htbp]
    \centering
    \includegraphics[width=\linewidth]{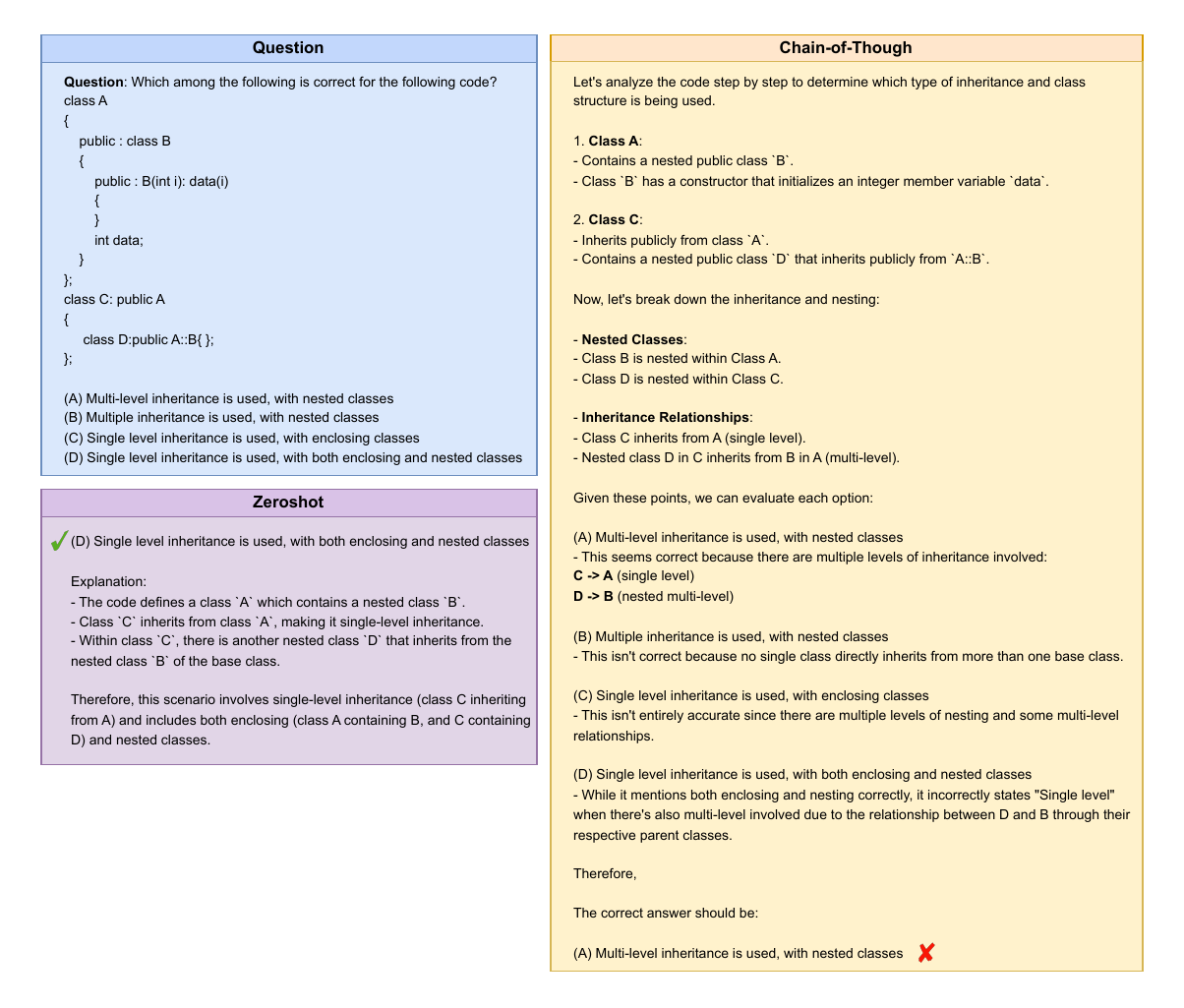}
    \label{fig:appendix:cot_example_1}
\end{figure}

\begin{figure}[htbp]
    \centering
    \includegraphics[width=\linewidth]{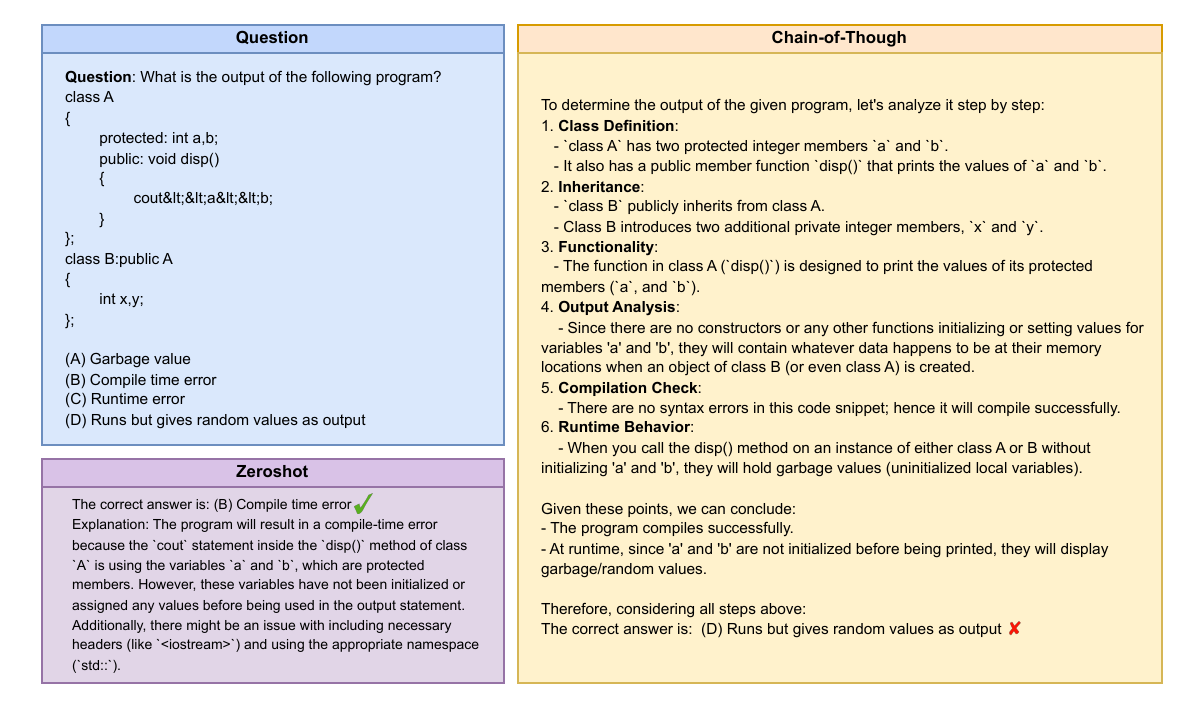}
    \label{fig:appendix:cot_example_3}
\end{figure}

\subsection{Experiment Results}
\label{appendix:full_result}

We provide the full experiment results of 43 LLMs from 15 families in Table \ref{tab:full_exp}, \ref{tab:full_exp_2}, \ref{tab:full_exp_3}.

% Table generated by Excel2LaTeX from sheet '_zero'
\newpage
\begin{landscape}
\begin{longtable}[c]
{p{4.25cm}|p{1.4cm}p{1.25cm}|p{1.25cm}p{1.25cm}p{1.25cm}|p{1.25cm}p{1.4cm}p{1.25cm}p{1.4cm}}
    \caption{\textbf{CodeMMLU zeroshot experimental results.}} 
\label{tab:full_exp} \\
    \textbf{Model} & \textbf{API \& Frameworks} & \textbf{PL Syntax} & \textbf{Software principle} & \textbf{DBMS \& SQL} & \textbf{Others} & \textbf{Code Completion} & \textbf{Fill in the blank} & \textbf{Code Repair} & \textbf{Execution Prediction} \\
    \midrule
    \midrule
    Claude3.7 Sonnet & 88.45 &	48.76 &	77.85 & 83.72 &	70.82 &	79.27 &	67.87 &	73.68 &	57.79 \\
    Claude3.5 Sonnet & 89.44 &	48.04 &	74.55 &	78.12 &	69.80 &	87.80 &	62.24 &	61.84 &	56.41 \\
    Claude3.5 Haiku & 84.59 &	45.26 &	69.23 &	70.99 &	65.28 &	75.00 &	69.94 &	50.00 &	53.16 \\
    Claude3 Sonnet & 89.02 & 45.42 & 62.27 & 72.52 & 63.46 & 4.88  & 87.98 & 56.86 & 3.32 \\
    \midrule
    GPT o3-mini	& 88.45 & 49.10 &	77.46 &	81.68 &	69.66 &	87.20 &	73.27 &	67.11 &	58.33 \\
    GPT 4o &	87.02 &	46.08 &	69.93 &	70.99 &	65.21 &	67.68 &	54.39 &	59.21 &	51.27 \\
    GPT 4o-mini &	88.45 &	44.18 &	50.12 &	70.23 &	63.82 &	65.85 &	39.50 &	46.05 &	11.97 \\
    GPT 3.5-turbo & 84.88 & 38.47 & 51.06 & 58.52 & 51.35 & 76.83 & 29.31 & 38.15 & 36.74 \\
    \midrule
    CodeLlama 7B Instruct & 72.04 & 28.23 & 36.80 & 47.84 & 39.02 & 0.00  & 6.95  & 7.90  & 4.26 \\
    CodeLlama 7b Python & 54.21 & 23.18 & 35.28 & 43.77 & 31.58 & 42.68 & 17.90 & 10.74 & 6.11 \\
    CodeLlama 13B Instruct & 72.04 & 29.12 & 39.09 & 50.89 & 39.02 & 0.00  & 39.97 & 1.35  & 0.79 \\
    CodeLlama 13B Python & 0.00  & 25.79 & 14.65 & 0.00  & 0.00  & 51.22 & 70.60 & \multicolumn{1}{l}{-} & 7.38 \\
    CodeLlama 13B & 72.04 & 29.12 & 39.08 & 50.89 & 38.88 & 1.22  & 20.43 & 0.00  & 4.96 \\
    CodeLlama 34B Instruct & 79.17 & 34.44 & 41.70 & 52.93 & 46.17 & 37.80 & 24.80 & 27.82 & 3.77 \\
    CodeLlama 34B Python & 0.00  & 31.14 & 16.02 & 0.00  & 0.00  & 12.20 & 8.31  & 14.66 & 2.30 \\
    \midrule
    Llama3 70B & 86.02 & 40.74 & 52.92 & 63.10 & 56.89 & 21.95 & 92.02 & 24.91 & 2.28 \\
    Llama3 70B Instruct & 87.59 & 42.20 & 58.83 & 67.68 & 62.36 & 78.05 & 76.00 & 82.64 & 6.66 \\
    Llama3 8B & 75.89 & 32.38 & 42.71 & 53.94 & 46.75 & 48.17 & 94.65 & 66.70 & 5.85 \\
    Llama3 8B Instruct & 81.31 & 34.89 & 33.43 & 60.31 & 50.91 & 60.37 & 55.52 & 34.30 & 3.33 \\
    Llama3.1 405B Instruct &	91.01 &	46.29 &	72.34 &	74.55 &	68.56 &	88.41	& 54.58 &	59.21 &	57.11 \\
    Llama3.1 70B & 87.02 & 41.16 & 53.44 & 65.65 & 57.91 & 3.05  & 4.93  & 22.42 & 2.48 \\
    % Llama3.1 70B Instruct & 87.45 & 41.38 & 58.24 & 69.21 & 59.30 & 43.90 & 84.64 & 92.17 & 3.72 \\
    Llama3.1 70B Instruct &	87.30 &	42.40 &	62.71 &	67.18 &	61.20 &	76.83 &	54.49 &	46.05 &	52.11 \\
    Llama3.1 8B & 75.04 & 32.86 & 41.94 & 54.71 & 47.63 & 35.37 & 33.26 & 52.60 & 5.11 \\
    Llama3.1 8B Instruct & 78.32 & 34.76 & 47.17 & 53.44 & 50.47 & 11.59 & 71.54 & 55.79 & 3.88 \\
    Llama3.3 70B Instruct &	87.59 &	39.44 &	50.72 &	62.09 &	54.34 &	81.71 &	63.27 &	38.16 &	18.03 \\
    \midrule
    Mistral 7B Instruct (v1) & 74.75 & 30.74 & 41.87 & 53.44 & 43.18 & 59.76 & 79.66 & 23.63 & 2.90 \\
    Mistral 7B Instruct (v2) & 71.90 & 32.38 & 42.92 & 54.96 & 46.02 & 43.90 & 10.38 & 47.61 & 2.15 \\
    Mistral 7B Instruct (v3) & 75.46 & 33.38 & 47.77 & 57.76 & 48.21 & 56.10 & 16.30 & 52.74 & 2.25 \\
    Mixtral 8x7B Instruct & 85.02 & 37.32 & 49.96 & 61.83 & 52.88 & 3.05  &       & 33.09 & 4.32 \\
    Codestral 22B & 82.17 & 38.50 & 47.62 & 58.52 & 50.18 & 74.39 & 25.69 & 48.82 & 2.53 \\
    \midrule
    Phi-4 & 87.02 &	40.64 &	56.08 &	67.43 &	57.48 &	75.61 &	58.43 &	60.53 &	43.14 \\
    Phi-4 Mini Instruct &	87.02 &	36.86 &	48.86 &	61.83 &	54.34 &	23.17 &	23.53 &	35.53 &	18.12 \\
    Phi3 Medium Instruct (128k) & 79.89 & 37.19 & 51.44 & 58.78 & 53.46 & 67.68 & 40.82 & 39.57 & 3.48 \\
    Phi3 Medium Instruct (4k) & 80.03 & 36.82 & 52.53 & 62.09 & 54.41 & 72.56 & 44.76 & 51.66 & 3.68 \\
    Phi3 Mini Instruct (128k) & 74.18 & 31.83 & 44.99 & 54.20 & 46.75 & 42.07 & 31.38 & 14.45 & 1.55 \\
    Phi3 Mini Instruct (4k) & 76.32 & 33.14 & 40.70 & 53.94 & 45.30 & 53.66 & 24.00 & 30.53 & 2.35 \\
    Phi3 Small Instruct (8k) & 77.89 & 37.32 & 54.40 & 64.12 & 52.22 & 62.80 & 21.42 & 21.07 & 1.95 \\
    \midrule
    PhindCL 34B v2 & 79.89 & 35.26 & 44.63 & 51.91 & 45.88 & 53.05 & 27.43 & 18.50 & 3.06 \\
    \midrule
    Qwen2.5 14B Instruct &	86.59 &	41.85 &	57.83 &	66.92 &	58.13 &	80.49 &	76.66 &	38.16 &	41.94 \\
    QwQ 38B &	78.32 &	39.05 &	52.93 &	59.29 &	51.71 &	35.37 &	40.86 &	36.84 &	42.79 \\
    Qwen2 0.5B Instruct & 52.92 & 24.23 & 36.45 & 43.26 & 32.90 & 28.66 & 13.72 & 74.74 & 1.07 \\
    Qwen2 1.5B Instruct & 73.04 & 30.03 & 43.52 & 54.20 & 44.78 & 9.76  & 21.28 & 27.20 & 2.49 \\
    Qwen2 57B-A14B Instruct & 84.88 & 37.80 & 51.87 & 65.14 & 55.43 & 58.54 & 24.52 & 35.90 & 2.96 \\
    Qwen2 7B & 82.88 & 33.73 & 48.12 & 62.85 & 54.70 & 75.00 & 56.93 & 60.64 & 4.63 \\
    Qwen2 7B Instruct & 82.74 & 37.06 & 54.41 & 62.85 & 53.98 & 69.51 & 50.78 & 46.05 & 4.31 \\
    \midrule
    QwenCoder2.5 32B Instruct &	86.88 &	46.54 &	71.15 &	73.28 &	65.35 &	81.71 &	68.58 &	38.16 &	48.12 \\
    QwenCoder2.5 14B Instruct &	87.45 &	41.58 &	58.08 &	63.10 &	55.51 &	82.93 &	73.41 &	55.26 &	31.35 \\
    CodeQwen1.5 7B & 74.47 & 30.56 & 38.89 & 51.65 & 40.41 & 39.63 & 5.40  & \multicolumn{1}{l}{-} & 5.92 \\
    CodeQwen1.5 7B Chat & 67.62 & 31.70 & 38.25 & 46.06 & 39.24 & 89.02 & 37.06 & \multicolumn{1}{l}{-} & 3.38 \\
    \midrule
    Yi1.5 34B Chat & 79.60 & 37.03 & 52.93 & 61.32 & 52.52 & 67.68 & 35.09 & 55.30 & 3.00 \\
    Yi-1.5 6B Chat & 75.75 & 34.45 & 45.19 & 58.02 & 49.53 & 53.05 & 41.90 & 36.66 & 2.67 \\
    Yi1.5 9B Chat & 75.46 & 35.82 & 52.32 & 60.05 & 52.81 & 60.98 & 32.27 & 49.90 & 5.43 \\
    \midrule
    DeepSeek-R1	& 71.47 &	39.12 &	56.12 &	63.87 &	56.09 &	56.71 &	33.26 &	42.11 &	39.23 \\
    DeepSeek-V3 & 87.02 &	43.94 &	54.85 &	66.92 &	60.39	 & 56.10 &	41.33 &	51.32 &	46.00 \\
    \midrule
    DSCoder 33B & 0.00  & 0.00  & 15.98 & 0.00  & 0.00  & 6.10  & 13.62 & 18.64 & 5.83 \\
    DSCoder 33B Instruct & 74.89 & 32.40 & 41.24 & 52.16 & 42.89 & 8.54  & 31.94 & 40.06 & 5.32 \\
    DSCoder 6.7B & 70.19 & 28.71 & 36.96 & 49.36 & 39.10 & 16.46 & 0.09  & 2.63  & 0.02 \\
    DSCoder 6.7B Instruct & 71.75 & 29.84 & 38.01 & 52.67 & 38.15 & 26.83 & 3.66  & 0.00  & 3.69 \\
    DSCoder 7B (v1.5) & 81.60 & 35.98 & 47.21 & 57.51 & 46.32 & 0.00  & 27.62 & 41.13 & 0.00 \\
    DSCoder 7B Instruct (v1.5) & 79.74 & 33.60 & 43.29 & 55.47 & 44.93 & 13.41 & 31.09 & 65.49 & 3.85 \\
    \midrule
    DSMoE 16B & 56.49 & 23.47 & 31.81 & 43.00 & 34.87 & 0.61  & 26.21 & 46.05 & 1.25 \\
    DSMoE 16B Chat & 41.94 & 21.54 & 33.44 & 41.48 & 31.36 & 18.90 & 50.54 & 37.21 & 2.65 \\
    \midrule
    DSCoderV2 Lite & 83.88 & 35.00 & 47.34 & 59.29 & 48.50 & 40.85 & 19.96 & 32.88 & 0.17 \\
    DSCoderV2 Lite Instruct & 81.74 & 38.07 & 49.85 & 64.38 & 50.04 & 46.34 & 45.04 & 39.57 & 3.53 \\
    \midrule
    InternLM2.5 20B Chat & 78.03 & 37.67 & 51.85 & 60.05 & 54.63 & 57.93 & 24.47 & 38.22 & 1.12 \\
    InternLM2.5 7B Chat & 79.32 & 35.32 & 50.93 & 56.74 & 51.71 & 59.15 & 18.69 & 26.47 & 5.41 \\
    \midrule
    StarCoder2 15B Instruct & 78.74 & 34.41 & 46.94 & 52.42 & 47.85 & 73.17 & 42.04 & 52.53 & 3.40 \\
    StarCoder2 7B & 63.77 & 27.97 & 34.19 & 48.85 & 36.25 & 30.49 & 67.78 & 6.62  & 4.81 \\
    \bottomrule
\end{longtable}%
\end{landscape}

% Table generated by Excel2LaTeX from sheet '_zero'
\newpage
\begin{landscape}
\begin{longtable}[c]
{p{4.25cm}|p{1.4cm}p{1.25cm}|p{1.25cm}p{1.25cm}p{1.25cm}|p{1.25cm}p{1.4cm}p{1.25cm}p{1.4cm}}
    \caption{\textbf{CodeMMLU few-shot experimental results.}} 
\label{tab:full_exp_2} \\
\toprule
\textbf{Model} & \textbf{PL syntax} & \textbf{API \& Frameworks} & \textbf{DBMS \& SQL} & \textbf{Software principles} & {\textbf{Others}} & {\textbf{Code Completion}} & {\textbf{Fill in the blank}} & {\textbf{Code Repair}} & {\textbf{Execution Prediction}} \\
\midrule
\midrule
Claude3 Sonnet & 44.58 & 86.31 & 70.99 & 64.73 & 63.68 & 11.59 & 40.11 & 52.67 & 3.11 \\
GPT-4o & 42.11 & 79.32 & 67.68 & 56.83 & 56.02 & 86.59 & 79.05 & 63.76 & 38.28 \\
GPT-3.5-turbo & 38.47 & 84.88 & 58.52 & 52.93 & 50.04 & 70.12 & 54.06 & 35.38 & 34.45 \\
\midrule
CodeLlama 13B Instruct & 29.12 & 72.04 & 50.89 & 39.09 & 39.02 & 62.80 & 17.61 & 25.71 & 3.03 \\
CodeLlama 13B Python & 25.79 & 62.62 & 44.02 & 35.16 & 33.41 & 0.61  & 0.00  & 1.28  & 0.03 \\
CodeLlama 13B & 29.12 & 72.04 & 50.89 & 42.04 & 38.88 & 0.61  & 0.19  & 0.00  & 0.00 \\
CodeLlama 34B Instruct & 34.44 & 79.17 & 52.93 & 44.51 & 46.17 & 10.37 & 0.00  & 27.03 & 0.07 \\
CodeLlama 34B Python & 31.14 & 74.18 & 50.13 & 39.94 & 42.52 & 1.22  & 0.00  & 1.28  & 0.69 \\
CodeLlama 7B Instruct & 28.23 & 72.04 & 47.84 & 39.56 & 39.02 & 1.22  & 44.20 & 6.69  & 0.02 \\
CodeLlama 7B Python & 23.49 & 54.07 & 44.27 & 37.23 & 31.80 & 0.00  & 0.00  & 0.00  & 0.12 \\
\midrule
Llama3 70B & 40.74 & 86.02 & 63.10 & 55.47 & 56.89 & 67.07 & 25.50 & 12.02 & 0.80 \\
Llama3 70B Instruct & 42.20 & 87.59 & 67.68 & 61.04 & 62.36 & 50.00 & 60.40 & 47.61 & 1.93 \\
Llama3 8B & 32.38 & 75.89 & 53.94 & 45.52 & 46.75 & 93.90 & 40.44 & 92.31 & 3.62 \\
Llama3 8B Instruct & 34.77 & 81.74 & 60.81 & 50.30 & 50.62 & 27.44 & 87.69 & 17.36 & 0.28 \\
Llama3.1 70B & 41.16 & 87.02 & 65.65 & 56.49 & 57.91 & 48.78 & 2.07  & 18.16 & 0.75 \\
Llama3.1 70B Instruct & 41.38 & 87.45 & 69.21 & 60.98 & 59.30 & 6.10  & 30.39 & 80.15 & 4.31 \\
Llama3.1 8B & 32.86 & 75.04 & 54.71 & 45.13 & 47.63 & 1.22  & 0.28  & 2.63  & 0.03 \\
Llama3.1 8B Instruct & 34.76 & 78.32 & 53.44 & 48.73 & 50.47 & 50.00 & 67.36 & 57.52 & 3.53 \\
\midrule
Mistral 7B Instruct (v1) & 30.74 & 74.75 & 53.44 & 44.76 & 43.18 & 14.02 & 7.84  & 44.53 & 2.13 \\
Mistral 7B Instruct (v2) & 32.38 & 71.90 & 54.96 & 45.93 & 46.02 & 23.78 & 5.31  & 7.97  & 0.75 \\
Mistral 7B Instruct (v3) & 33.38 & 75.46 & 57.76 & 50.27 & 48.21 & 19.51 & 2.35  & 9.32  & 0.84 \\
Mixtral 8x7B Instruct & 37.33 & 84.74 & 61.07 & 53.24 & 53.32 & 25.00 & 7.23  & 9.32  & 1.83 \\
Codestral 22B & 38.50 & 82.17 & 58.52 & 50.35 & 50.18 & 31.71 & 14.61 & 21.00 & 0.97 \\
\midrule
Phi3 Medium Instruct (128k) & 37.19 & 79.89 & 58.78 & 53.27 & 53.46 & 10.37 & 4.04  & 2.70  & 0.22 \\
Phi3 Medium Instruct (4k) & 36.82 & 80.03 & 62.09 & 54.92 & 54.41 & 7.32  & 6.34  & 29.66 & 0.50 \\
Phi3 Mini Instruct (128k) & 31.83 & 74.18 & 54.20 & 47.29 & 46.75 & 1.22  & 0.38  & 2.56  & 0.07 \\
Phi3 Mini Instruct (4k) & 33.14 & 76.32 & 53.94 & 44.01 & 45.30 & 0.00  & 0.19  & 0.00  & 0.05 \\
Phi3 Small Instruct (8k) & 37.32 & 78.60 & 64.12 & 56.83 & 52.22 & 56.10 & 56.65 & 38.77 & 1.90 \\
\midrule
Qwen2 57B-A14B Instruct & 37.80 & 84.88 & 65.14 & 55.19 & 55.43 & 42.07 & 19.26 & 14.59 & 1.14 \\
Qwen2 7B & 36.38 & 82.88 & 61.58 & 53.06 & 54.49 & 7.32  & 2.54  & 1.28  & 0.00 \\
Qwen2 7B Instruct & 37.01 & 82.74 & 62.85 & 56.35 & 54.27 & 35.98 & 43.68 & 10.60 & 2.91 \\
CodeQwen1.5 7B & 30.56 & 74.04 & 51.15 & 42.46 & 40.34 & 0.00  & 0.14  & 0.00  & 0.07 \\
CodeQwen1.5 7B Chat & 31.70 & 67.62 & 46.06 & 40.17 & 39.10 & 1.83  & 3.05  & 8.04  & 1.58 \\
\midrule
Yi1.5 34B Chat & 37.03 & 80.17 & 62.34 & 54.93 & 52.81 & 21.95 & 13.01 & 21.28 & 0.25 \\
Yi-1.5 6B Chat & 34.34 & 75.75 & 58.02 & 48.51 & 49.53 & 0.61  & 5.26  & 15.73 & 1.13 \\
Yi1.5 9B Chat & 35.74 & 75.46 & 60.05 & 54.19 & 52.81 & 0.61  & 24.99 & 3.98  & 0.07 \\
\midrule
DSCoder 33B & 29.26 & 74.75 & 48.60 & 40.93 & 40.41 & 0.61  & 0.00  & 1.28  & 0.00 \\
DSCoder 33B Instruct & 32.35 & 75.32 & 51.91 & 43.73 & 43.03 & 14.63 & 9.02  & 2.56  & 1.22 \\
DSCoder 6.7B & 28.71 & 70.19 & 49.36 & 40.06 & 39.10 & 1.22  & 0.28  & 1.28  & 0.00 \\
DSCoder 6.7B Instruct & 29.84 & 71.75 & 52.67 & 41.68 & 38.15 & 5.49  & 0.89  & 1.28  & 0.02 \\
DSCoder 7B (v1.5) & 36.08 & 81.31 & 57.51 & 49.79 & 46.46 & 37.80 & 63.74 & 23.22 & 2.60 \\
DSCoder 7B Instruct (v1.5) & 33.46 & 79.60 & 54.71 & 45.86 & 44.93 & 0.00  & 0.00  & 0.00  & 0.02 \\
DSMoE 16B & 23.89 & 56.49 & 43.51 & 35.59 & 34.72 & 0.00  & 0.09  & 0.00  & 0.00 \\
DSMoE 16B Chat & 21.54 & 43.37 & 41.73 & 35.87 & 31.15 & 0.61  & 0.05  & 0.00  & 0.08 \\
DSCoderV2 Lite & 35.00 & 83.88 & 59.29 & 50.33 & 48.50 & 0.61  & 0.00  & 1.28  & 0.00 \\
DSCoderV2 Lite Instruct & 38.07 & 81.74 & 64.38 & 53.48 & 50.04 & 7.93  & 14.09 & 2.63  & 0.17 \\
\midrule
InternLM2.5 20B Chat & 37.67 & 78.03 & 60.05 & 53.90 & 54.63 & 25.61 & 1.03  & 1.35  & 0.02 \\
\midrule
StarCoder2 15B Instruct & 34.41 & 78.89 & 52.93 & 47.85 & 47.70 & 4.27  & 4.23  & 2.63  & 0.17 \\
\bottomrule

\end{longtable}%
\end{landscape}

\begin{landscape}
\begin{longtable}[c]
{p{4.25cm}|p{1.4cm}p{1.25cm}|p{1.25cm}p{1.25cm}p{1.25cm}|p{1.25cm}p{1.4cm}p{1.25cm}p{1.4cm}}
    \caption{\textbf{CodeMMLU Chain-of-Though with zeroshot experimental results.}} 
\label{tab:full_exp_3} \\
\toprule
{\textbf{Model}} & {\textbf{PL syntax}} & {\textbf{API \& Frameworks}} & {\textbf{DBMS \& SQL}} & {\textbf{Software principles}} & \multicolumn{1}{l|}{\textbf{Others}} & {\textbf{Code Completion}} & {\textbf{Fill in the middle}} & {\textbf{Code Repair}} & {\textbf{Execution Prediction}} \\
\midrule
\midrule
Claude3 Sonnet & 42.60 & 41.94 & 71.76 & 72.81 & 61.20 & 0.61  & 61.34 & 48.61 & 3.26 \\
GPT-4o & 27.83 & 37.09 & 30.03 & 34.69 & 34.06 & 24.39 & 44.76 & 38.46 & 57.44 \\
GPT-3.5-turbo & 0.79  & 85.73 & 60.81 & 49.00 & 50.47 & 69.51 & 39.36 & 48.82 & 41.95 \\
\midrule
CodeLlama 7B Instruct & 27.30 & 69.47 & 45.55 & 38.43 & 37.93 & 37.20 & 35.23 & 29.18 & 2.70 \\
CodeLlama 7B Python & 24.49 & 59.49 & 42.24 & 30.03 & 30.49 & 56.71 & 0.00  & 51.35 & 0.00 \\
CodeLlama 13B Instruct & 30.06 & 76.89 & 50.64 & 37.65 & 40.04 & 20.12 & 40.02 & 65.70 & 2.45 \\
CodeLlama 13B Python & 27.17 & 69.33 & 46.06 & 30.03 & 36.11 & 0.00  & 0.00  & 100.00 & 0.00 \\
CodeLlama 13B & 30.05 & 76.75 & 50.89 & 37.60 & 40.04 & 0.61  & 0.52  & 3.92  & 0.03 \\
CodeLlama 34B Instruct & 34.00 & 77.32 & 54.45 & 42.76 & 45.37 & 79.27 & 41.94 & 18.16 & 2.32 \\
CodeLlama 34B Python & 23.87 & 74.75 & 51.91 & 32.06 & 40.34 & 0.00  & 1.69  & 30.46 & 0.77 \\
\midrule
Llama3 70B & 30.80 & 72.90 & 50.64 & 38.03 & 40.70 & 39.63 & 19.16 & 0.00  & 0.53 \\
Llama3 70B Instruct & 30.85 & 68.19 & 45.04 & 38.12 & 36.11 & 89.63 & 46.97 & 98.72 & 2.84 \\
Llama3 8B & 39.84 & 85.73 & 63.10 & 52.23 & 55.73 & 19.51 & 92.02 & 100.00 & 3.53 \\
Llama3 8B Instruct & 41.64 & 86.16 & 67.18 & 55.36 & 60.25 & 73.17 & 76.00 & 85.76 & 3.98 \\
Llama3.1 70B & 31.48 & 76.60 & 56.74 & 41.05 & 46.32 & 11.59 & 94.65 & 79.87 & 2.65 \\
Llama3.1 70B Instruct & 33.54 & 78.60 & 55.98 & 46.06 & 49.02 & 59.76 & 13.81 & 33.02 & 2.15 \\
Llama3.1 8B & 40.50 & 86.16 & 64.12 & 52.21 & 55.87 & 19.51 & 4.93  & 45.15 & 1.62 \\
Llama3.1 8B Instruct & 40.10 & 78.89 & 65.14 & 53.50 & 57.55 & 85.98 & 84.64 & 82.99 & 3.70 \\
\midrule
Mistral 7B Instruct (v1) & 31.98 & 75.04 & 55.98 & 43.99 & 46.24 & 23.17 & 17.57 & 38.15 & 0.00 \\
Mistral 7B Instruct (v2) & 33.41 & 78.74 & 54.20 & 46.02 & 46.54 & 72.56 & 71.54 & 44.70 & 2.81 \\
Mistral 7B Instruct (v3) & 26.53 & 63.91 & 50.64 & 35.32 & 33.84 & 75.00 & 62.61 & 74.95 & 4.53 \\
Mixtral 8x7B Instruct & 31.45 & 69.90 & 51.65 & 41.07 & 43.84 & 37.80 & 23.44 & 47.68 & 1.92 \\
Codestral 22B & 32.11 & 70.47 & 52.67 & 39.97 & 44.27 & 57.32 & 26.30 & 55.37 & 2.13 \\
\midrule
Phi3 Medium Instruct (128k) & 35.16 & 79.60 & 55.73 & 46.01 & 47.78 & 33.54 & 60.40 & 81.70 & 3.21 \\
Phi3 Medium Instruct (4k) & 38.22 & 79.32 & 58.52 & 49.15 & 48.94 & 80.49 & 46.17 & 60.78 & 2.45 \\
Phi3 Mini Instruct (128k) & 37.30 & 79.32 & 57.76 & 50.21 & 51.79 & 85.98 & 47.11 & 73.87 & 3.28 \\
Phi3 Mini Instruct (4k) & 35.69 & 78.03 & 57.76 & 47.43 & 49.60 & 85.37 & 45.89 & 60.64 & 2.46 \\
Phi3 Small Instruct (8k) & 30.21 & 65.05 & 50.64 & 40.43 & 40.92 & 67.07 & 30.62 & 60.57 & 2.97 \\
\midrule
PhindCL 34B v2 & 30.74 & 62.77 & 47.33 & 37.82 & 40.04 & 57.93 & 29.92 & 31.60 & 2.50 \\
\midrule
Qwen2 0.5B Instruct & 36.74 & 78.60 & 64.89 & 56.41 & 53.10 & 43.29 & 41.99 & 43.83 & 3.03 \\
Qwen2 1.5B Instruct & 35.23 & 80.60 & 55.22 & 45.82 & 45.51 & 0.61  & 36.59 & 27.89 & 1.42 \\
Qwen2 57B-A14B Instruct & 22.14 & 48.50 & 40.97 & 31.37 & 29.61 & 37.20 & 13.72 & 21.28 & 1.27 \\
Qwen2 7B & 28.15 & 68.76 & 53.69 & 41.37 & 41.50 & 27.44 & 21.28 & 36.66 & 3.53 \\
Qwen2 7B Instruct & 31.93 & 72.33 & 56.74 & 41.35 & 45.15 & 72.56 & 40.16 & 57.73 & 3.70 \\
\midrule
CodeQwen1.5 7B & 36.96 & 82.88 & 64.63 & 51.45 & 54.12 & 75.00 & 77.64 & 58.14 & 4.18 \\
CodeQwen1.5 7B Chat & 32.32 & 77.60 & 57.25 & 47.22 & 48.29 & 52.44 & 76.00 & 60.64 & 5.31 \\
\midrule
Yi1.5 34B Chat & 35.69 & 77.46 & 54.45 & 45.61 & 47.92 & 87.20 & 76.33 & 68.75 & 4.66 \\
Yi-1.5 6B Chat & 32.19 & 70.04 & 58.52 & 42.17 & 45.08 & 65.85 & 68.11 & 28.76 & 3.25 \\
Yi1.5 9B Chat & 34.45 & 73.47 & 56.49 & 52.42 & 49.45 & 71.95 & 75.95 & 19.58 & 3.08 \\
\midrule
DSCoder 6.7B & 26.51 & 60.49 & 44.78 & 30.98 & 31.58 & 0.00  & 0.05  & 0.00  & 0.00 \\
DSCoder 6.7B Instruct & 29.44 & 68.90 & 49.11 & 34.37 & 36.47 & 85.37 & 47.39 & 77.51 & 3.10 \\
DSCoder 7B (v1.5) & 35.24 & 79.32 & 54.96 & 44.78 & 42.30 & 0.00  & 27.38 & 55.72 & 2.08 \\
DSCoder 7B Instruct (v1.5) & 33.22 & 79.32 & 52.42 & 43.82 & 45.08 & 50.00 & 45.66 & 29.11 & 3.29 \\
DSCoder 33B & 28.50 & 71.33 & 47.84 & 38.36 & 38.07 & 0.00  & 0.00  & 3.92  & 0.44 \\
DSCoder 33B Instruct & 31.54 & 75.75 & 49.11 & 38.84 & 41.58 & 45.12 & 28.51 & 31.46 & 2.48 \\
\midrule
DSMoE 16B & 21.25 & 49.50 & 38.93 & 33.48 & 29.98 & 0.00  & 0.00  & 0.00  & 0.13 \\
DSMoE 16B Chat & 23.81 & 60.77 & 48.09 & 37.37 & 36.83 & 29.27 & 36.17 & 76.65 & 2.20 \\
\midrule
DSCoderV2 Lite & 33.78 & 83.59 & 57.51 & 47.19 & 46.10 & 29.88 & 0.80  & 64.62 & 0.00 \\
DSCoderV2 Lite Instruct & 22.22 & 38.66 & 28.24 & 25.64 & 20.50 & 91.46 & 55.47 & 62.99 & 2.63 \\
\midrule
InternLM2.5 20B Chat & 35.00 & 75.75 & 57.76 & 45.70 & 50.04 & 81.10 & 75.90 & 64.48 & 3.58 \\
InternLM2.5 7B Chat & 31.30 & 67.76 & 50.38 & 42.03 & 42.01 & 79.27 & 49.98 & 60.64 & 3.75 \\
\midrule
StarCoder2 15B Instruct & 34.44 & 83.31 & 56.23 & 46.99 & 46.61 & 75.00 & 53.31 & 24.91 & 2.30 \\
StarCoder2 7B & 27.77 & 63.77 & 47.33 & 36.95 & 34.72 & 1.22  & 0.05  & 1.35  & 0.08 \\
\bottomrule
\end{longtable}
\end{landscape}
\lstdefinestyle{examples}{
    commentstyle=\color{magenta},
    keywordstyle=\color{blue},
    numberstyle=\scriptsize\bfseries\color{gray},
    basicstyle=\fontfamily{cmtt}\scriptsize,
    % basicstyle=\footnotesize,
    breakatwhitespace=true,         
    breaklines=true,    
    breakindent=0pt,
    captionpos=b,                    
    keepspaces=true,                 
    numbersep=5pt,                  
    showspaces=false,                
    showstringspaces=false,
    showtabs=false,                  
    tabsize=2,
    % escapeinside={*@}{@*},
    postbreak=\mbox{\textcolor{red}{$\hookrightarrow$}\space},
    % belowskip=-\baselineskip,
    % aboveskip=- 0.5\baselineskip,
    language=Python
}
\lstset{style=examples}

\newpage
\subsection{CodeMMLU Example}
\label{appendix:example}

\textbf{General knowledge MCQ example:}
\begin{tcolorbox}[colback=gray!5!white,colframe=gray!75!black,breakable]
The following are multiple-choice questions (with answers) about debugging a programming problem.

\textbf{Question:} Suppose we have an O(n) time algorithm that finds the median of an unsorted array. Now consider a QuickSort implementation where we first find the median using the above algorithm, then use the median as a pivot. What will be the worst-case time complexity of this modified QuickSort?

{\color{red!75!black}{(A)}} $O(n^2 \log n)$

{\color{red!75!black}{(B)}} $O(n^2)$

{\color{red!75!black}{(C)}} $O(n\log n \log n)$

{\color{green!75!black}{(D)}} $O(n \log n)$

\end{tcolorbox}

\textbf{Code Completion example:}

\begin{tcolorbox}[colback=gray!5!white,colframe=gray!75!black,breakable]

    The following are multiple-choice questions (with answers) about programming problems.
    
    \textbf{Question:} Which solution below is the most likely to complete the following code to achieve the desired goal?
    
    \begin{lstlisting}
    from typing import List
    
    def has_close_elements(numbers: List[float], threshold: float) -> bool:
        """ Check if in given list of numbers, are any two numbers closer to each other than given threshold.
        >>> has_close_elements([1.0, 2.0, 3.0], 0.5)
        False
        >>> has_close_elements([1.0, 2.8, 3.0, 4.0, 5.0, 2.0], 0.3)
        True
        """
    \end{lstlisting}
    
    {\color{red!75!black}{(A)}}
    \begin{lstlisting}
        for i in range(len(numbers)):  # Change range to len(numbers)
            for j in range(i + 1, len(numbers)):
                if abs(numbers[i] - numbers[j]) < threshold:
                    return True
            return False
    \end{lstlisting}
    {\color{red!75!black}{(B)}}
    \begin{lstlisting}
        return any(abs(a - b) < threshold for a, b \
            in zip(numbers, numbers[1:]))
    \end{lstlisting}
    {\color{red!75!black}{(C)}}
    \begin{lstlisting}
        for i in range(len(numbers) - 1):
            for j in range(i + 1, len(numbers)):
                if abs(numbers[i] - numbers[j]) > threshold:
                    return False
            return True
    \end{lstlisting}
    {\color{green!75!black}{(D)}}
    \begin{lstlisting}
        for idx, elem in enumerate(numbers):
            for idx2, elem2 in enumerate(numbers):
                if idx != idx2:
                    distance = abs(elem - elem2)
                    if distance < threshold:
                        return True
    
        return False
    \end{lstlisting}
    
    \textbf{Answer}: 
\end{tcolorbox}

\textbf{Fill in the blank example:}

\begin{tcolorbox}[colback=gray!5!white,colframe=gray!75!black,breakable]
The following are multiple-choice questions (with answers) about a programming problem with incomplete solution.

\textbf{Problem statement:} You are given an array of intervals, where intervals[i] = [starti, endi] and each starti is unique. The right interval for an interval i is an interval j such that startj >= endi and startj is minimized. 
Note that i may equal j. Return an array of right interval indices for each interval i. If no right interval exists for interval i, then put -1 at index i.

\textbf{Incomplete Solution:}
\begin{lstlisting}
def find_right_interval(intervals):
    n = len(intervals)
    res = [-1] * n
    for i in range(n):
        intervals[i].append(i)

    def binary_search(ele):
        left, right = 0, n-1
        ans = float('inf')
        while left <= right:
            mid = (left + right) // 2
            if intervals[mid][0] >= ele:
                ans = min(ans, mid)
                right = mid - 1
            else:
                left = mid + 1
        return ans
            
    intervals.sort()
    for i in intervals:
        _________________
        if val != float('inf'):
            res[i[2]] = intervals[val][2]
    return res
\end{lstlisting}

\textbf{Question:} The provided solution is missing a part, which option below is the most likely to
complete the solution and achieve the desired goal?

{\color{green!75!black}{(A)}}
\begin{lstlisting}
    val = binary_search(i[1])
\end{lstlisting}
{\color{red!75!black}{(B)}}
\begin{lstlisting}
    if val != float('inf'): 
\end{lstlisting}
{\color{red!75!black}{(C)}}
\begin{lstlisting}
    val = binary_search(i[1])
\end{lstlisting}
{\color{red!75!black}{(D)}}
\begin{lstlisting}
    if val != float('inf'): 
      res[i[2]] = intervals[val][2]
\end{lstlisting}
\textbf{Answer}: 
\end{tcolorbox}

\textbf{Code Repair example:}

\begin{tcolorbox}[colback=gray!5!white,colframe=gray!75!black,breakable]
The following are multiple-choice questions (with answers) about debugging a programming problem.

\textbf{Question:} The following code snippet is producing incorrect results; Which solution below correctly identifies the bug and repairs it to achieve the desired goal?

\begin{lstlisting}[language=Java, numbers=left, xleftmargin=2pt]
import java.util.*;
public class DETECT_CYCLE {
    public static boolean detect_cycle(Node node) {
        Node hare = node;
        Node tortoise = node;
        while (true) {
            if (hare.getSuccessor() == null)
                return false;
            tortoise = tortoise.getSuccessor();
            hare = hare.getSuccessor().getSuccessor();
            if (hare == tortoise)
                return true;
        }
    }
}
\end{lstlisting}

{\color{red!75!black}{(A)}} Modify line 6:
\begin{lstlisting}[language=Java]
        for (; ; ) {
\end{lstlisting}

{\color{red!75!green}{(B)}} Modify line 7:
\begin{lstlisting}[language=Java]
           if (null==hare ||hare.getSuccessor() == null)
\end{lstlisting}

{\color{red!75!black}{(C)}} Modify line 12:
\begin{lstlisting}[language=Java]
        return hare.getSuccessor() != null && hare == tortoise;
\end{lstlisting}

{\color{green!75!black}{(D)}} Modify line 11:
\begin{lstlisting}[language=Java]
            if (Objects.equals(hare, tortoise))
\end{lstlisting}
    
\end{tcolorbox}

\textbf{Execution Prediction example:}
\begin{tcolorbox}[colback=gray!5!white,colframe=gray!75!black,breakable]

The following are multiple-choice questions (with answers) about programming problem.

\textbf{Question:} Given a code snippet below, which behavior most likely to occur when running the solution?

\begin{lstlisting}[language=Java]
    import java.util.*;
    public class Main {
        public static void main(string[] args) {
            Scanner sc = new Scanner(System.in);
            int A = sc.nextInt();
            int B = sc.nextInt();
            int T = sc.nextInt();
            int S = T/A System.out.println(s*b);
        }
    }
\end{lstlisting}

{\color{red!75!black}{(A)}} Memory Limit Exceeded

{\color{red!75!black}{(B)}} Runtime Error

{\color{green!75!black}{(C)}} Compile Error

{\color{red!75!black}{(D)}} No abnormally found

\end{tcolorbox}

\section{Models Setup}
\label{appendix:models}
In our experiment and study, we consider GPT-4o \citep{openai2024gpt4technicalreport}, GPT-3.5 \citep{openai-2023}, Claude-3.5, Claude-3 \citep{TheC3}, MetaLlama 3.1 \citep{dubey2024llama3herdmodels}, MetaLlama3 \citep{dubey2024llama3herdmodels}, CodeLLaMA \citep{rozière2024codellamaopenfoundation}, DeepSeek AI, DeepSeek Coder, DeepSeek Coder V2 \citep{guo2024deepseek, deepseekai2024deepseekcoderv2breakingbarrierclosedsource, guo2024deepseekcoderlargelanguagemodel}, MistralAI, Codetral \citep{jiang2024mixtralexperts}, Qwen2 \citep{yang2024qwen2technicalreport}, CodeQwen1.5 \citep{qwen}, Yi \citep{ai2024yiopenfoundationmodels}, StarCoder2 \citep{lozhkov2024starcoder2stackv2}, InternLM \citep{cai2024internlm2technicalreport}, Phind \citep{phind-2023}.

% Table generated by Excel2LaTeX from sheet 'model_sheet'
\begin{longtable}{p{1.75cm}|p{4.25cm}|p{2.5cm}|p{3.25cm}}
% \label{tab:addlabel}%
  \caption{Language Models Description} \\
    \toprule
          & \textbf{Model ID} & \textbf{Short Name} & \textbf{Link} \\
    \midrule
    \midrule
    \multirow{2}[1]{*}{OpenAI} 
    & GPT-o1-2024-12-17 & GPT o1 & -  \\
    & GPT-o3-mini-2025-01-31 & GPT o3-mini & - \\
    & GPT-4o-2024-05-13 & GPT 4o & - \\
          & GPT-4o-mini-2024-07-18 & GPT 4o mini & - \\
          & GPT-3.5-turbo-16k-0613 & GPT-3.5-turbo & - \\
    \midrule
    \multirow{5}[0]{*}{Anthropic} & Claude-3.5-sonnet-20241022 & Claude3.5 Sonnet & - \\
          & Claude-3.7-sonnet-20250219 & Claude3.7 Sonnet & - \\
          & Claude-3.5-haiku-20241022 & Claude3.5 Haiku & - \\
          & Claude-3-haiku-20240307 & Claude3 Haiku & - \\
          & Claude-3-sonnet-202402029 & Claude3 Sonnet & - \\
          & Claude-3-opus-20240229 & Claude3 Opus & - \\
    \midrule
    \multirow{7}[0]{*}{CodeLlama} & codellama/CodeLlama-13b-Instruct-hf & CodeLlama 13B Instruct & \href{https://huggingface.co/codellama/CodeLlama-13b-Instruct-hf}{codellama/CodeLlama-13b-Instruct-hf} \\
          & codellama/CodeLlama-13b-Python-hf & CodeLlama 13B Python & \href{https://huggingface.co/codellama/CodeLlama-13b-Python-hf}{codellama/CodeLlama-13b-Python-hf} \\
          & codellama/CodeLlama-13b-hf & CodeLlama 13B & \href{https://huggingface.co/codellama/CodeLlama-13b-hf}{codellama/CodeLlama-13b-hf} \\
          & codellama/CodeLlama-34b-Instruct-hf & CodeLlama 34B Instruct & \href{https://huggingface.co/codellama/CodeLlama-34b-Instruct-hf}{codellama/CodeLlama-34b-Instruct-hf} \\
          & codellama/CodeLlama-34b-Python-hf & CodeLlama 34B Python & \href{https://huggingface.co/codellama/CodeLlama-34b-Python-hf}{codellama/CodeLlama-34b-Python-hf} \\
          & codellama/CodeLlama-7b-Instruct-hf & CodeLlama 7B Instruct & \href{https://huggingface.co/codellama/CodeLlama-7b-Instruct-hf}{codellama/CodeLlama-7b-Instruct-hf} \\
          & codellama/CodeLlama-7b-Python-hf & CodeLlama 7B Python & \href{https://huggingface.co/codellama/CodeLlama-7b-Python-hf}{codellama/CodeLlama-7b-Python-hf} \\
    \midrule
    \multirow{8}[0]{*}{MetaLlama} & meta-llama/Meta-Llama-3-70B & Llama3 70B & \href{https://huggingface.co/meta-llama/Meta-Llama-3-70B}{meta-llama/Meta-Llama-3-70B} \\
          & meta-llama/Meta-Llama-3-70B-Instruct & Llama3 70B Instruct & \href{https://huggingface.co/meta-llama/Meta-Llama-3-70B-Instruct}{meta-llama/Meta-Llama-3-70B-Instruct} \\
          & meta-llama/Meta-Llama-3-8B & Llama3 8B & \href{https://huggingface.co/meta-llama/Meta-Llama-3-8B}{meta-llama/Meta-Llama-3-8B} \\
          & meta-llama/Meta-Llama-3-8B-Instruct & Llama3 8B Instruct & \href{https://huggingface.co/meta-llama/Meta-Llama-3-8B-Instruct}{meta-llama/Meta-Llama-3-8B-Instruct} \\
          & meta-llama/Meta-Llama-3.1-70B & Llama3.1 70B & \href{https://huggingface.co/meta-llama/Meta-Llama-3.1-70B}{meta-llama/Meta-Llama-3.1-70B} \\
          & meta-llama/Meta-Llama-3.1-70B-Instruct & Llama3.1 70B Instruct & \href{https://huggingface.co/meta-llama/Meta-Llama-3.1-70B-Instruct}{meta-llama/Meta-Llama-3.1-70B-Instruct} \\
          & meta-llama/Meta-Llama-3.1-8B & Llama3.1 8B & \href{https://huggingface.co/meta-llama/Meta-Llama-3.1-8B}{meta-llama/Meta-Llama-3.1-8B} \\
          & meta-llama/Meta-Llama-3.1-8B-Instruct & Llama3.1 8B Instruct & \href{https://huggingface.co/meta-llama/Meta-Llama-3.1-8B-Instruct}{meta-llama/Meta-Llama-3.1-8B-Instruct} \\
          & meta-llama/Meta-Llama-3.1-405B-Instruct & Llama3.1 405B Instruct & \href{https://huggingface.co/meta-llama/Meta-Llama-3.1-405B-Instruct}{meta-llama/Meta-Llama-3.1-405B-Instruct} \\
          & meta-llama/Meta-Llama-3.3-70B-Instruct & Llama3.3 70B Instruct & \href{https://huggingface.co/meta-llama/Meta-Llama-3.3-70B-Instruct}{meta-llama/Meta-Llama-3.3-70B-Instruct} \\
    \midrule
    \multirow{5}[0]{*}{Mistral} & mistralai/Mistral-7B-Instruct-v0.1 & Mistral 7B Instruct (v1) & \href{https://huggingface.co/mistralai/Mistral-7B-Instruct-v0.1}{mistralai/Mistral-7B-Instruct-v0.1} \\
          & mistralai/Mistral-7B-Instruct-v0.2 & Mistral 7B Instruct (v2) & \href{https://huggingface.co/mistralai/Mistral-7B-Instruct-v0.2}{mistralai/Mistral-7B-Instruct-v0.2} \\
          & mistralai/Mistral-7B-Instruct-v0.3 & Mistral 7B Instruct (v3) & \href{https://huggingface.co/mistralai/Mistral-7B-Instruct-v0.3}{mistralai/Mistral-7B-Instruct-v0.3} \\
          & mistralai/Mixtral-8x7B-Instruct-v0.1 & Mixtral 8x7B Instruct & \href{https://huggingface.co/mistralai/Mixtral-8x7B-Instruct-v0.1}{mistralai/Mixtral-8x7B-Instruct-v0.1} \\
          & mistralai/Codestral-22B-v0.1 & Codestral 22B & \href{https://huggingface.co/mistralai/Codestral-22B-v0.1}{mistralai/Codestral-22B-v0.1} \\
    \midrule
    \multirow{5}[0]{*}{Phi} 
        & microsoft/phi-4 & Phi-4 & \href{https://huggingface.co/microsoft/phi-4}{microsoft/phi-4} \\
        & microsoft/Phi-4-mini-instruct & Phi-4-mini-instruct & \href{https://huggingface.co/microsoft/Phi-4-mini-instruct}{microsoft/Phi-4-mini-instruct} \\
        & microsoft/Phi-3-medium-128k-instruct & Phi3 Medium Instruct (128k) & \href{https://huggingface.co/microsoft/Phi-3-medium-128k-instruct}{microsoft/Phi-3-medium-128k-instruct} \\
          & microsoft/Phi-3-medium-4k-instruct & Phi3 Medium Instruct (4k) & \href{https://huggingface.co/microsoft/Phi-3-medium-4k-instruct}{microsoft/Phi-3-medium-4k-instruct} \\
          & microsoft/Phi-3-mini-128k-instruct & Phi3 Mini Instruct (128k) & \href{https://huggingface.co/microsoft/Phi-3-mini-128k-instruct}{microsoft/Phi-3-mini-128k-instruct} \\
          & microsoft/Phi-3-mini-4k-instruct & Phi3 Mini Instruct (4k) & \href{https://huggingface.co/microsoft/Phi-3-mini-4k-instruct}{microsoft/Phi-3-mini-4k-instruct} \\
          & microsoft/Phi-3-small-8k-instruct & Phi3 Small Instruct (8k) & \href{https://huggingface.co/microsoft/Phi-3-small-8k-instruct}{microsoft/Phi-3-small-8k-instruct} \\
    \midrule
    PhinD & Phind/Phind-CodeLlama-34B-v2 & PhindCL 34B v2 & \href{https://huggingface.co/Phind/Phind-CodeLlama-34B-v2}{Phind/Phind-CodeLlama-34B-v2} \\
    \midrule
    \multirow{2}[0]{*}{CodeQwen} & Qwen/CodeQwen1.5-7B & CodeQwen1.5 7B & \href{https://huggingface.co/Qwen/CodeQwen1.5-7B}{Qwen/CodeQwen1.5-7B} \\
          & Qwen/CodeQwen1.5-7B-Chat & CodeQwen1.5 7B Chat & \href{https://huggingface.co/Qwen/CodeQwen1.5-7B-Chat}{Qwen/CodeQwen1.5-7B-Chat} \\
    \midrule
    \multirow{5}[0]{*}{Qwen} 
    & Qwen/Qwen2.5-Coder-32B-Instruct & QwenCoder2.5 32B Inst & \href{https://huggingface.co/Qwen/Qwen2.5-Coder-32B-Instruct}{Qwen/Qwen2.5-Coder-32B-Instruct} \\
    & Qwen/Qwen2.5-Coder-14B-Instruct & QwenCoder2.5 14B Inst & \href{https://huggingface.co/Qwen/Qwen2.5-Coder-14B-Instruct}{Qwen/Qwen2.5-Coder-14B-Instruct} \\
    & Qwen/QwQ-32B-Preview & QwQ 32B  & \href{https://huggingface.co/Qwen/Qwen2-0.5B-Instruct}{Qwen/Qwen2-0.5B-Instruct} \\
    & Qwen/Qwen2-0.5B-Instruct & Qwen2 0.5B Instruct & \href{https://huggingface.co/Qwen/Qwen2-0.5B-Instruct}{Qwen/Qwen2-0.5B-Instruct} \\
          & Qwen/Qwen2-1.5B-Instruct & Qwen2 1.5B Instruct & \href{https://huggingface.co/Qwen/Qwen2-1.5B-Instruct}{Qwen/Qwen2-1.5B-Instruct} \\
          & Qwen/Qwen2-57B-A14B-Instruct & Qwen2 57B-A14B Instruct & \href{https://huggingface.co/Qwen/Qwen2-57B-A14B-Instruct}{Qwen/Qwen2-57B-A14B-Instruct} \\
          & Qwen/Qwen2-7B & Qwen2 7B & \href{https://huggingface.co/Qwen/Qwen2-7B}{Qwen/Qwen2-7B} \\
          & Qwen/Qwen2-7B-Instruct & Qwen2 7B Instruct & \href{https://huggingface.co/Qwen/Qwen2-7B-Instruct}{Qwen/Qwen2-7B-Instruct} \\
    \midrule
    \multirow{3}[0]{*}{Yi} & 01-ai/Yi-1.5-34B-Chat & Yi1.5 34B Chat & \href{https://huggingface.co/01-ai/Yi-1.5-34B-Chat}{01-ai/Yi-1.5-34B-Chat} \\
          & 01-ai/Yi-1.5-6B-Chat & Yi-1.5 6B Chat & \href{https://huggingface.co/01-ai/Yi-1.5-6B-Chat}{01-ai/Yi-1.5-6B-Chat} \\
          & 01-ai/Yi-1.5-9B-Chat & Yi1.5 9B Chat & \href{https://huggingface.co/01-ai/Yi-1.5-9B-Chat}{01-ai/Yi-1.5-9B-Chat} \\
    \midrule
    \multirow{2}[0]{*}{\makecell{DeepSeek}} & deepseek-ai/DeepSeek-R1 & DeepSeek R1 & \href{https://huggingface.co/deepseek-ai/DeepSeek-R1}{deepseek-ai/R1} \\
          & deepseek-ai/deepseek-V3 & DeepSeek V3 & \href{https://huggingface.co/deepseek-ai/DeepSeek-V3}{deepseek-ai/DeepSeek-V3} \\
    \midrule
    \multirow{7}[0]{*}{\makecell{DeepSeek\\ Coder}} & deepseek-ai/deepseek-coder-33b-base & DSCoder 33B & \href{https://huggingface.co/deepseek-ai/deepseek-coder-33b-base}{deepseek-ai/deepseek-coder-33b-base} \\
          & deepseek-ai/deepseek-coder-33b-instruct & DSCoder 33B Instruct & \href{https://huggingface.co/deepseek-ai/deepseek-coder-33b-instruct}{deepseek-ai/deepseek-coder-33b-instruct} \\
          & deepseek-ai/deepseek-coder-6.7b-base & DSCoder 6.7B & \href{https://huggingface.co/deepseek-ai/deepseek-coder-6.7b-base}{deepseek-ai/deepseek-coder-6.7b-base} \\
          & deepseek-ai/deepseek-coder-6.7b-instruct & DSCoder 6.7B Instruct & \href{https://huggingface.co/deepseek-ai/deepseek-coder-6.7b-instruct}{deepseek-ai/deepseek-coder-6.7b-instruct} \\
          & deepseek-ai/deepseek-coder-7b-base-v1.5 & DSCoder 7B (v1.5) & \href{https://huggingface.co/deepseek-ai/deepseek-coder-7b-base-v1.5}{deepseek-ai/deepseek-coder-7b-base-v1.5} \\
          & deepseek-ai/deepseek-coder-7b-instruct-v1.5 & DSCoder 7B Instruct (v1.5) & \href{https://huggingface.co/deepseek-ai/deepseek-coder-7b-instruct-v1.5}{deepseek-ai/deepseek-coder-7b-instruct-v1.5} \\
          & deepseek-ai/DeepSeek-Coder-V2-Lite-Base & DSCoderV2 Lite & \href{https://huggingface.co/deepseek-ai/DeepSeek-Coder-V2-Lite-Base}{deepseek-ai/DeepSeek-Coder-V2-Lite-Base} \\
          & deepseek-ai/DeepSeek-Coder-V2-Lite-Instruct & DSCoderV2 Lite Instruct & \href{https://huggingface.co/deepseek-ai/DeepSeek-Coder-V2-Lite-Instruct}{deepseek-ai/DeepSeek-Coder-V2-Lite-Instruct} \\
    \midrule
    \multirow{2}[0]{*}{\makecell{DeepSeek\\ MoE}} & deepseek-ai/deepseek-moe-16b-base & DSMoE 16B & \href{https://huggingface.co/deepseek-ai/deepseek-moe-16b-base}{deepseek-ai/deepseek-moe-16b-base} \\
          & deepseek-ai/deepseek-moe-16b-chat & DSMoE 16B Chat & \href{https://huggingface.co/deepseek-ai/deepseek-moe-16b-chat}{deepseek-ai/deepseek-moe-16b-chat} \\
    \midrule
    \multirow{2}[0]{*}{InternLM} & internlm/internlm2\_5-20b-chat & InternLM2.5 20B Chat & \href{https://huggingface.co/internlm/internlm2_5-20b-chat}{internlm/internlm2\_5-20b-chat} \\
          & internlm/internlm2\_5-7b-chat & InternLM2.5 7B Chat & \href{https://huggingface.co/internlm/internlm2_5-7b-chat}{internlm/internlm2\_5-7b-chat} \\
    \midrule
    \multirow{2}[1]{*}{StarCoder2} & bigcode/starcoder2-15b-instruct-v0.1 & StarCoder2 15B Instruct & \href{https://huggingface.co/bigcode/starcoder2-15b-instruct-v0.1}{bigcode/starcoder2-15b-instruct-v0.1} \\
          & bigcode/starcoder2-7b & StarCoder2 7B & \href{https://huggingface.co/bigcode/starcoder2-7b}{bigcode/starcoder2-7b} \\
    \bottomrule
\end{longtable}%

\newpage

\subsection{Prompt Libarary}
\label{appendix:filtering_prompts}
\paragraph{Filtering prompts:} LLM-based filtering for ranking questions' completeness, coherence, and clarity.

{\small
\begin{longtable}{p{\textwidth}}
    \toprule
    \textbf{Quality filtering prompt:}\\ \midrule
    \begin{lstlisting}[language=Python,]
Please rate the following question based on three criteria, with a score from 1 to 5 for each criterion (where 1 is the lowest and 5 is the highest). No explanation needed:

	1.	Completeness: 
	- Does the question stand alone and provide enough information independently? 
	- Avoid including any images, links, or external references.
	2.	Coherence and Clarity: 
	- Is the question phrased clearly, with proper grammar? 
	- Is there any ambiguity or confusion in the wording?
	3.	Relevance: 
	- Is the question directly related to software development or programming issues? 
	- Does it involve technical challenges, concepts, or tools commonly used in software or programming?
	
Question:
"""{}"""
    \end{lstlisting} \\
    \bottomrule
\end{longtable}
}

\label{appendix:prompts}

\paragraph{Data creation prompts:} Prompt used for synthesis distractor for real-world task:

{\small
\begin{longtable}{p{\textwidth}}
    \toprule
    \textbf{Code Repair distractor creation prompts:}\\ \midrule
    After extracting statement from buggy version, we use LLMs to rewrite a new version of that statement. 
    We command LLMs to assume the bug is located in the assigned line and their target is correct that line. 
    
    Here is the prompt: 
    \begin{lstlisting}[language=Python,]
        Given a buggy Python code snippet, you will be asked to debugging the code.
        ```
        def truncate_number(number: float) -> float:
            return number * (number % 1)
        ```
        Let assume the bug is located in this line:
        ```    return number * (number % 1)```
        Adjust this line in order to solve the bug.
        The re-written line must be syntactic correct, executable and wrapped in ``` ``` brace.
        Don't give any details.
        ### Rewritten line:
        ```    return number % 1.0```
        
        Given a buggy Java code snippet, you will be asked to debugging the code.
        ```{code}```
        Let assume the bug is located in this line:
        ```{line}```
        Adjust this line in order to solve the bug.
        The re-written line must be syntactic correct, executable and wrapped in ``` ``` brace.
        Don't give any details.
        ### Rewritten line:
    \end{lstlisting} 
    We executing the problem with given test cases. Our target is to create reasonable false answer that would require deep interpretation. Follow by an LLMs based filter to pick from pool of negative answer the most likely able to solve the buggy problem. This result a set of confusing negative answer. Those reasonable false sample with executable (and if they can pass through few testcases) is golden negative answer.
    \\ \midrule
    \newpage
    \textbf{\textit{Fill in the blank distractor creation prompt:}} \\ \midrule
    From correct solution from leetcode, we randomly mask a line/a block of code and generate false answer (for multiple choice) from LLMs:
    \begin{lstlisting}[language=Python,]
        Following this code:
        {code}
        I prepare some multiple choice questions answering 
        so i want to make small change on this line 
        but it still look true of this line : {line}
        help me generate 3 version change in this code and each output should in ``` ``` brace and code only. 
        Don't give any details
    \end{lstlisting} \\
    \bottomrule
\end{longtable}
}

\paragraph{Experimental prompts:} Prompt used in CodeMMLU evaluation.
\label{appendix:prompt-collection}

{\small
\begin{longtable}{p{\textwidth}}
    \toprule
    \textbf{Zero-shot prompts}\\ \midrule
    \textbf{\textit{General knowledge MCQ test set:}}
    \begin{lstlisting}[language=Python,]
        The following are multiple-choice questions (with answers) about software development.
        
        Question: {question}
        {multiple_choices}
        
        Answer:
    \end{lstlisting} \\ \midrule
    \textbf{\textit{Code completion:}}
    \begin{lstlisting}[language=Python,]
        The following are multiple-choice questions (with answers) about software development.
        
        Question: {question}
        {multiple_choices}
        
        Answer:
    \end{lstlisting} \\ \midrule
    \textbf{\textit{Fill in the blank:}}
    \begin{lstlisting}[language=Python,]
        The following are multiple-choice questions (with answers) about a programming problem with an incomplete solution.
        
        Problem statement: {question}
        
        Incomplete Solution:
        {codebase}
        
        Question: The provided solution is missing a part, Which option below is the most likely to complete the solution and achieve the desired goal?
        
        {multiple_choices}
        
        Answer:
    \end{lstlisting} \\ \midrule
    \textbf{\textit{Code Repair:}}
    \begin{lstlisting}[language=Python,]
        The following are multiple-choice questions (with answers) about debugging a programming problem.

        Question: The implementation below is producing incorrect results. Which solution below correctly identifies the bug and repairs it to achieve the desired goal?
        {question}
        
        {multiple_choices}
        
        Answer:
    \end{lstlisting} \\ \midrule
    \textbf{\textit{Defect Detection:}}
    \begin{lstlisting}[language=Python,]
        The following are multiple-choice questions (with answers) about programming problems.
        
        Question: Given a code snippet below, which behavior most likely to occur when execute it?
        {question}
        
        {multiple_choices}
        
        Answer:
    \end{lstlisting} \\
    \bottomrule
\end{longtable}
}

\newpage
{\small
\begin{longtable}{p{\textwidth}}
    \toprule
    \textbf{Few-shot prompt} \\ \midrule
    \textbf{\textit{General knowledge MCQ test set:}}
    \begin{lstlisting}
        The following are multiple choice questions (with answers) about software development.

        Question: If a sorted array of integers is guaranteed to not contain duplicate values, in order to search a for a specific value which of the following algorithms is the most efficient for this task?
        
        (A) Bubble Sort (B) Linear Search (C) Insertion Sort (D) Binary Search
        
        Answer: The answer is (D).
        
        Question: {question}
        {multiple_choices}
        
        Answer:
    \end{lstlisting} \\ \midrule
    \textbf{\textit{Code completion:}} \\
    \begin{lstlisting}
        The following are multiple-choice questions (with answers) about programming problems.
        
        Question: Which solution below is the most likely completion the following code snippet to achieve the desired goal?
        ```python
        from typing import List
        
        def two_sum(nums: List[int], target: int) -> List[int]:
            """
            Given an array of integers nums and an integer target, return indices of the two numbers such that they add up to target. 
            You may assume that each input would have exactly one solution, and you may not use the same element twice.
            
            >>> two_sum([2,7,11,15], 9) 
            [0,1]
            >>> two_sum([3,2,4], 6) 
            [1,2]
            >>> two_sum([3,3], 6) 
            [0,1]
            """
        ```
    
        (A) ```python
            n = len(nums)
            for i in range(n - 1):
                for j in range(i + 1, n):
                    if nums[i] + nums[j] == target:
                        return [i, j]
            return []
        ```
        (B) ```python
            for num in nums:
                if target - num in nums:
                    return [nums.index(num), nums.index(target - num)]
            return []
        ```
    \end{lstlisting} \\
    \begin{lstlisting}
        (C) ```python
            for i in range(len(nums)):
                if nums[i] * 2 == target:
                    return [i, i]
            return []
        ```
        (D) ```python
            num_dict = {}
            for i, num in enumerate(nums):
                if target - num in num_dict:
                    return [num_dict[target - num], i]
                num_dict[i] = num
            return []
        ```
        Answer: The answer is A.
        
        Question: Which solution below is the most likely completion the following code snippet to achieve the desired goal?
        ```python
        {question}
        ```
        
        {multiple_choices}
        
        Answer:'''
    \end{lstlisting} \\ \midrule
    \textbf{\textit{Fill in the blank:}}
    \begin{lstlisting}[language=Python,]
        The following are multiple-choice questions (with answers) about a programming problem with incomplete solution.
        
        Problem statement: You are given an array of intervals, where intervals[i] = [starti, endi] and each starti is unique. 
        The right interval for an interval i is an interval j such that startj >= endi and startj is minimized. 
        Note that i may equal j. Return an array of right interval indices for each interval i. 
        If no right interval exists for interval i, then put -1 at index i.
        
        Incomplete Solution:
        python```
        def find_right_interval(intervals):
            n = len(intervals)
            res = [-1] * n
            for i in range(n):
                intervals[i].append(i)
        
            def binary_search(ele):
                left, right = 0, n-1
                ans = float('inf')
                while left <= right:
                    mid = (left + right) // 2
                    if intervals[mid][0] >= ele:
                        ans = min(ans, mid)
                        right = mid - 1
                    else:
                        left = mid + 1
                return ans
                    
            intervals.sort()
            for i in intervals:
                _________________
        
            return res
        ```
        Question: The provided solution is missing a part, Which option below is the most likely to complete the solution and achieve the desired goal?
    \end{lstlisting} \\
    \begin{lstlisting}
        
        (A) ```python
            val = binary_search(i[1])
            if val != float('inf'):
                res[i[2]] = intervals[val][2]
        ```
        (B) ```python
            if val != float('inf'): 
                res[i[2]] = intervals[val][2]
            else:
                continue
        ```
        (C) ```python
            val = binary_search(i[1])
        		if val != float('inf'): res[i[2] + 1] = intervals[val][2]
        ```
        (D) ```python
            if val != float('inf'): 
        			  res[i[2]] = intervals[val][2]
        		else:
        		  continue
        ```
        Answer: The answer is (A).
        
        Problem statement: {question}
        
        Incomplete Solution:
        {codebase}
        
        Question: The provided solution is missing a part, Which option below is the most likely to complete the solution and achieve the desired goal?
        
        {multiple_choices}
        
        Answer:
    \end{lstlisting} \\ \midrule
    \textbf{\textit{Code Repair:}}
    \begin{lstlisting}[language=Python,]
        The following are multiple-choice questions (with answers) about debugging a programming problem.

        Question: The implementation below is producing incorrect results. 
        Which solution below correctly identifies the bug and repairs it to achieve the desired goal?
        
        1 def two_sum(nums, target):
        2     complement_map = {{}}    
        3     for i, num in enumerate(nums):
        4         complement = target - num
        5         complement_map[num] = i
        6         if complement in complement_map:
        7             return [complement_map[complement], i]  
        8     return None
        
        (A) Remove line 5.
        
        (B) Remove line 5. Add at line 7:
        ```        complement_map[num] = i```
        
        (C) Modify line 7:
        ```         return [i, complement_map[complement]]```
        
        (D) Remove line 5. Add at line 7:
        ```     if i == len(nums) - 1:
                    return None
                complement_map[num] = i```
        
        Answer: The answer is (B).
        
        Question: The implementation below is producing incorrect results. 
        Which solution below correctly identifies the bug and repairs it to achieve the desired goal?
        {question}
        
        {choices}
        
        Answer: 
    \end{lstlisting} \\ \midrule
    \textbf{\textit{Defect Detection:}}
    \begin{lstlisting}[language=Python,]
        The following are multiple choice questions (with answers) about programming problem.
        
        Question: Given a code snippet below, which behavior most likely to occurr when execute it?
        ```python
        def chkPair(A, size, x):
            for i in range(0, size - 1):
                for j in range(i + 1, size):
                    if (A[i] + A[j] == x):
                        return 1
            return 0
        
        ```
        
        (A). The code contain no issue.
        (B). Memory Limit Exceeded
        (C). Compile error
        (D). Runtime Error
    \end{lstlisting} \\
    \begin{lstlisting}
        
        Answer: The answer is (A).
        
        Question: Given a code snippet below, which behavior most likely to occurr when execute it?
        {question}
        
        {multiple_choices}
        
        Answer:
    \end{lstlisting} \\
    \bottomrule
\end{longtable}
}

\newpage
{\small
\begin{longtable}{p{\textwidth}}
    \toprule
    \textbf{Chain-of-Thought zero-shot prompts}\\ \midrule
    \textbf{\textit{General knowledge MCQ test set:}}
    \begin{lstlisting}
        The following are multiple choice questions (with answers) about software devopment.
 
        Question: {question}
        {multiple_choices}
        
        Answer: Let's think step by step. 
    \end{lstlisting} \\ \midrule
    \textbf{\textit{Code completion:}}
    \begin{lstlisting}[language=Python,]
        The following are multiple choice questions (with answers) about programming problems.
        
        Question: Which solution below is the most likely completion the following code snippet to achieve the desired goal?
        ```python
        {question}
        ```
        {multiple_choices}
        
        Answer: Let's think step by step.
    \end{lstlisting} \\ \midrule
    \textbf{\textit{Fill in the blank:}}
    \begin{lstlisting}[language=Python,]
        The following are multiple-choice questions (with answers) about a programming problem with uncomplete solution.
        
        Problem statement: {question}
        
        Incomplete Solution:
        {codebase}
        
        Question: The provided solution is missing a part, Which option below is the most likely to
        complete the solution and achieve the desired goal?
        
        {multiple_choices}
        
        Answer: Let's think step by step.
    \end{lstlisting} \\ \midrule
    \textbf{\textit{Code Repair:}}
    \begin{lstlisting}[language=Python,]
        The following are multiple-choice questions (with answers) about debugging a programming problem.

        Question: The implementation below is producing incorrect results. 
        Which solution below correctly identifies the bug and repairs it to achieve the desired goal?
        {question}
        
        {multiple_choices}
        
        Answer: Let's think step by step.
    \end{lstlisting} \\ \midrule
    \textbf{\textit{Defect Detection:}}
    \begin{lstlisting}[language=Python,]
        The following are multiple-choice questions (with answers) about debugging a programming problem.

        The algorithm implementation below is producing incorrect results;
        Which solution below correctly identifies the bug and repairs it to achieve the desired goal?
        {question}
        
        {multiple_choices}
        
        Answer: Let's think step by step.
    \end{lstlisting} \\
    \bottomrule
\end{longtable}
}

\newpage
{\small
\begin{longtable}{p{\textwidth}}
    \toprule
    \textbf{Chain-of-Thought few-shot prompts}\\ \midrule
    \textbf{\textit{General knowledge MCQ test set:}}
    \begin{lstlisting}
        The following are multiple choice questions (with answers) about software devopment.

        Question: If a sorted array of integers is guaranteed to not contain duplicate values, in order to search a for a specific value which of the following algorithms is the most efficient for this task?
         
        (A) Bubble Sort (B) Linear Search (C) Insertion Sort (D) Binary Search
         
        Answer: Let's think step by step. Binary Search is a divide-and-conquer algorithm that works by repeatedly dividing the search interval in half and searching for the value in the appropriate half. Since the array is already sorted and does not contain any duplicate value, this algorithm is optimal to find the desired value. The answer is (D).
         
        Question: {question}
        {multiple_choices}
        
        Answer: Let's think step by step.
    \end{lstlisting} \\ \midrule
    
    \textbf{\textit{Code completion:}}
    \begin{lstlisting}[language=Python,]
        The following are multiple choice questions (with answers) about programming problem.
        
        Question: Which solution below is the most likely completion the following code snippet to achieve the desired goal?
        ```python
        def is_vowel(char: str) -> bool:
        	  """
        	  Checks if the input character is a vowel.
        	  """
        ```
        
        (A) ```python
        		return char.lower().is_vowel()
        ```
        (B) ```python
        		vowels = set("aeiou")
        	  return char.lower() in vowels
        ```
        (C) ```python
        		vowels = set("aeiou")
        	  return char.upper() in vowels
        ```
        (D) ```python
        		vowels = "aeiou"
        		return char.count(vowels) > 0
        ```
        
        Answer: Let's think step by step. The goal is to write a function is_vowel(char: str) -> bool that checks if the input character char is a vowel. The solution B correctly converts the input character to lowercase and checks if it is in the set of vowels. 
        The answer is (B).
        
        Question: Which solution below is the most likely completion the following code snippet to achieve the desired goal?
        ```python
        {question}
        ```
        {multiple_choices}
        
        Answer: Let's think step by step.
    \end{lstlisting} \\ \midrule
    \textbf{\textit{Fill in the blank:}}
    \begin{lstlisting}[language=Python,]
        The following are multiple-choice questions (with answers) about a programming problem with uncomplete solution.
        
        Problem statement: You are given an array of intervals, where intervals[i] = [starti, endi] and each starti is unique. 
        The right interval for an interval i is an interval j such that startj >= endi and startj is minimized. 
        Note that i may equal j. Return an array of right interval indices for each interval i. 
        If no right interval exists for interval i, then put -1 at index i.
        
        Incomplete Solution:
        python```
        def find_right_interval(intervals):
            n = len(intervals)
            res = [-1] * n
            for i in range(n):
                intervals[i].append(i)
            def binary_search(ele):
                left, right = 0, n-1
                ans = float('inf')
                while left <= right:
                    mid = (left + right) // 2
                    if intervals[mid][0] >= ele:
                        ans = min(ans, mid)
                        right = mid - 1
                    else:
                        left = mid + 1
                return ans
                    
            intervals.sort()
            for i in intervals:
                _________________
        
            return res
        ```
        
        
        Question: The provided solution is missing a part, Which option below is the most likely to
        complete the solution and achieve the desired goal?
        
        (A) ```python
            val = binary_search(i[1])
            if val != float('inf'):
                res[i[2]] = intervals[val][2]
        ```
        (B) ```python
            if val != float('inf'): 
                res[i[2]] = intervals[val][2]
            else:
                continue
        ```
        (C) ```python
            val = binary_search(i[1])
        		if val != float('inf'): res[i[2] + 1] = intervals[val][2]
        ```
        (D) ```python
            if val != float('inf'): 
        			  res[i[2]] = intervals[val][2]
        		else:
        		  continue
        ```
    \end{lstlisting} \\
    \begin{lstlisting}
        
        Answer: Let's think step by step. The incomplete solution first sorts the intervals and then iterates over the sorted intervals. For each interval, it finds the right interval using a binary search.
        This option (A) finds the right interval index using the binary search and updates the result array accordingly.
        The option (B) is similar to (A), but it does not increment the index when finding the right interval index. This could lead to incorrect results.
        The option (C) increments the index when finding the right interval index. However, this is incorrect because the problem statement asks for the index of the right interval, not the offset from the original index.
        The option (D) uses the same index for both the original interval and the right interval, which could lead to incorrect results.
        The answer is (A).

        Problem statement: {question}
        
        Incomplete Solution:
        {codebase}
        
        Question: The provided solution is missing a part, Which option below is the most likely to
        complete the solution and achieve the desired goal?
        
        {multiple_choices}
        
        Answer: Let's think step by step.
    \end{lstlisting} \\ \midrule
    \textbf{\textit{Code Repair:}}
    \begin{lstlisting}[language=Python,]
        The following are multiple-choice questions (with answers) about debugging a programming problem.

        Question: The implementation below is producing incorrect results. 
        Which solution below correctly identifies the bug and repairs it to achieve the desired goal?
        
        1 def two_sum(nums, target):
        2     complement_map = {{}}  
        3     for i, num in enumerate(nums):
        4         complement = target - num
        5         complement_map[num] = i
        6         if complement in complement_map:
        7             return [complement_map[complement], i]  
        8     return None
        
        (A) Remove line 5.
        
        (B) Remove line 5. Add at line 7:
        ```        complement_map[num] = i```
        
        (C) Modify line 7:
        ```         return [i, complement_map[complement]]```
        
        (D) Remove line 5. Add at line 7:
        ```     if i == len(nums) - 1:
                    return None
                complement_map[num] = i```
        
        Answer: Let's think step by step. The bug in the code occurs because the current number is added to the complement_map before checking if its complement already exists, which can lead to incorrectly matching a number with itself. To fix this, the number should only be added to the map after checking for its complement. Solution (B) does exactly this by moving the line that adds the current number to the map after the complement check, ensuring the logic works as intended without self-matching errors.
        The answer is (B).
        
        Question: The implementation below is producing incorrect results. 
        Which solution below correctly identifies the bug and repairs it to achieve the desired goal?
        {question}
        
        {choices}
        
        Answer: Let's think step by step. 
    \end{lstlisting} \\ \midrule
    \textbf{\textit{Defect Detection:}}
    \begin{lstlisting}[language=Python,]
        The following are multiple choice questions (with answers) about programming problem.
        
        Question: Given a code snippet below, which behavior most likely to occurr when execute it?
        ```python
        def chkPair(A, size, x):
            for i in range(0, size - 1):
                for j in range(i + 1, size):
                    if (A[i] + A[j] == x):
                        return 1
            return 0
        
        ```
        (A). The code contain no issue.
        (B). Memory Limit Exceeded
        (C). Compile error
        (D). Runtime Error
    
        Answer: Let's think step by step. The code appears to have no issues with typical valid inputs and will function as expected. It correctly checks for pairs of elements whose sum is x.
        The answer is (A).
        
        Question: Given a code snippet below, which behavior most likely to occurr when execute it?
        {question}
        {multiple_choices}
        
        Answer: Let's think step by step.
    \end{lstlisting} \\
    \bottomrule
\end{longtable}
}

\end{document}